\documentclass[prd,aps,amssymb,amsmath,tightenlines,nofootinbib]{revtex4}
\usepackage{graphicx,epsfig}
\usepackage{slashed}
\usepackage[colorinlistoftodos]{todonotes}
\usepackage{appendix}
\usepackage{hyperref}

\newcommand{\be}{\begin{equation}}
\newcommand{\ee}{\end{equation}}

\def\cP{$\mathcal P$}

\def\cT{$\mathcal T$}
\def\cPT{$\mathcal {PT}$}

\usepackage{subcaption}

\usepackage{tikz}
\usetikzlibrary{calc,patterns,angles,quotes}

\usepackage{amsmath}

\begin{document}

\preprint[{\leftline{KCL-PH-TH/2022-{\bf 50}}

\title{\boldmath Renormalisation group flows connecting a $4-\epsilon$ dimensional Hermitian field theory to a \cPT-symmetric theory for a fermion coupled to an axion}

\author{Lewis Croney$^{a}$}

\author{Sarben Sarkar$^a$ }

\affiliation{$^a$Theoretical Particle Physics and Cosmology, King's College London, Strand, London, WC2R 2LS, UK}

\begin{abstract}
The renormalisation group flow of a Hermitian field theory is shown to have trajectories which lead to a non-Hermitian Parity-Time (\cPT) symmetric field theory for an axion coupled to a fermion in spacetime dimensions $D=4-\epsilon$, where $\epsilon >0 $. In this renormalisable  field theory, the Dirac fermion field has a Yukawa coupling $g$ to a pseudoscalar (axion) field and there is quartic pseudoscalar self-coupling $u$. The robustness of this finding is established by considering flows between $\epsilon$ dpependent Wilson-Fisher fixed points and also by working to \emph{three loops} in the Yukawa coupling and  to \emph{two loops} in the quartic scalar coupling.
The flows in the neighbourhood of the non-trivial fixed points are calculated using perturbative analysis, together with the $\epsilon$ expansion. The global flow pattern indicates flows from positive $u$ to negative $u$; there are no flows between real and imaginary $g$. Using summation techniques  we demonstrate a possible non-perturbative $\mathcal{PT}$-symmetric saddle point for $D=3$.

\end{abstract}


\maketitle

\section{Introduction }
\label{sec:intro}

Non-Hermitian \cPT-symmetric field theories are effective theories, which may describe aspects of Beyond-the-Standard Model  physics (BSM)~\cite{R3.1, R3.2, R3.3, R3.4, R3.5, R3.6, R3.7, R3.8, R3.9, R3.10, R3.11, R3.12, R3.13, R3.14, R3.15}.  \cP ~is a linear operator (such as parity) and  \cT~is an anti-linear operator (such as time-reversal). A quantum mechanical system with unbroken \cPT-symmetry~\cite{R1,R2} has a completely real spectrum which leads to unitary dynamics \cite{R17}. Our aim is not to pursue phenomenological aspects of BSM physics, but to investigate in depth an intriguing behaviour noticed in  a recent study of a field theory developed for gravitational axion phenomenology and dynamical mass generation~\cite{R3.13,R3.14,R3a,R3c}. We noticed a renormalisation group flow~\cite{R3a} from Hermitian values of the coupling to those of a non-Hermitian but \cPT-symmetric version of the field theory in a one-loop analysis. We examine here the robustness of these findings by working with beta functions with non-zero~$\epsilon$ and by working to \emph{three loops} in the Yukawa coupling and \emph{two loops} in the quartic scalar coupling \cite{R10, Pickering:2001aq,Poole:2019kcm,Bednyakov:2021qxa,Davies:2021mnc}.
The quantum theory is performed using path integrals~\cite{Rivers:2011zz}. The issues dealing with path integrals for \cPT-symmetric theories has been studied at length recently~\cite{R3a,R3b}. 

In spacetime dimensions $D$, Hermitian quantum mechanical systems are  treated either in the language of path integrals \cite{Rivers:1987hi} or of operators acting on a Hilbert space~\cite{Bjorken:100769}. The bridge between path integrals and operator descriptions is understood for Hermitian theories \cite{Swanson:1992cz, Dowker:2010ng}. For \cPT-symmetric quantum theories in $D=1$ the observables are self-adjoint with respect to an inner product~\cite{R1,R2,R2a} which is different from the usual Dirac inner product and is specific to the theory being considered. The path integral formulation of \cPT-symmetric theories in $D=1$  has been shown in detailed examples to give the the same Green's functions \cite{Jones:2009br, R3a, Bender_2006} as the operator treatment. The general argument~\cite{Jones:2009br} justifying this in $D=1$ is extended to $D>1$ in \cite{R3a}.~In ~\cite{R3a,R3b} it was shown that the Feynman rules which describe the weak coupling behaviour of the theory around the trivial saddle point of the path integral follow just from the Lagrangian of the theory and produce the correct asymptotoic series at weak coupling of the theory.
%Most work on the quantum aspects of \cPT symmetry is in $D=1$ spacetime dimension and uses the Schr\"odinger equation. For BSM we require $D=4$ and a field theoretic formulation that incorporates renormalisation.~Renormalisable field theory in $D=4$ is not treated with a functional generalisation of the Schr\"odinge but rather with path integrals. }

An early example providing an indication that a Hermitian field theory, when renormalised, may need a reinterpretation as a \cPT-symmetric field theory~\cite{R5, R4} is provided by the Lee model~\cite{R6.1}. The Lee model has been solved explicitly in $D=1$ and $D=4$. It has mass, wave function and coupling constant renormalisation in $D=4$. However, the model does not have crossing symmetry and the particles in the model do not obey the spin-statistics theorem~\cite{Streater:1989vi}.  An important feature of the model is that the bare coupling has a square root singularity in terms of the renormalised coupling.~This nonanalyticity leads to ghost states in  a conventional interpretation.~In a \cPT-symmetric interpretation the Hamiltonian is self-adjoint with respect to a different inner product~\cite{R5}. A second example is the emergence of unstable but \cPT-symmetric effective potential for the Higgs field in the Standard Model (discussed in a $D=1$ approximation~\cite{R3a}). This effective potential arises from renormalised one-loop effects~\cite{R7.4,Isidori:2001bm}.
%A third example is a preliminary  analysis~\cite{R3a} of renormalisation group flow between a Hermitian and a \cPT-symmetric version of a model (in $D=4$), which arises in the study of dynamical mass generation of axions.~The model of the third example is studied in great detail in this paper.

It is known that there is an asymptotic  weak coupling perturbation theory~\cite{R3a,R3b} of a \cPT-symmetric field theory in $D=4$. The key to this is the existence of path integrals in \cPT-symmetric theories, which are steepest descent paths and are associated with boundary conditions on the complex-valued paths or Lefscchetz thimbles~\cite{Behtash:2015loa, R32w}  used in the path integral. 
%A rigorous definition of path integrals does not exist even for Hermitian field theories~\cite{Albeverio:1976je,Swanson:1992cz}.~The $D=0$ case illustrates the form of the complex deformation of contours and the associated \emph{steepest descent} contours (passing through saddle points) necessary to define \cPT-symmetric quantum field theories. The $D=1$ case illustrates the nature of \cPT-symmetric paths in function space and functional saddle points in a semiclassical analysis.
When we come to consider $D=4$, we have the additional issues of regularisation and renormalisation associated with Feynman perturbation theory around the trivial saddle point. Dimensional regularisation with $D=4-\epsilon$, where $\epsilon>0$~enables the study of Wilson-Fisher fixed points~\cite{R11}. Flow between such fixed points remain perturbatively small because $\epsilon$ is small.
We consider a renormalisable field-theory for axion physics, which is a massive Yukawa model~\cite{R3.13,R3.14} and is also one of the simplest renormalisable field theories~\cite{R14a}. The interaction terms have a conventional form but can be tuned to have values which render the QFT no longer Hermitian, but still \cPT-symmetric (as in \cite{R5}). The model provides a framework for studying the interplay of renormalisation and \cPT~symmetry in the presence of a fermion and a pseudoscalar near four dimensions.~Unlike the Lee model~\cite{R6.1, R6.2} this model is a conventional crossing-symmetric field theory.
%~One of the early successes of the \cPT-symmetric approach addressed renormalisation of the field-theoretic Hermitian Lee model~\cite{R6.1, R6.2} at strong coupling, which led to a non-Hermitian but \cPT-symmetric \cite{R5,R4} theory
%\footnote{A different non-perturbative approach to renormalisation based on path integrals has also recently appeared \cite{R4, branchina}.However this approach is not promising since the} .
%The Hermitian Lee model is not a conventional crossing-symmetric field theory and so the significance of the emergence of $\mathcal{PT}$ symmetry as a result of renormalisation is not clear in a wider context\footnote{More recently this issue has been discussed in toy one-dimensional models based on effective theory~\cite{R8} using quantum mechanics.}. 
Our principal aim is to understand,~in a \emph{controlled} way, the interplay of renormalisation and $\mathcal {PT}$ symmetry in a relativistic four-dimensional QFT model, starting with a~\emph{Hermitian} theory.
%it is necessary to introduce two tools: the $\epsilon$ expansion~\cite{R11} in $D=4-\epsilon$ spacetime dimensions, and the perturbative renormalisation group~\cite{Hollowood:2009eh} which are developed for Hermitian theories.  When \cPT symmetry emerges as a result of renormalisation we need to understand in what sense the theory is a QFT.
The massive Yukawa model we consider is given by the bare Lagrangian~\cite{R3a} in $3$-space and $1$-time dimensions in terms of bare parameters with subscript $0$ \footnote{Our Minkowski-metric signature convention is $(+,-,-,-)$.}

\be
\label{e1}
{\mathcal{L}}= \frac{1}{2} \partial_{\mu} \phi_0 \partial^{\mu} \phi_0 -
\frac{M_{0}^2}{2} \phi_{0}^2 + {\bar{\psi}}_{0} \left( i \slashed{\partial} - m_{0} \right)
\psi_0 - i g_0 \bar{\psi}_0 \gamma^5 \psi_0 \phi_0 - \frac{u_0}{4!} \phi_0^4.
\ee
$\mathcal{L}$ is renormalised in four dimensions through mass, coupling constant and wavefunction renormalisations; the scalar self-interaction is obtained from continuation of $\delta$ to 2 in the manifestly \cPT-symmetric deformation~\cite{R1,R2}
\begin{align}\label{ptnh}
\frac{u_0}{4!} \, \phi_{0}^{2}{(i\phi_{0})}^{\delta}~,
\end{align}
 for $u_0, \delta>0$, in any spacetime dimension $D$. To be clear, the parameter being continued  is $\delta$ and not $u_0$; this is essential for \cPT~symmetry as will become clear when the reality of path integrals is considered.
%because we will use dimensional regularisation we will consider the spacetime dimensionality $D$  to be $4-\epsilon$, where $\epsilon >0$ is a small parameter. 
This is the simplest non-trivial renormalisable
model of a Dirac fermion field $\psi_0$ interacting with a pseudoscalar field $\phi_0$. In the Dirac representation of $\gamma$ matrices the standard discrete  transformations~\cite{R20} on $\psi_0$ are
\begin{align}\label{pandt}
\mathcal{P}\psi_0 (t,\vec{x})\mathcal{P^{\it{-1}}}=\gamma^{0}\psi_0 (t,-\vec{x}), \quad \mathcal{T}\psi_0 (t,\vec{x})\mathcal{T^{\it{-1}}}=i\gamma^{1}\gamma^{3}\psi_0(-t,\vec{x}), 
%\quad \mathcal{C} \psi \left(t,\overrightarrow{x}\right) \mathcal{C}^{-1}=i\gamma^{2}\psi^{\dagger}\left(t,\overrightarrow{x}\right).
\end{align}
 $\mathcal{T}$ is an anti-linear operator. Moreover, under the action of $\mathcal P$ and $\mathcal T$, the pseudoscalar field $\phi_{0}\left(t,\vec{x}\right)$ transforms as
 \begin{equation}
\label{par}
 {\mathcal P}\phi_{0} \left( t,\vec{x} \right)  \mathcal P^{-1}=-\phi_{0} \left( t,-\vec{x} \right)  ,\  \  \ {\mathcal T}\phi_{0} \left( t,\vec{x} \right)  \mathcal T^{-1}=\phi_{0} \left( -t,\vec{x} \right).
 \end{equation}
These definitions go through in $D$ dimensions with the Dirac gamma algebra given in \eqref{l1}. In dimensional regularisation, expressions for Green functions from covariant perturbation theory, which are valid for integer $D$, are analytically continued in $D$~\cite{RevModPhys.47.849}. Lorentz covariants  such as $\gamma_{\mu } ,p_{\mu },g_{\mu \nu }$ are treated as formal entities~\cite{Breitenlohner:1977hr} that obey prescribed algebraic identities. The specific values of indices are not used\footnote{These calculations differ from those required for the energy eigenvalues of a Dirac equation in general integer dimensions where the explicit representations of the gamma matrices are used.}. However the definition of $\gamma_{5}$ requires special consideration (see \ref{DimReg}). 

If $g_0$ is real, then the Yukawa term in \eqref{e1} is Hermitian and $g_{0}^{2}>0$. If $g_0$ is imaginary, then the Yukawa term is non-Hermitian but is \cPT-symmetric and so $g_{0}^{2}<0$. $u_0$ is real but it can be positive (Hermitian) or negative (\cPT-symmetric). 

 The plan of this paper is as follows:
\begin{enumerate}
    \item We briefly review the role of renormalisability in $\mathcal{PT}$-symmetric quantum field theory and the subtleties in defining the corresponding path integrals \cite{R3a,R3b, R2}. In particular we note:
\begin{itemize}
    \item In the Lee model~\cite{R6.1,R6.2, R4}, a model of historical importance in the study of renormalisation, the bare coupling has a non-analytic dependence on the renormalised coupling. Moreover, the non-analyticity is in terms of a branch cut. 
   %in D=1 for our Yukawa model. This is an example of the behaviour found discussed in \cite{R4}. Our thesis that a Hermitian theory may through renormalisation lead to nonHermitian \cPT symmetric model will be illustrated by the progenitor quantum mechanical Lee model.
   %which will be introduced. The  in the first instance The possibility that renormalisation may lead froma Hermitian theory to a nonHermitan We will motivate our model through a study of a progenitor, a version of the Lee model. 
   The Lee model is a quantum mechanical Hermitian model which allows for (an exact treatment of) renormalisation starting with  a Hilbert space  with the conventional Dirac inner product. The well-known ghost problem~\cite{R5}, which develops due to renormalisation, is removed by interpreting the model with a new inner product related to the $C$ operator of $\mathcal{PT}$ symmetry \cite{Bender:2004zz}.

\item
In order to understand \cPT-symmetric path integrals  it is instructive to consider $D=0$ \cPT-symmetric integrals using standard complex analysis techniques. The related analysis of $D\ge 1$ can be found in~\cite{R3a, R3b}.  The presence of fermions does not change this analysis qualitatively since massive fermions can be integrated out (at one loop) to give an effective potential contribution \cite{Coleman:1973jx, Ellis:2020ivx, Manohar:2020nzp} to the scalar self-interaction, in terms of logarithmic factors.

\end{itemize}

\item 
Perturbation theory using Feynman diagrams is applied to the Yukawa model. This gives an asymptotic series in the couplings that is valid near the trivial saddle point. The contributions from the non-trivial saddle points (due to bounces) are asymptotically subdominant in the weak coupling limit~\cite{R21}. However, the bounce (instanton) solutions give rise to imaginary contributions to odd point Green's functions which would otherwise vanish~\cite{R3b, R21}. Hence our approach, which ignores the subdominant contributions from non-trivial saddle points, is based on perturbation theory around the trivial saddle point, which is valid for renormalisation group flows around all sufficiently weak-coupling fixed points. We also comment on the subtleties of using dimensional regularisation in non-integer dimensions. Using a general purpose Mathematica program RGBeta~\cite{R10}, the perturbation theory is performed to three loops in $g$ and two loops in $u$. RGBeta has the feature that it also accepts complex couplings. Beta functions of the renormalisation group flow~\cite{R14a, Hollowood:2009eh} can be calculated. We solve for the fixed points and determine their stability. Going from $\epsilon=0$ to non-zero $\epsilon$ leads to the trivial fixed point spawning three new $\epsilon$-dependent fixed points, whose magnitudes are directly controlled by $\epsilon$. Furthermore, the flow in the neighbourhood of the fixed points is joined together to give a more global flow picture. From this picture, we can see how the Hermitian and non-Hermitian fixed points interact with each other i.e. how the flow is organised around these fixed points. For one \emph{non-Hermitian} fixed point the $\epsilon$ expansion is stable, i.e. the coefficients do not increase rapidly with order, so resummation techniques using Pad\'e approximants leads to a genuine fixed point in $D=3$, which is not sensitive to variations in the form of Pad\'e approximants used. This fixed point has the stability of a saddle point.

\item We examine some aspects of applying finite loop-order perturbation theory, and compare our model to that presented in \cite{R9}, where similar analysis is performed. 

\item In the conclusions we discuss and summarise our results. Furthermore, there are appendices giving some additional details on our findings; we give some checks of robustness of our main results related to the effects of finite loop order in perturbation theory.
\end{enumerate}

\section{The Lee model} 
The Lee model (LM)  is a class of soluble simplified field theories~\cite{R6.1} used to study  renormalisation, which can be carried out exactly.~LM\footnote{A  $D=1$ version of the Lee model suffices to show the essential effect of renormalisation present in the $D=4$ model~\cite{R5}.} involves fermionic particles $N$ and $V$ with operators $\psi_{N}$ and $\psi_{V}$ and a bosonic particle $\theta$ with operator $a$ (in $D=1$). The interactions in the model allow
\be
\label{E2}
V\rightarrow N+\theta
\ee
and also  the reverse process
\be
\label{E3}
N+\theta \rightarrow V.
\ee
Because the field theory does not have crossing symmetry the process $N\rightarrow V+\bar{\theta } $ is not allowed where~$\bar{\theta }$ is the antiparticle of $\theta$.~The fermions $N$ and $V$ do not have spin and so the spin-statistics theorem~\cite{Streater:1989vi} is not satisfied. 
The interactions imply conservation rules for $B$ and $Q$ where
\begin{itemize} 
    \item$ B=n_{N}+n_{V}$
    \item $Q=n_{V}-n_{\theta },$
\end{itemize}
and $n_{N}$, $n_{V}$ and $n_{\theta }$ are the number of quanta for $N, V$ and $\theta$ respectively. This simplification facilitates the ability to solve the model~\cite{R5}. In $D=1$ the Hamiltonian $\mathcal{H}$  is $\mathcal {H}=\mathcal{H}_{0}+\mathcal{H}_{1}$
where 
\be
\label{E4}
\mathcal{H}_{0}=m_{V}\psi^{\dag }_{V}\psi_{V}+m_{N}\psi^{\dag }_{N}\psi_{N}+\mu a^{\dag }a
\ee
and 
\be
\label{E5}
\mathcal{H}_{1}=\delta m_{V}\psi^{\dag }_{V}\psi_{V}+g_{0}\left( \psi^{\dag }_{V}\psi_{N}a+a^{\dag }\psi^{\dag }_{N}\psi_{V}\right).  
\ee
The sector with $B=1$ and $Q=0$ is spanned by the states $|1,0,0\rangle \  {\rm and}\  |0,1,1\rangle $. The eigenstates of $\mathcal H$ are denoted by $|V\rangle \  {\rm and}\  |N\theta \rangle $, with associated eigenvalues $m_V$ and $E_{N \theta}$ given by 

\be
\begin{split}
m_{V} & =\frac{1}{2} \left( m_{N}+\mu +m_{V_{0}}-\sqrt{M^{2}_{0}+4g^{2}_{0}} \right) \nonumber \\
E_{N \theta} & =\frac{1}{2} \left( m_{N}+\mu +m_{V_{0}}+\sqrt{M^{2}_{0}+4g^{2}_{0}} \right)
\end{split}
\ee
where $M_{0} \equiv m_{N}+\mu-m_{V_{0}}$ and $m_{V_{0}}\equiv m_{V}+\delta m_{V}$. The wave-function renormalisation constant $Z_V$ is determined~\cite{R5} through the relation
\be
\label{E6}
\sqrt{Z_{V}} =\langle 0|\psi_{V} |V\rangle 
\ee
which leads to~\cite{R5}
\be
\label{E7}
Z_{V}=\frac{2g^{2}_{0}}{\sqrt{M^{2}_{0}+4g^{2}_{0}} \left( \sqrt{M^{2}_{0}+4g^{2}_{0}} -M_{0}\right)  } .
\ee
The renormalised coupling constant $g$ satisfies 
\be
\label{ren1}
g^{2}=Z_{V}g^{2}_{0}.
\ee
In terms of $M \equiv m_{N}+\mu-m_{V}$, a renormalised quantity, it is straightforward to see that 
\be
\label{ren2}
M_{0}=M-\frac{g^{2}_{0}}{M} .
\ee
From \eqref{ren1} and \eqref{ren2} we can deduce the non-perturbative result that 
\be
g^{2}_{0}=\frac{{}g^{2}}{\left( 1-\frac{g^{2}}{M^{2}} \right)  } .
\ee
Hence $g_{0}$ is related to $g$ by a square root singularity with a branch cut between $-M$ and $M$.
If $g^{2}>M^{2}$, then the bare coupling can become imaginary and the Hamiltonian is non-Hermitian, but \cPT-symmetric~\cite{R5}. Explicitly the transformations 
due to $\mathcal{P}$ are
\be
\label{E8}
\begin{matrix}\mathcal{P}V\mathcal{P}=-V&\  \mathcal{P}N\mathcal{P}=-N&\  \mathcal{P}a\mathcal{P}=-a\\ \mathcal{P}V^{\dag }\mathcal{P}=-V^{\dag }&\  \mathcal{P}N^{\dag }\mathcal{P}=-N^{\dag }&\  \mathcal{P}a^{\dag }\mathcal{P}=-a^{\dag }\end{matrix} 
\ee
and due to $\mathcal{T}$ are
\be
\label{E9}
\begin{matrix}\mathcal{T}V\mathcal{T}=V&\  \mathcal{T}N\mathcal{T}=N&\  \mathcal{T}a\mathcal{T}=a\\ \mathcal{T}V^{\dag }\mathcal{T}=V^{\dag }&\  \mathcal{T}N^{\dag }\mathcal{T}=N^{\dag }&\  \mathcal{T}a^{\dag }\mathcal{T}=a^{\dag }.\end{matrix} 
\ee

The non-Hermiticity of the Hamiltonian leads to states with energies that are not real. Because of the $\mathcal{PT}$-symmetry, a new inner product was constructed for the Hilbert space which removed ghost states from the spectrum \cite{R5}\footnote{An analogue version of the Lee model in the nonHermitian region has also been proposed±~\cite{PhysRevA.85.012112}.}. The Lee model has some similarities with $\mathcal{L}$ in~\eqref{e1}.
The massive Yukawa model has the trilinear interaction between fermions and bosons as in the Lee model but it has also a quartic boson self-interaction. It has crossing symmetry and the spin-statistics connection, features which are essential for any realistic fundamental theory.  \cPT~symmetry in the Lee model emerges for a non-weak coupling strength. Non-Hermiticity in the massive Yukawa model occurs for small couplings and hence is amenable to a renormalisation group analysis.

 \section{\cPT-symmetric path integrals} 
In the modern study of  field theory, quantum aspects can be explored through path integrals where the Hilbert space structure is not paramount \cite{R14a}. In non-Hermitian (but $\mathcal{PT}$-symmetric) field theory, this advantage persists and simplifies calculations at weak coupling \cite{Bender_2006}. We  concentrate on the modification in $D=0$ of paths for the existence of path integrals in \cPT-symmetric framework. The discussion of semi-classical analysis and steepest descent paths  can be found in \cite{R3a, R3b}.

 \par  We shall focus on the bosonic part of the path integral for $\mathcal{L}$~\cite{R3a}\footnote{The fermions in the model give a logarithmic correction to the quartic self-interaction when integrated out \cite{Coleman:1973jx} of the path integral and does not cause a significant change in the discussion.}.  and consider two forms of the bosonic path integral, one which preserves manifest $\mathcal{PT}$ symmetry and the other which does not 
  \be
 \label{E10}
Z_{i}=\int \mathcal{D}\phi \  \exp \left( -S_{i}\left[ \phi \right]  \right),  \quad i=1,2
\ee
where $\mathcal{D}\phi$ is the path integral measure and the action is given by
  \be
  \label{E11}
S_{i}\left[ \phi \right]  =\int d^{D}x\,\left( \frac{1}{2}  \left( \partial_{\mu } \phi \right)^{2}  +V_{i}\left( \phi \right)  \right)  
\ee
and 
\begin{align}
\label{E11a}
   V_{1}\left( \phi \right) &=\frac{1}{2}m^2\phi^2+\frac{u}{4} \phi^{2} \left( i\phi \right)^{\delta }, \\
   V_{2}\left( \phi \right)  &=\frac{1}{2}m^2\phi^2+\frac{u}{4} \exp{i\zeta}\,\phi^4
\end{align}
where we consider monotonic continuations in the parameters, with $\delta \rightarrow 2$ in the first case and $\zeta \rightarrow \pm \pi$ in the second case. In both cases we need the path integral to converge and the contours of paths have to be chosen appropriately. Although the limiting form of $V_i$ in the parameter continuations are 
\be
V\left( \phi \right)  =\frac{1}{2}m^2\phi^2 - \frac{u}{4} \phi^{4} 
\ee
the contours required with the different deformations are \emph{distinct} and we will see that $Z_{1}\ne Z_{2}$ in their imaginary parts. The first deformation is \cPT-symmetric whereas the second deformation is not since under \cP~and \cT~we require
\begin{itemize}
  \item $\mathcal{P}:\  \phi \longrightarrow -\phi$\,;
  \item $\mathcal{T}:\  \phi \longrightarrow \phi^{\ast }\,,\ \{\it i \longrightarrow -{\it i }$\}.\label{Ttransfm}
\end{itemize}
The $\delta$ deformation is central to \cPT~symmetry. We shall show that the $\delta$ deformation keeps the partition function real while the coupling deformation leads to a $Z$ with imaginary parts.

\subsection{$D=0$}

We consider the $D=0$ case\footnote{This case is an example of a trivial field theory at a single spacetime point. It is useful in understanding the nature of the deformations which are necessary to have a \cPT-symmetric path integral.}
to illustrate the importance of \cPT-preserving deformations. Then we have
\be
\label{E12}
Z_{1}=\int_{{\mathcal{C}}}
 {\rm d}z \exp\left(-\left(\frac{1}{2}m^2 z^2 + \frac{u}{4} z^{2}\left( iz\right)^{\delta }  \right)\right)
\ee
where $\mathcal{C}$ is a contour in the complex plane, which is a deformation of the real line interval $\left( -\infty,\infty \right) $ such that $Z$ is finite and $\phi$ has been replaced by the variable $z$. The path integral has become an integral whose convergence is determined by the term proportional to $u$. On writing  $z=r\exp(i \chi)$ we have
\be
\label{E13}
z^{2}\left( iz\right)^{\delta }  =r^{2+\delta}\exp \left( i\left[ 2\chi +\delta \left( \chi +\frac{\pi }{2} \right)  \right]  \right)  
\ee
and the integral for $Z$ converges when 
\be
\label{E14}
2n\pi -\frac{\pi }{2} <\left( \delta +2\right)  \chi +\delta \frac{\pi }{2} <\frac{\pi }{2} +2n\pi 
\ee
where $n$ is an integer defining Stokes wedges which defines an opening in $\chi$
\be\chi_{l} <\chi <\chi_{u}  \ee
where $\chi_{u} =\frac{\frac{\pi }{2} (1-\delta )+2n\pi }{\delta +2} $ and $\chi_{l} =\frac{2n\pi -\frac{\pi }{2} (1+\delta )}{\delta +2} .$ There are four distinct wedges labelled by $n=0,1,2,3$. The $n=0$ and $n=3$ form a \cPT-symmetric set. By Cauchy's theorem, any contour in a wedge is equivalent to any other in its contribution to the integral. Our choice will be to take the contour through the \emph{centre} of the wedge. We shall call this particular contour ${\mathcal{C}}_{PT}$, see Figure \ref{Tikz1}. 

\begin{figure}[!h]
\centering
\begin{tikzpicture}

\fill[fill=blue!40!white] (0,0) -- (5,-2.07) -- (5,-5) -- (2.07,-5) -- cycle;
\fill[fill=blue!40!white] (0,0) -- (-5,-2.07) -- (-5, -5) -- (-2.07,-5) -- cycle;
\draw[green, ultra thick] (5,-5) -- (0,0);
\draw[green, ultra thick] (-5,-5) -- (0,0);
\draw[green, ultra thick,->] (0,0) -- (2.5,-2.5);
\draw[green, ultra thick,->] (-5,-5) -- (-2.5,-2.5);
\draw[red, thick, dashed] (5,-2.07) -- (0,0);
\draw[red, thick, dashed] (-5,-2.07) -- (0,0);
\draw[red, thick, dashed] (2.07,-5) -- (0,0);
\draw[red, thick, dashed] (-2.07,-5) -- (0,0);
\draw (-2.2,-2.8) node {$\mathcal{C}_{\mathcal{PT}}$};

\coordinate (a) at (0,0);
\coordinate (b) at (5,-2.07);
\coordinate (c) at (2.07,-5);
\draw pic[draw, ->, "$\pi/4$" shift={(-3mm,3mm)}, angle radius=2cm, angle eccentricity=1.5] {angle = c--a--b};

\coordinate (d) at (5,-5);
\draw pic[draw, ->, "$\pi/8$" shift={(-0.8mm,0.8mm)}, angle radius=1cm, angle eccentricity=1.5] {angle = d--a--b};

\draw[thick,->] (-5.0,0) -- (5.0,0) node[anchor=north] {$\textrm{Re}(z)$};
\draw[thick,->] (0,-5.0) -- (0,5.0) node[anchor=east] {$\textrm{Im}(z)$};
\end{tikzpicture}
\caption{Stokes wedges (shown in blue, boundaries in red) and contour $\mathcal{C}_{\mathcal{PT}}$ (shown in green) for $\delta \rightarrow 2$ in $Z_1$.} \label{Tikz1}
\end{figure}
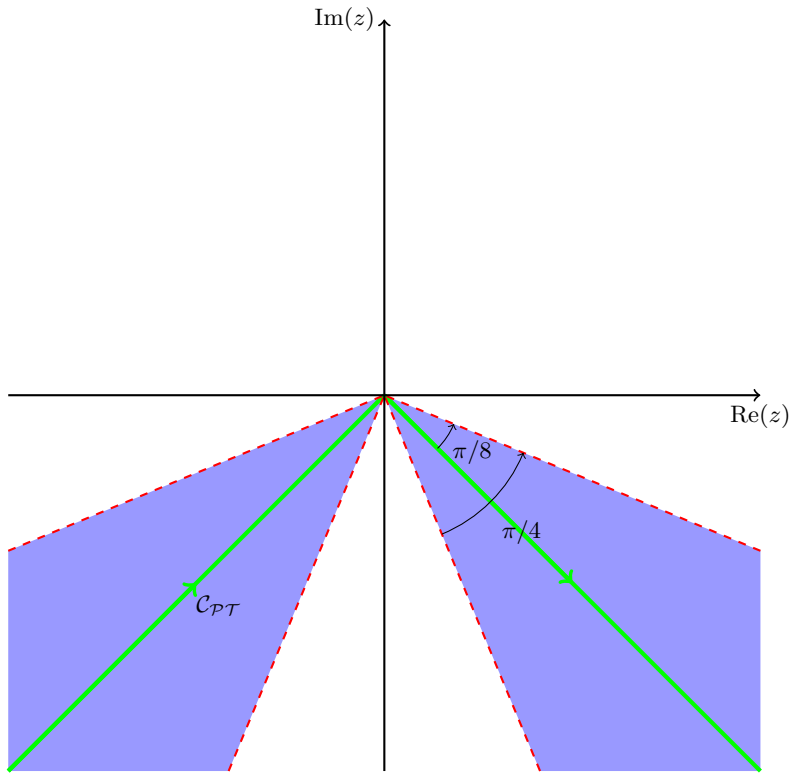

It is convenient to rescale $z \to z/\sqrt{u}$, for the case $\delta=2$, which leads to 
\be
\label{E17}
Z_{1}=\int_{\mathcal{C}_{PT}} dz\exp \left( -\frac{1}{u} \left[ \frac{1}{2} m^{2}z^{2}-\frac{z^{4}}{4} \right]  \right) . 
\ee
We will now evaluate $Z_1$ over the $\mathcal{C}_{\mathcal{PT}}$ contour (for $\delta = 2$) to show that it is real. We find
\begin{align}
  Z_1&=\exp \left( -\frac{i\pi }{4} \right)  \int^{\infty }_{0} dr\left[ \cos \left( \frac{m^{2}\  r^{2}}{2u} \right)  +i\sin \left( \frac{m^{2}\  r^{2}}{2u} \right)  \right]  \exp \left( -\frac{r^{4}}{4u} \right)  + c.c. \\ 
   &=\frac{m\pi }{{}2^{\frac{3}{2} }} \exp \left( -\frac{m^{4}}{8u} \right)  \left( I_{-\frac{1}{4} }\left( \frac{m^{4}}{8u} \right)  +I_{\frac{1}{4} }\left( \frac{m^{4}}{8u} \right)  \right).
    \end{align}
where $c.c.$ refers to complex conjugation and the $I_{\nu}(x)$ are the modified Bessel functions of the first kind. $Z_{\mathcal{C}_{PT}}$  is real and has a nonzero small $u$ expansion since $I_{\nu }\left( x\right)  \sim \frac{\exp (x)}{\sqrt{2\pi x} } \left[ 1-\frac{4\nu^{2} -1}{8x} \right]$ as $x \to \infty$ and the exponential pieces cancel.

\vspace{.1cm}
We will compare with the $D=0$ version of $Z_{2}$, given by
\be
\label{E15}
Z_{2}=\int_{\mathcal{C}} \exp \left( -\frac{1}{2} m^{2}z^{2}-\frac{u}{4}e^{i\zeta }z^{4}\right)  dz
\ee
Similarly, we let $z=re^{i\theta }$. The integral in $r$ converges if 
\be
\label{E16}
-\frac{\pi }{8} \left( 4n+1\right)  -\frac{\zeta }{4} <\theta <\frac{\pi }{8} \left( 1-4n\right)  -\frac{\zeta }{4} 
\ee
The distinct Stokes wedges are for $n=0$ and $n=1$ when $\zeta=\pi$. This wedge pair is not \cPT-symmetric. We shall call this particular contour ${\mathcal{C}}_{\textrm{rotation}}$, see Figure \ref{Tikz2}. The Hermitian case is $\zeta=0$ and $\mathcal{C}=\left( -\infty ,\infty \right)$.

\begin{figure}[!h]
\centering
\begin{tikzpicture}

\fill[fill=blue!40!white] (0,0) -- (5,-2.07) -- (5,-5) -- (2.07,-5) -- cycle;
\fill[fill=blue!40!white] (0,0) -- (-5,2.07) -- (-5, 5) -- (-2.07,5) -- cycle;
\draw[green, ultra thick] (5,-5) -- (0,0);
\draw[green, ultra thick] (-5,5) -- (0,0);
\draw[green, ultra thick,->] (0,0) -- (2.5,-2.5);
\draw[green, ultra thick,->] (-5,5) -- (-2.5,2.5);
\draw[red, thick, dashed] (5,-2.07) -- (0,0);
\draw[red, thick, dashed] (-5,2.07) -- (0,0);
\draw[red, thick, dashed] (2.07,-5) -- (0,0);
\draw[red, thick, dashed] (-2.07,5) -- (0,0);
\draw (-2.0,3) node {$\mathcal{C}_{\textrm{rotation}}$};

\coordinate (a) at (0,0);
\coordinate (b) at (5,-2.07);
\coordinate (c) at (2.07,-5);
\draw pic[draw, ->, "$\pi/4$" shift={(-3mm,3mm)}, angle radius=2cm, angle eccentricity=1.5] {angle = c--a--b};

\coordinate (d) at (5,-5);
\draw pic[draw, ->, "$\pi/8$" shift={(-0.8mm,0.8mm)}, angle radius=1cm, angle eccentricity=1.5] {angle = d--a--b};

\draw[thick,->] (-5.0,0) -- (5.0,0) node[anchor=north] {$\textrm{Re}(z)$};
\draw[thick,->] (0,-5.0) -- (0,5.0) node[anchor=east] {$\textrm{Im}(z)$};
\end{tikzpicture}
\caption{Stokes wedges (shown in blue, boundaries in red) and contour $\mathcal{C}_{\textrm{rotation}}$ (shown in green) for $\zeta \rightarrow \pi$ in $Z_2$.} \label{Tikz2}
\end{figure}

On consideration of $Z_2$ for the $\zeta=\pi$ wedge pair, we find that it is complex
   \be
   \label{E18}
   Z_{2}=\frac{m\pi }{2\sqrt{2} } \left( 1-i\right)  \exp \left( -\frac{m^{4}}{8u} \right)  \left[ I_{-1/4}\left( \frac{m^{4}}{8u} \right)  +i I_{1/4}\left( \frac{m^{4}}{8u} \right)  \right] . 
   \ee
We therefore see how the choice of contours is crucial for defining a \cPT-symmetric theory and ensuring that the path integrals are real.

Furthermore, we note that Green's functions for odd functions of $\phi$ are purely imaginary in the $\mathcal{PT}$-deformed theory, which is characteristic of $\mathcal{PT}$ symmetry. Explicitly we have
   \be
   \label{E19}
   \left< z^{2n+1}\right>  =i^{n+1}\left\{ \int^{\infty }_{0} dr\  r^{2n+1}\exp \left( -\frac{r^{4}}{4 u} \right)  \left[ \left( -1\right)^{n+1}  \exp \left( i\frac{m^{2}r^{2}}{2 u} \right)  -\exp \left( -i\frac{m^{2}r^{2}}{2 u} \right)\right]  \right\}  
   \ee
   where $n=0,1,2, \cdots$. These integrals can be written in terms of modified Bessel functions.
   
 The partition function and Green's functions cannot be calculated exactly for $D>0$. However, we are interested in weak coupling expansions of the \cPT~field theories. A way of analysing weak coupling expansions of partition functions is through a saddle point analysis of the path integral which is discussed in~\cite{R3a, R3b}.
 We have defined path integrals in~\cite{R3a, R3b} appropriate for $\mathcal{PT}$ symmetry in weak coupling using the method of steepest descents. The formal arguments have been illustrated in a specific case~\cite{Jones:2009br, Bender_2006} where the 
Hamiltonian is
\be
\label{E35}
H=\frac{1}{2} \left( p^{2}+x^{2}\right)  +i\lambda x^{3}
\ee
and Greens functions are also calculated using operator methods. The two methods agree for $D=1$.
The findings of this concrete calculation have been supported more generally by an  argument for  $D=1$~\cite{Jones:2009br} (based on the Schwinger construction~\cite{Rivers:1987hi} of the partition function in the operator theory). It was also stated in \cite{Jones:2009br}, without an explicit proof, that the arguments go through for $D>1$. The details of the generalisation for $D>1$ are given in~\cite{R3a}.

%%%%%%%%%%%%%%%%%%%%%%%%%%%%%%%%%%%%%%%%%%%%%%%%%%%%%
\section{The Yukawa model} \label{YukawaSec}
We have the basis for applications of path integral quantisation to our \cPT-symmetric model. The path integral is defined using complex deformation of paths or thimbles in complex Morse theory~\cite{R32w, Witten:2010cx, Behtash:2015loa} which ensures that the integral converges. In $D=4$ a closely related path-integral method was used to study false vacuum decay in \cite{R34cc, PhysRevD.15.2929}. The feature missing from these earlier treatments is the requirement of $\mathcal{PT}$ symmetry.

We are interested in the leading small coupling \emph{asymptotic} expansion~\cite{R15} using Feynman rules for the Yukawa model. The perturbation expansion around the trivial saddle point needs regularisation and renormalisation because of well-known infinities of loop Feynman diagrams~\cite{R14a}. The regularisation is achieved by going to $D=4-\epsilon$ where $\epsilon>0$, i.e. by using the method of dimensional regularisation~\cite{RevModPhys.47.849}. The renormalisation is achieved through minimal subtraction.

\subsection{Dimensional regularisation in scalar/fermionic theories} \label{DimReg}
Although dimensional regularisation is a well-established technique, there are subtleties such as the consistent treatment of chiral anomalies and evanescent operators~\cite{DiPietro:2017vsp} in $D$ dimensions. These, however,  have been well investigated~\cite{Jegerlehner:2000dz, Breitenlohner:1977hr}.

For our application, since we are not dealing with chiral \emph{gauge} theories, the procedures we adopt are mathematically consistent. For Hermitian theories it is generally accepted that the continuation in dimension \emph{preserves} unitarity and causality. Our treatment of \cPT~theories involves an analytic continuation in the coupling or in a deformation parameter in the scalar self-interaction. Moreover we are following a flow from a Hermitian theory to a non-Hermitian theory and so we assume that our conclusions about flow to non-Hermitian theories is unaffected by subtle issues in dimensional regularisation.

The validity of the quantum action principle~\cite{cmp/1103900439} within the framework of dimensional regularisation allows the study of symmetries  of Greens functions.~The consequences of symmetries such as Lorentz and gauge invariance are preserved.~Non-anomalous symmetry breaking is removed by the use of evanescent operators. Explicitly for vector gauge theories, gauge invariance is preserved by dimensional regularisation~\cite{THOOFT1972189}. 

From the early days of dimensional regularisation it was noticed that it is impossible to require the relations 
\begin{align}
    \left\{ \gamma_{\mu } ,\gamma_{\nu } \right\}  &=2g_{\mu \nu },\;\;\mu=1, \cdots ,D \label{l1}\\
    \left\{ \gamma_{5} ,\gamma_{\nu } \right\}  &=0,\;\;\mu=1, \cdots ,D \label{l2}
\end{align}
since they imply
\be
Tr\left( \gamma_{5} \gamma_{\mu } \gamma_{\nu } \gamma_{\rho } \gamma_{\sigma } \right)  =0,\:\: D\ne 0, 2,4.
\ee
This result cannot be continued to $D=4$ where
\be
Tr\left( \gamma_{5} \gamma_{\mu } \gamma_{\nu } \gamma_{\rho } \gamma_{\sigma } \right)  =4\epsilon_{\mu \nu \rho \sigma }. 
\ee
We follow the resolution proposed by 't Hooft and Veltman~\cite{THOOFT1972189} by defining 
\be
\label{gamma5}
\gamma_{5} =\frac{1}{4!} \epsilon_{\mu_{1} \mu_{2} \mu_{3} \mu_{4} } \gamma_{\mu_{1} } \ldots \gamma_{\mu_{4} } 
\ee
where the indices take values in $(0, 1, 2, 3)$. This ensures the validity of \eqref{gamma5}; however now 
\begin{gather}
\left\{ \gamma_{5} ,\gamma_{\mu } \right\}  =0, \; \; \mu=1, \cdots, 4 \\
\left[ \gamma_{5} ,\gamma_{\mu } \right]  =0, \;\;\mu =5,\ldots ,D
\end{gather}
This scheme is algebrically consistent. The work in \cite{Breitenlohner:1977hr} has shown that Ward identities are preserved, at least when chiral gauge theories are not involved\footnote{Even for chiral gauge theories the scheme can be modified with nongauge invariant finite counterterms~\cite{Bonneau:1990xu}}. This is the relevant situation for us; for our Yukawa model different schemes of dimensional regularisation have been explcitly shown to be consistent \cite{Schubert:1989xu}. 

\subsection{Renormalisation of the Yukawa model}
 
 Corresponding to the bare Lagrangian of the Yukawa model, the associated renormalised Lagrangian~$\mathcal{L}$ (in terms of  renormalised parameters  without the subscript $0$ and with counterterms) is 
 \begin{eqnarray}
\mathcal{L}&=& \frac{1}{2} (1 + \delta Z_{\phi}) \partial_{\mu} \phi
\partial^{\mu} \phi - \frac{M_0^2}{2} (1 + \delta Z_{\phi}) \phi^2 + (1 +
\delta Z_{\psi}) \bar{\psi} \left( i \slashed{\partial} - m_0 \right) \psi \nonumber\\
& & - i g_0(1 + \delta Z_{\psi}) \sqrt{1 + \delta Z_{\phi}} \bar{\psi} \gamma^5 \psi \phi
- \frac{u_0}{4!} (1 + \delta Z_{\phi})^2 \phi^4 \label{e3},
\end{eqnarray}
 where we have introduced the multiplicative renormalisations $Z_{\phi}$, $Z_{\psi}$, $Z_{g}$, $Z_{u}$, $Z_{m}$, and $Z_{M}$ defined through
\begin{eqnarray}
\phi_0 &=& \sqrt{ Z_{\phi}} \phi \equiv \sqrt{1 + \delta Z_{\phi}} \phi,\label{e5}\\
\psi_0 &=& \sqrt{ Z_{\psi}} \psi\equiv \sqrt{1 + \delta Z_{\psi}} \psi,\label{e6}\\
M^2_0 Z_{\phi} &=& M^2 + \delta M^2\equiv M^2 Z_{M}, \label{e7}\\
m_0 Z_{\psi} &=& m + \delta m \equiv m Z_{m}, \label{e8} \\
g_0 Z_{\psi} \sqrt{Z_{\phi}} &=& g + \delta g\equiv g Z_{g}, \label{e9}\\
u_0 (Z_{\phi})^2 &=& u + \delta u\equiv u Z_{u}.\label{e10}
\end{eqnarray}

We  use dimensional regularisation to evaluate the counterterms, taking $D=4-\epsilon$ and $\mu$ as the renormalisation scale. This leads to the perturbative renormalisation group (see, for example, \cite{R14a}). From the discussion in Section \ref{sec:intro}, the perturbative renormalisation group is unaffected by the non-trivial saddle points~\cite{R21}, which give asymptotically subdominant contributions. 

The field theoretic action $S$ generally depends on these $\mu$ dependent couplings such that 
\begin{equation}
\label{e1a}
S\left[ Z\left( \mu \right)^{1/2}  \Phi ;\mu ,g_{i}\left( \mu \right)  \right]  =S\left[ Z\left( \mu^{\prime } \right)^{1/2}  \Phi ;\mu^{\prime } ,g_{i}\left( \mu^{\prime } \right)  \right]  
\end{equation}
where $Z\left( \mu \right)$ is the wave function renormalisation (generally a matrix)  of the generic field  $\Phi$. As an example, for a scalar field theory, we can write
\begin{equation}
\label{e2a}
S\left[ \phi ;\mu ,g_{i}\right]  =\int d^{D}x\left( -\frac{1}{2} \partial_{\mu } \phi \partial^{{}\mu } \phi +\sum_{i} \mu^{D-d_{i}} g_{i}O_{i}\left( x\right)  \right)
 \end{equation}
where $O_{i}\left( x\right)$ is a local operator of mass dimension $d_i$ and $g_i$ is dimensionless. The $\mu$ dependence of $g_i$ is determined through functions $\beta_{i} \left( \left\{ g_{j}\right\}  \right)$
\begin{equation}
\label{e3a}
\mu \frac{dg_{i}\left( \mu \right)  }{d\mu } =\beta_{i} \left( \left\{ g_{j}\right\}  \right) ,
\end{equation}
which are the renormalisation group equations.

\subsection{Coupling constant analyticity}
We have noted that in the Lee model, the bare coupling  has a square root singularity in the renormalised coupling. The Lee model was constructed in such a way that renormalisation could be performed exactly. In realistic theories, we cannot expect to obtain exact information about renormalisation. We use a renormalisation (or subtraction point $\mu$) to define our theory. If we could calculate to all orders in perturbation theory then it is expected that results for physical quantities would be independent of $\mu$. The renormalisation group enforces this condition on quantities calculated to low orders in the loop expansion. In this sense, some of the important features of an exact analysis are incorporated. However, the situation is more complicated since the perturbation series are believed not to be convergent, but only asymptotic \cite{LeGuillou:1990nq,PhysRev.85.631,Dunne:2002rq}.
This led to investigations of the analyticity properties of physical quantities such as  the ground state energy (related to the partition function) as a function of couplings (e.g. $u$)~\cite{Bender:1969si, Simon:1970mc, Simon1991FiftyYO, LeGuillou:1990nq} using large orders in perturbation theory.

We conjecture that square root singularities of the type found in the Lee model may contribute to the emergence of  $\mathcal{PT}$ theories starting with a Hermitian theory.~Such a result would be extremely hard to prove. The presence of a square root singularity implies that the coupling has a different sign on either side of the cut.
For the anharmonic oscillator Bender and Wu~\cite{Bender:1969si} found an accumulation of square root singularities in the complex coupling constant Riemann sheets for the energy levels arbitrarily close to the origin. %There were an infinite number of such singularities centred symmetrically around the $\arg{u}=\pm  3\pi/2$ in the complex $u$ plane. In any open neighbourhood around $u=0$ such singularities were shown to exist using WKB analysis. The energy levels also form one function in an (infinitely) multi-sheeted Riemann surface. In this way, it is possible to go from the ground state energy functions to a sheet for the first excited state, and so on. These results are found through WKB analysis and cannot be proved rigorously~\cite{Simon:1970mc}.  Indeed for many-component anharmonic oscillators in $D=1$, it has only been shown rigorously shown that origin $u=0$ remains an accumulation point for singularities. 

However, on general grounds, it may be expected that square root singularities will also be present in field theories. Higher $D$ field theories are much more complicated than the $D=1$ anharmonic oscillator and so square root singularities will not be expected to appear in the same way as in the single component anharmonic oscillator~\cite{Simon:1970mc}. Eigenvalue problems are ubiquitous in field theory and it is argued persuasively\footnote{See Chapter 7, Section $7.5$ of \cite{R15} for a comprehensive discussion.} in~\cite{R15}  that square root singularities are generically the most likely singularities of eigenvalues as functions of couplings continued to the complex plane.

\subsection{The renormalisation group analysis}

In terms of $t={\rm log} \, \mu$ and $h=g^2$ the renormalisation group beta functions for $h \geq 0$ are
\be
\label{basicbeta}
\frac{dh}{dt}=\beta_{h} \left( h,u\right) \textrm{ and } \frac{du}{dt}=\beta_{u} \left( h,u\right) \ee
where
\be
\label{betah}
\begin{split}
\beta_{h} \left( h,u\right) = & - \epsilon h + \frac{1}{(4\pi)^2} 10h^2 + \frac{1}{(4\pi)^4} \left(-\frac{57}{2} h^3 - 4 h^2 u + \frac{1}{6} h u^2 \right) \\ & + \frac{1}{(4\pi)^6} \left(\left[-\frac{339}{8} + 222 \; \zeta(3)\right] h^4 + 72 h^3 u + \frac{61}{24} h^2 u^2 - \frac{1}{8} h u^3 \right) 
\end{split}
\ee
and
\be
\label{betau}
\beta_{u} \left( h,u\right) = - \epsilon u + \frac{1}{(4\pi)^{2} } \left( -48h^{2} + 8 h u + 3 u^{2}\right)  +\frac{1}{(4\pi)^{4} } \left( 384 h^{3} + 28 h^{2}u-12 hu^{2} - \frac{17}{3} u^{3} \right).
\ee
where $\zeta$ denotes the Riemann zeta function. These expressions for the beta functions have been found from a perturbative calculation to three loops for the Yukawa coupling and two loops for the quartic coupling using the Mathematica package RGBeta~\cite{R10} and are independent of $m$ and $M$\footnote{The flows for $m$ and $M$ are dependent on the flows for $h$ and $u$ however.}. When $g$ is pure imaginary, $h$ is negative and so $h$ positive or negative distinguishes between Hermitian and $\mathcal{PT}$-symmetric cases, respectively. The expressions for the beta functions given here are only applicable for $h \geq 0$ (the case for which $g$ is real). Our qualitative conclusions are unaffected by the sign of $h$, and the $h \geq 0$ and $h<0$ sectors do not mix, so for brevity \textbf{in the main text we restrict to} $\boldsymbol{h \geq 0}$ (the Hermitian case for $g$). However, we give the $h<0$ (non-Hermitian in $g$) results for completeness in Appendix \ref{Negative_h_appendix}.

In the next subsections we shall consider: 
\begin{enumerate}
  \item The zeros of the beta functions $\beta_{u}$ and $\beta_{h}$ which determine the fixed points of the renormalisation group.
  \item The stability of the fixed points, which can be determined from a linearised analysis around the fixed points (except for the trivial fixed point when $\epsilon =0$).
  \item The full non-linear flows connecting the different fixed points. These flows are instructive, especially for the epsilon-dependent fixed points emanating from the trivial fixed point.
  \item Once we have an $\epsilon$ expansion of the fixed points it is natural to enquire about any possible resummation to determine information about fixed points and their stability at $D=3$. We have used the method of Pad\'e approximants and made checks on the pole structure~\cite{R15} in the neighbourhood of $\epsilon =1$ to determine the trustworthiness of any $D=3$ fixed point determined this way.

\end{enumerate}

\subsubsection{Fixed points for $\epsilon=0$}
 
 It is customary to denote the fixed point of $h$ as $h^*$ and the fixed point of $u$ as $u^*$. However, in the main text, for clarity we will use $f_{i,h}$ (the fixed point value for $h$) and $f_{u,h}$ (the fixed point value for $u$) for our numerical results for the fixed points, given to three significant figures. When $\epsilon=0$, we have two fixed points
 \begin{enumerate}
  \item The trivial (or Gaussian) fixed point: $f_{1,h}=0,$ and  $f_{1,u}=0$.
  \item $f_{2,h}=0,$  and $f_{2,u}\simeq 83.6$ which corresponds to a quartic coupling $\simeq 3.48$ (rescaled by $1/4!$); since the $f_{2,h}$ and $f_{2,u}$ are non-negative  this is a Hermitian fixed point. 
\end{enumerate} 

The trivial fixed point is the progenitor of the fixed points for $\epsilon \ne 0$. We perform a linearised analysis first for the fixed point $f_2$.
A non-linear analysis is necessary for $f_1$.

\subsection{Stability analysis}
A linearised analysis around fixed points $h^*$ and $u^*$ consists of examining the evolution of $\delta h=h-h^{\ast }$ and 
$\delta u=u-u^{\ast }$. A linearised stability analysis~\cite{R12a} is determined by 
\be
\label{stability}
\frac{d}{dt} \left( \begin{matrix}\delta h\\ \delta u\end{matrix} \right)  ={\mathcal{M}}\left( h^{\ast },u^{\ast }\right)  \left( \begin{matrix}\delta h\\ \delta u\end{matrix} \right)  
\ee
where ${\mathcal{M}}$ is a $2 \times 2$ matrix\footnote{ ${\mathcal{M}}$ will also have a dependence on $\epsilon$ in $D=4-\epsilon$.}.~${\mathcal{M}}$ is diagonalised to obtain eigenvalues $(\lambda_{1}(h^{\ast},u^{\ast}),\lambda_{2}(h^{\ast},u^{\ast}))$ and corresponding eigenvectors  $(\vec{e}_{1}(h^{\ast},u^{\ast}),\vec{e}_{2}(h^{\ast},u^{\ast}))$.

Here, we summarise the eigenvectors and eigenvalues for $f_{2,h}$:
\begin{itemize}
  \item $\lambda_{1} \left( f_{2h},f_{2u}\right)  \approx -1.59$, and  $\vec{e}_{1} \left( f_{2h},f_{2u}\right)  =\left( \begin{matrix}0\\ 1\end{matrix} \right)  $
  \item $\lambda_{2} \left( f_{2h},f_{2u}\right)  \approx 0.0282$ and $\vec{e}_{2} \left( f_{2h},f_{2u}\right)  =\left( \begin{matrix}1.85\\ 1\end{matrix} \right)  $
\end{itemize}

\subsubsection*{Non-linear analysis around trivial fixed point}

The stability of the trivial fixed point requires a non-linear analysis, due to the vanishing of the eigenvalues of the linear stability matrix $M$.

\noindent{For the study of renormalisation group flows in the neighbourhood of the trivial fixed point, $\beta_{u} \left( h,u\right)$ and $\beta_{h} \left( h,u\right)$} can be simplified to
\be
\label{simu}
\beta_{u} \left( h,u\right)  \simeq \frac{1}{\pi^{2} } \left[ -3h^{2}+\frac{1}{2} hu+\frac{3}{16} u^{2}\right]  
\ee
and
\be
\label{simh}
\beta_{h} \left( h,u\right)  \simeq \frac{5}{8\pi^{2} } h^{2}.
\ee
The family of flows for $h$,  parameterised  with $h_{0}$ and $t_{0}$, is given by
\be
\label{flowh}
h\left( t\right)  =\frac{8\pi^{2} h_{0}}{8\pi^{2} -5h_{0}\left( t-t_{0}\right)  }. 
\ee
%The accompanying flow for $u$ is
%\be
%\label{flowu}
%u\left( t\right)  \simeq -\frac{16\pi^{2} h_{0}}{3f\left( t\right)  } a(t)  
%\ee
%where $f\left( t\right)  =8\pi^{2} -5h_{0}\left( t-t_{0}\right)  $ and $a(t)=\left[ \frac{-6.5\  f\left( t\right)^{2.4}  +5.5c}{f\left( t\right)^{2.4\  }  +c} \right] $. If $u_{0}=u(t_{0})$ then $c$ is determined by $u_{0}$ and $h_{0}$. The behaviour is complicated and when $h$ or $u$ becomes large (because of the Landau pole), the perturbative analysis will not be valid. For generic values of ${u_{0}}/{h_{0}}$, $c$ is large and so a crude approximation is $a(t) \approx 5.5$, which leads to
%\be
%\label{flowapprox}
%u\left( t\right)  \approx -\frac{11}{3} h\left( t\right).  
%\ee
 %This crude analysis captures some aspects of the negative slopes seen in the figure.
We define $f(t) = 8\pi^{2} -5h_{0}\left( t-t_{0}\right)  $ for convenience.
The accompanying flow for $u$ is
\be
u(t) = -\frac{8\pi^{2} h_{0}}{3f(t)  } \left[ \frac{-p \; f(t)^{n}  +q \; c}{f(t)^{n } +c} \right]  
\ee
where $c$ is an integration constant, $p = 1 + \sqrt{145} \approx 13$, $q = -1 + \sqrt{145} \approx 11$, $n = \sqrt{\frac{29}{5}} \approx 2.4$. The behaviour is complicated and when $h$ or $u$ becomes large, which occurs due to the presence of a Landau pole, the perturbative analysis is not valid. We can write $u(t)$ in terms of $h(t)$ directly as
\be \label{flowapprox}
u(t) = - \frac{1}{3} h(t) \left[ \frac{-p \; h_0^n  +q \; \tilde{c} \; h(t)^{n }}{h_0^n + \tilde{c} \; h(t)^{n }} \right] 
\ee
writing $c = (8 \pi^2)^n \; \tilde{c}$. This allows us to relate $\tilde{c}$ to $h_0$ and $u_0$ as
\be
u_0 = - \frac{1}{3} h_0 \left[ \frac{-p +q \; \tilde{c} }{1 + \tilde{c}} \right] 
\ee
If we define $k = \frac{u_0}{h_0}$, then we find
\be
\tilde{c} = \frac{p - 3k}{3k + q}
\ee
This suggests that if the $h_0$ and $u_0$ are sufficiently close to the origin, then any straight line through the origin is possible.

\subsection{Renormalisation group flows}

We shall examine the flow around the fixed points  $f_{ih}$ and $f_{iu}$, for $i=1,2$. 
For $\epsilon=0$  the dimensionless couplings are of $O(1)$ and are not small in any controlled fashion; hence the flows derived from perturbation theory can only be indicative  of possible features of renormalisation. Moreover, geometric methods are best suited to visualise the flows\footnote{Solving individual trajectories as a function of $t$ requires initial conditions and the description of flows requires a grid of initial conditions. A geometric method~\cite{R12b}, whereby tangents to the flows are pieced together as streamlines, is preferable.  }.

\begin{figure}[ht]
\centerline{\includegraphics[scale=0.7]{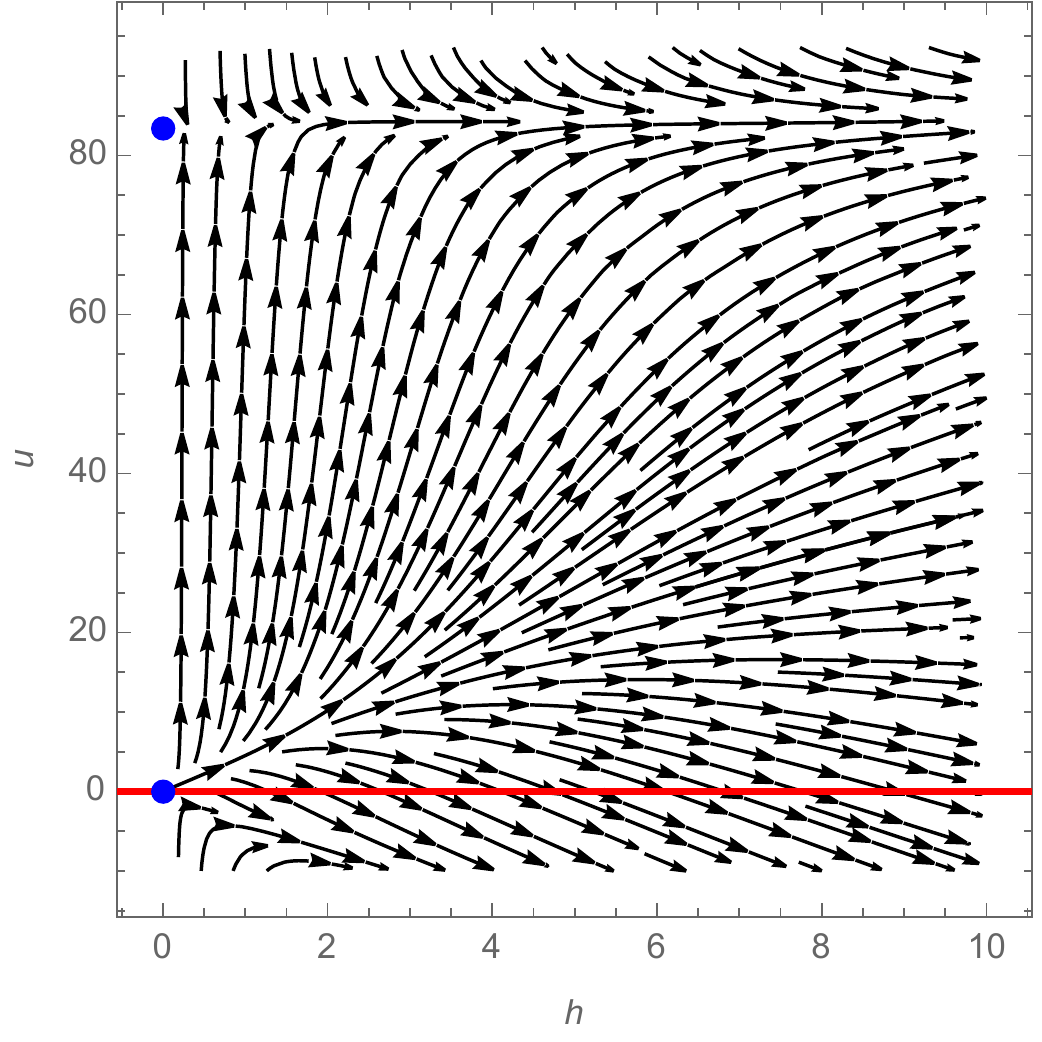}}
\caption{Global flow for $\epsilon=0$.}
\label{flow1}
\end{figure}

\begin{figure*}[ht]
        \centering
        \begin{subfigure}[t]{0.475\textwidth}
            \centering
            \includegraphics[scale=0.6]{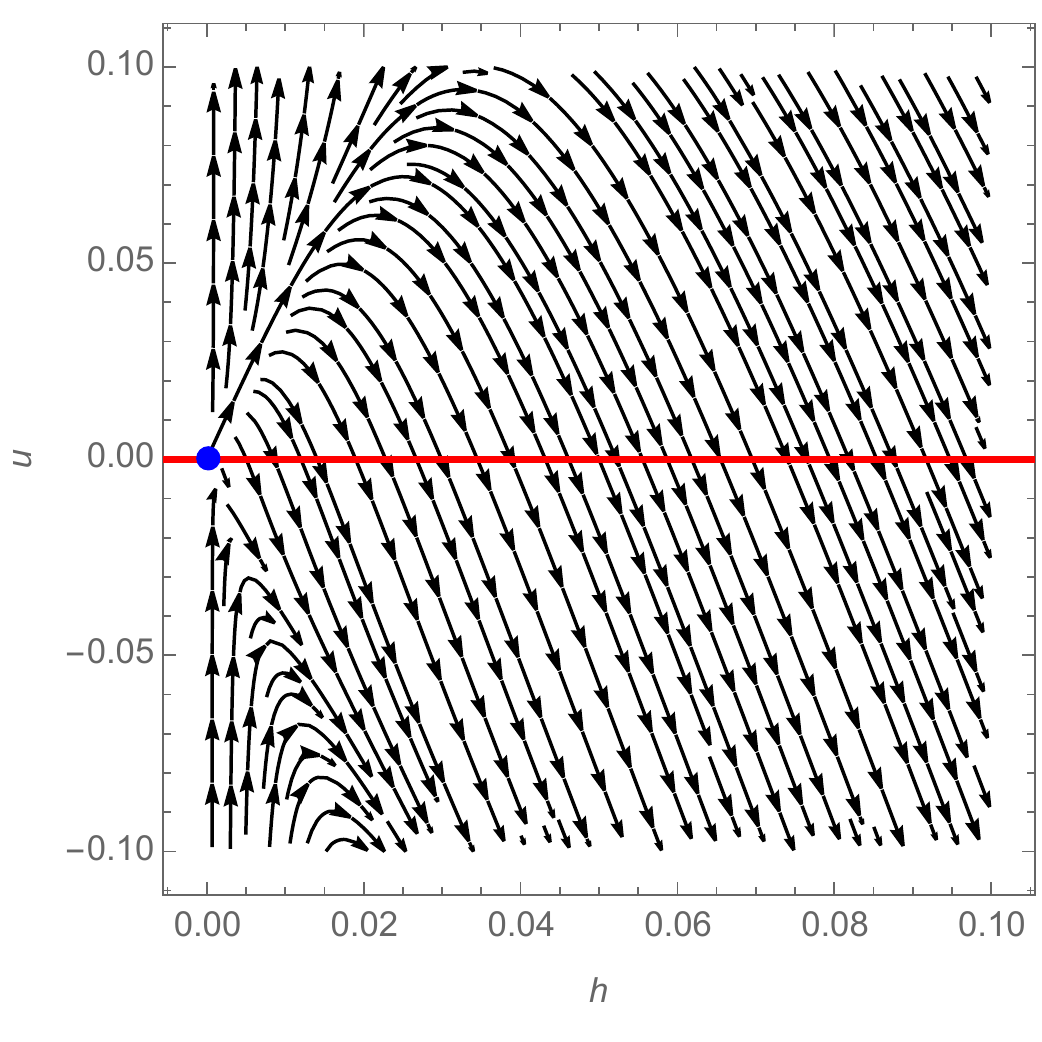}
            \caption[]%
            {{\small Non-linear saddle at trivial fixed point $f_1$.}}  
            \label{flow2}
        \end{subfigure}
        \hfill
        \begin{subfigure}[t]{0.475\textwidth}
            \centering 
            \includegraphics[scale=0.585]{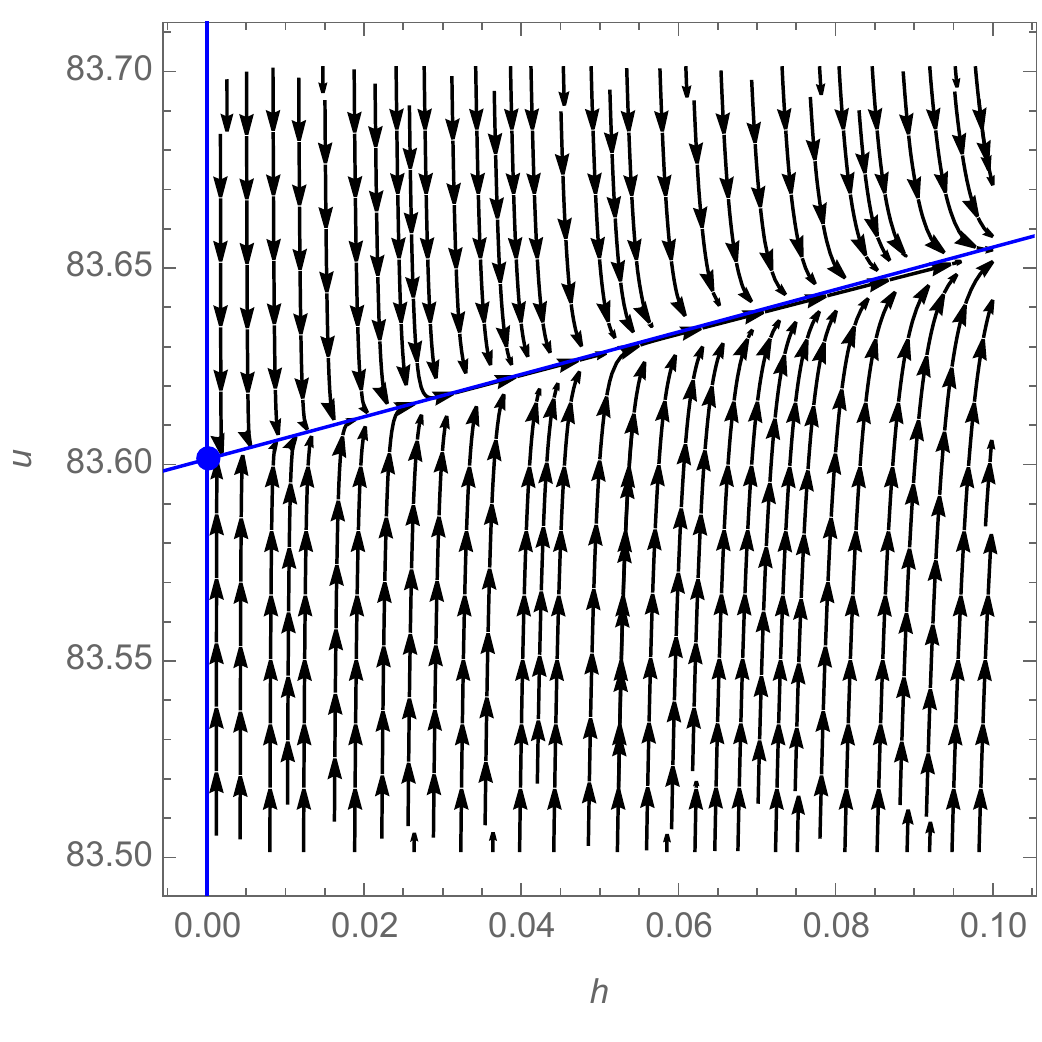}
            \caption[]%
            {{\small Hermitian saddle fixed point $f_2$.}}
            \label{flow3}
        \end{subfigure}
        \vskip\baselineskip
        \caption[]
        {\small The local flows around the fixed points for $\epsilon = 0$.}
        \label{LocalFlow0}
    \end{figure*}

In the figures, the vertical axis is the $u$-axis and the horizontal axis is the $h$-axis. The $h$-axis (where present) is shown in red, and any fixed points are shown in blue (colour online). ~Some features to be noted are:
\begin{itemize}
  \item There are no flows from positive to negative $h$ and vice versa\footnote{This has been verified by performing the analysis for $h<0$, see Appendix \ref{Negative_h_appendix}.}.
  \item  There are flows from positive $u$ to negative $u$, i.e. from a Hermitian to a $\mathcal{PT}$-symmetric region.
  \item The flows around the trivial fixed point $f_1$ do not show a simple source, sink or saddle point behaviour, but rather a non-linear flow. This flow is complicated but an approximate solution is given in \eqref{flowapprox}. In Figure \ref{flow1}, there are approximate lines of both positive and negative slope crossing the $h$-axis, which are an indication of this behaviour.
\end{itemize} 

Given that the analysis is based on perturbation theory,  flows in regions where the couplings are large compared to $1$ can only be misleading. However, near the trivial fixed point, we can see evidence for flows from positive to negative $u$, i.e. from Hermitian to $\mathcal{PT}$-symmetric behaviour. This type of behaviour is discussed and investigated below in much more detail for a situation where there are four fixed points which occur at small values of $u$ and $h$. In our context, this arises since there is a separate parameter which controls the size of the couplings and makes perturbation theory possible. This parameter is $\epsilon$.

\subsubsection{Fixed points for $\epsilon \ne 0$}
We consider $\epsilon >0$ and examine the flows of \eqref{basicbeta}. 
We have fixed points which we denote by $F_{i}$, $i=0,1,2,\ldots,4$. $F_{0}=f_1$ is the trivial fixed point. The remaining $F_{i}$ are given in terms of series which are not typically convergent but asymptotic as $\epsilon \to 0$. The expressions for the fixed points are given in Appendix \ref{nonzeroeps}.
These expressions allow tracking of fixed points as a function of $\epsilon$ and also, in some circumstances, an extrapolation to ${\epsilon}=1$ using the technique of Pad\'e approximants.
In the limit $\epsilon \to 0$, the fixed point ${F_{4}}\to{f_{2}}$, and the fixed points $F_{i}\to f_{1}$ for $i=1,2,3$. Hence the 
trivial fixed point becomes $4$ fixed points for $\epsilon \ne 0$: the trivial fixed point and $3$ further fixed points  ($F_{i},  i=1,2,3$) which are $O(\epsilon)$. For sufficiently small $\epsilon$, ~$F_{2}$ is a non-Hermitian ($\mathcal{PT}$-symmetric) fixed point whereas $F_{1}$ and $F_{3}$ are Hermitian.
 The renormalisation group flows in the neighbourhoods of $F_{i},  i=1,2,3$ and $f_{1}$ are described through perturbative analysis and are our main focus. Although near $F_{4}$ our analysis does indicate possible new behaviour (in terms of flows between Hermitian and $\mathcal{PT}$-symmetric regions in the $h$ coupling)   these latter findings can only remain conjectural since perturbation theory is unreliable for large couplings. As such, we ignore this point in most of our analysis below. However, it is worth noting that the emergence of $\mathcal{PT}$ symmetry in the Lee model is in terms of $h$~\cite{R5} and occurs at strong coupling.
 %\textendash  

\subsection{The stability of fixed points for $\epsilon \ne 0$} \label{stabnonzeroeps}

We follow the linear stability analysis of \eqref{stability} for the fixed points $F_{0}\equiv f_1$ and  $F_{j}, (j =1,2,3)$. $F_{\alpha}~  (\alpha =0,1,2,3$) has two components:  $F_{\alpha, u}$, the fixed point value for $u$ and  $F_{\alpha, h}$, the fixed point value for $h$. The eigenvalues of the stability matrix around $F_{\alpha}$, will be denoted by $\Lambda_{\alpha, j} ,  \  j=1,2$. The corresponding 2 component eigenvectors will be denoted by $\vec{E}_{\alpha j},\    j=1,2$.

\begin{figure}[ht]
\centerline{\includegraphics[scale=0.7]{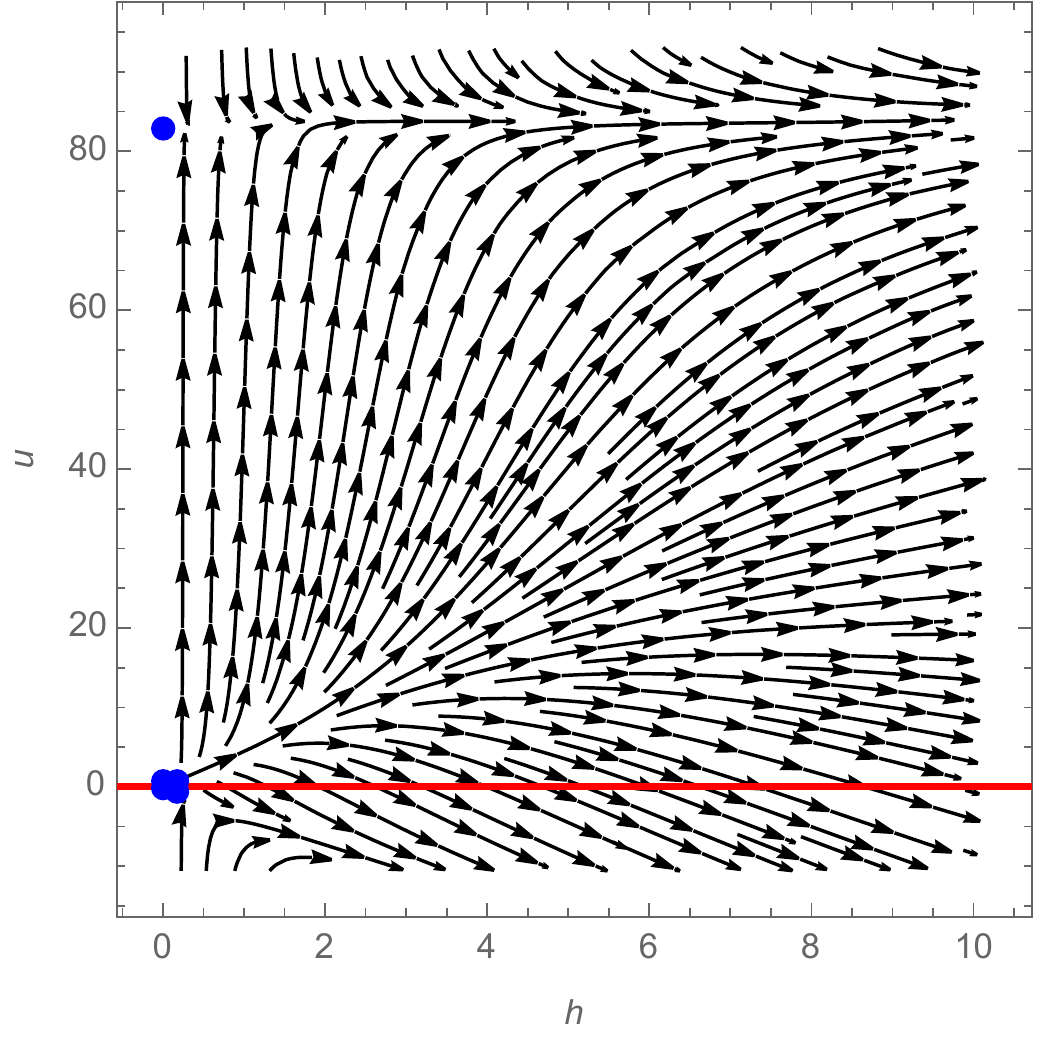}}
\caption{Global flow for $\epsilon=0.01$. There are a group of four fixed points that are close to the origin, and one high-$u$ fixed point that we ignore from concerns over its validity in perturbation theory.}
\label{flow4}
\end{figure}

\subsubsection{The renormalisation group flow between fixed points for $\epsilon \ne 0$}

The renormalisation group flows for $0 < \epsilon \lesssim 0.027$ are qualitatively the same and so we shall consider the case $\epsilon=0.01$ as a representative flow. The flows are organised by the different fixed points $F_{\alpha}$. We determine the flows numerically and non-perturbatively in $\epsilon$.

\begin{figure}[ht]
\centerline{\includegraphics[scale=0.7]{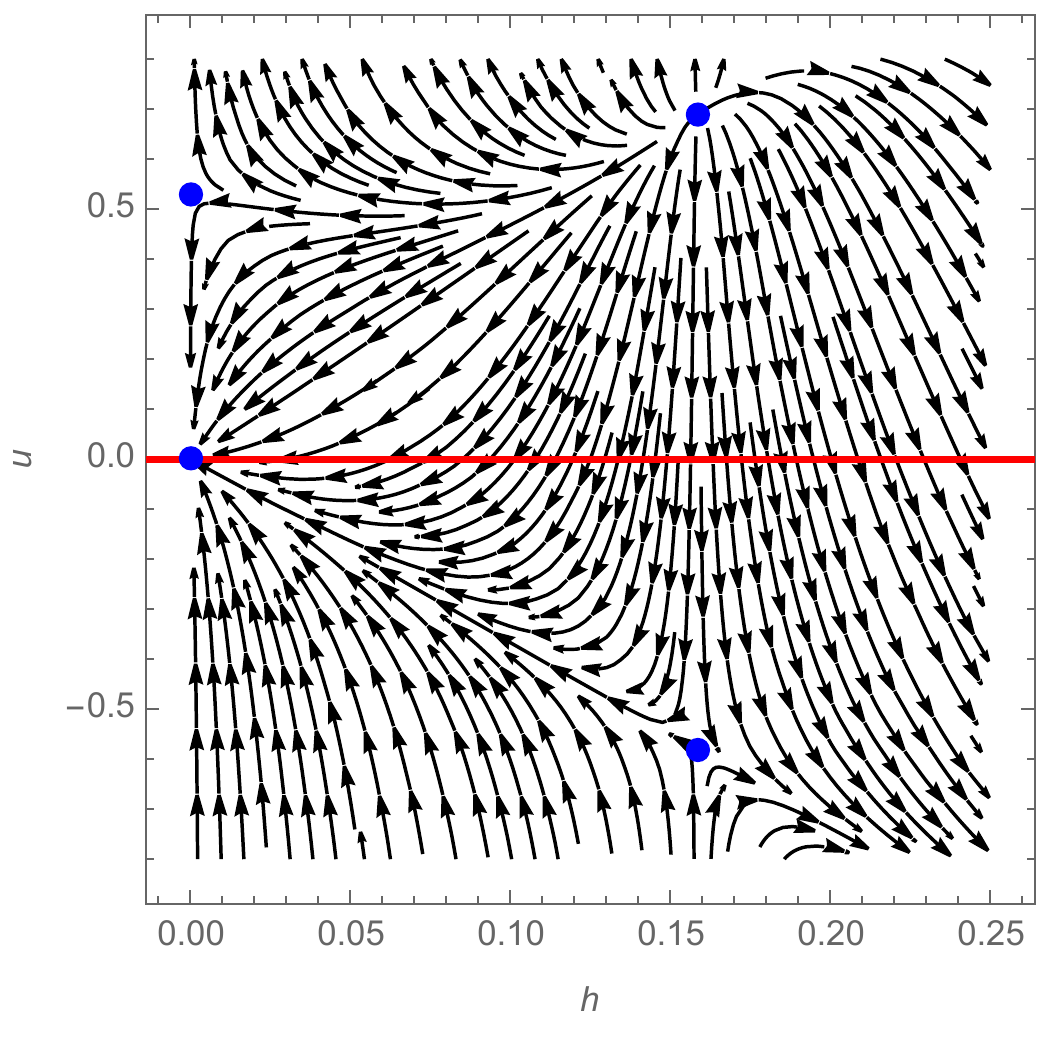}}
\caption{Flows around the group of fixed points near the origin for $\epsilon=0.01$.}
\label{flow5}
\end{figure}

\begin{figure*}[ht]
        \centering
        \begin{subfigure}[t]{0.475\textwidth}
            \centering
            \includegraphics[scale=0.6]{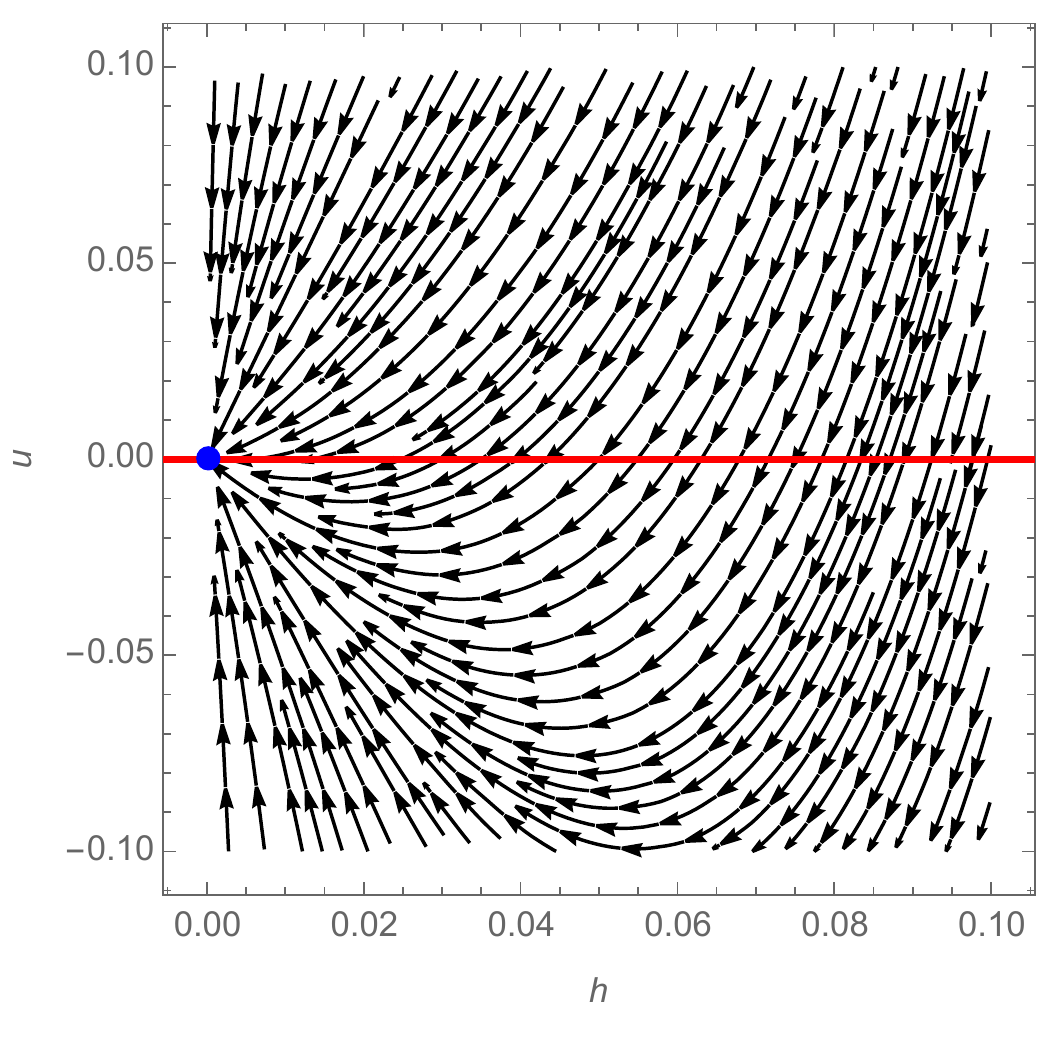}
            \caption[]%
            {{\small Ultraviolet stellar node at trivial fixed point $F_0$.}}    
            \label{flow6}
        \end{subfigure}
        \hfill
        \begin{subfigure}[t]{0.475\textwidth}  
            \centering 
            \includegraphics[scale=0.585]{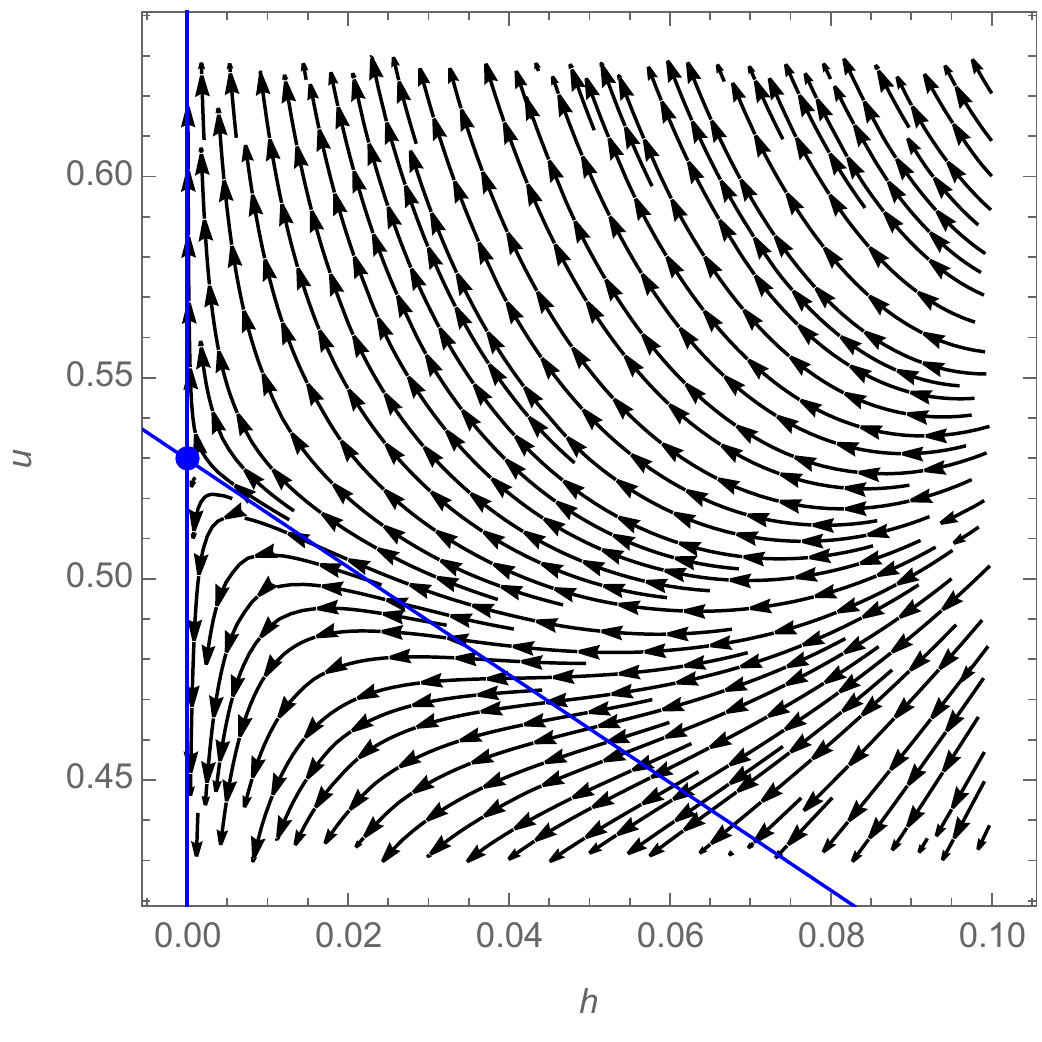}
            \caption[]%
            {{\small Hermitian saddle fixed point $F_1$.}}    
            \label{flow7}
        \end{subfigure}
        \vskip\baselineskip
        \begin{subfigure}[t]{0.475\textwidth}   
            \centering 
            \includegraphics[scale=0.6]{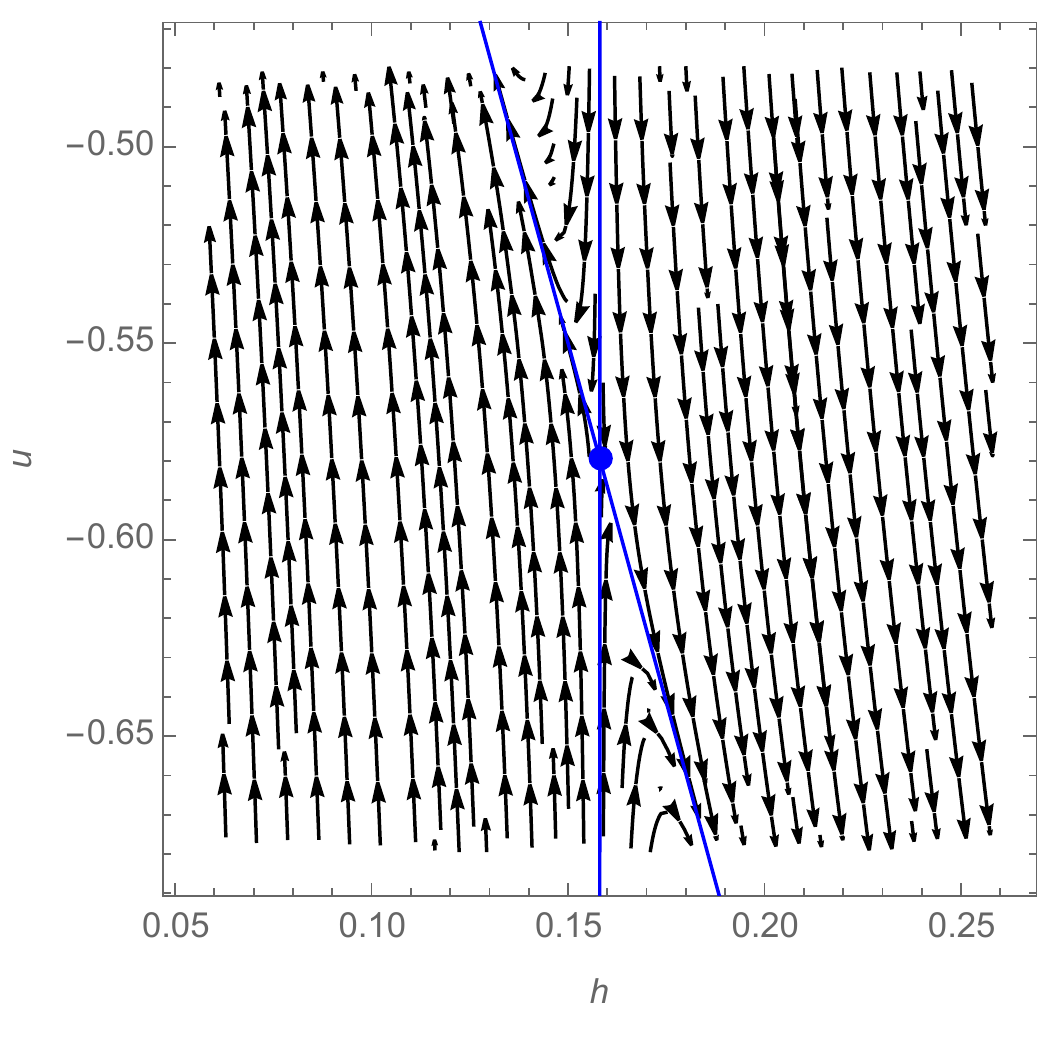}
            \caption[]%
            {{\small Non-Hermitian saddle fixed point $F_2$.}}    
            \label{flow8}
        \end{subfigure}
        \hfill
        \begin{subfigure}[t]{0.475\textwidth}   
            \centering 
            \includegraphics[scale=0.585]{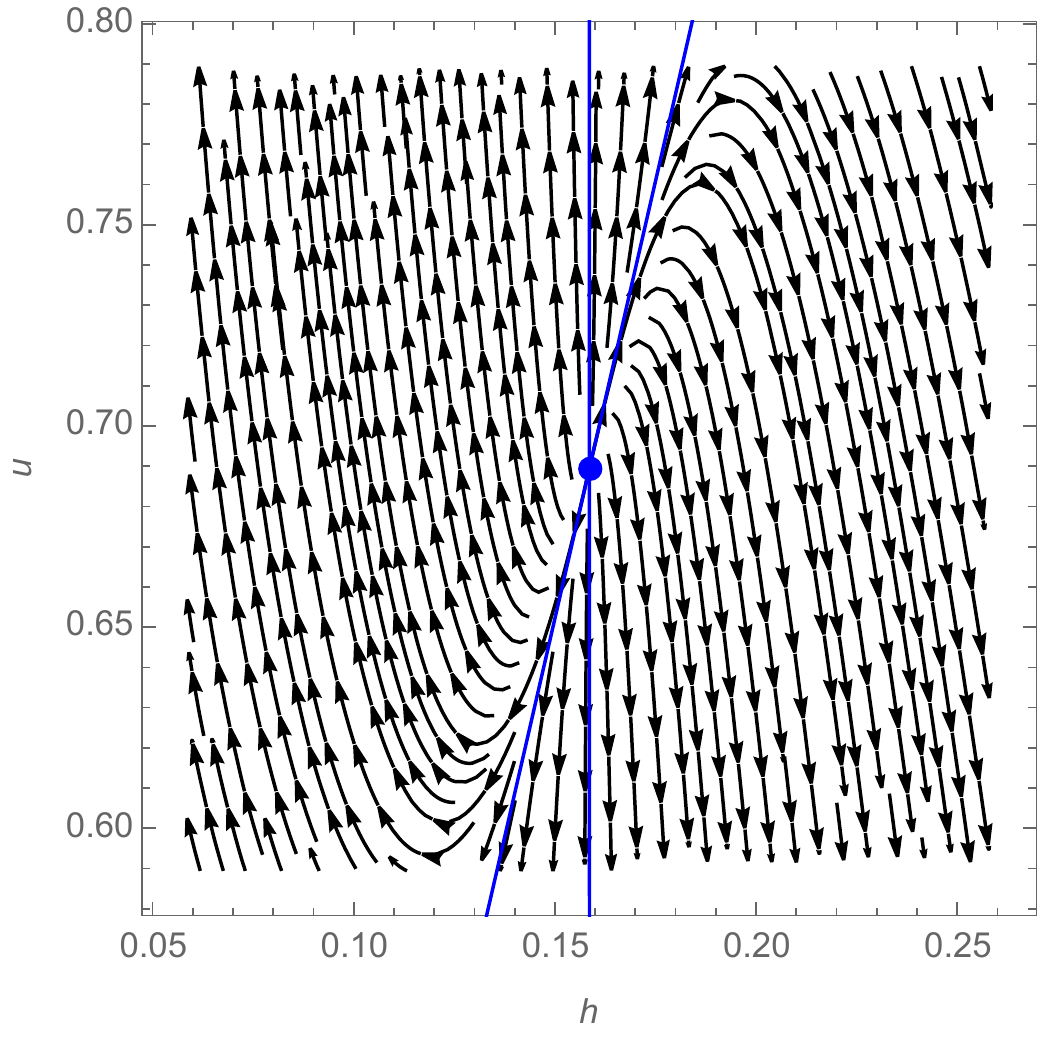}
            \caption[]%
            {{\small Hermitian infrared fixed point $F_3$.}}    
            \label{flow9}
        \end{subfigure}
        \caption[]
        {\small The four trustworthy (in perturbation theory) fixed points for $\epsilon=0.01$.} 
        \label{LocalFlow0.01}
    \end{figure*}

\eject
As expected, many of the features from the $\epsilon = 0$ case persist, particularly regarding flows across the coordinate axes. However, the non-zero $\epsilon$ ensures that the behaviour of the flow near the origin can now be characterised using linear stability analysis \cite{R12a}; we find an ultraviolet stable stellar node there (as shown in Figure \ref{flow6}). Furthermore, three additional points emanate from the origin as $\epsilon$ has increased. If we focus on the non-Hermitian (and $\mathcal{PT}$-symmetric) saddle fixed point $F_2$ (Figure \ref{flow8}), we note that (by examining Figure \ref{flow5}):

\begin{itemize}
    \item There is a flow that originates at the Hermitian infrared fixed point $F_3$ (Figure \ref{flow9}) in the IR (large negative $t$) limit, which can flow to the non-Hermitian saddle $F_2$ in the UV (large positive $t$) limit.

    \item There is a flow that originates at the stellar node at the origin $F_0$ (Figure \ref{flow6}) in the UV (large positive $t$) limit, which can flow to the non-Hermitian saddle $F_2$ in the IR (large negative $t$) limit. 
\end{itemize}

Some of these features have been noted previously in the literature in the context of the Hermitian theory (for example, in \cite{Degrassi:2012ry, R9}), but we are now able to interpret the flow to the non-Hermitian region for the coupling constants in the framework of $\mathcal{PT}$-symmetric theory \cite{R8}. Furthermore, we have additional control here from the use of the engineering dimension $\epsilon$.

As $\epsilon$ continues to increase, we reach a critical value $\epsilon_c \sim 0.027$ where the behaviour of the large-$u$ fixed point changes (in terms of the eigenvalues of the linear stability analysis). However, this is not significant for our interests here, since we cannot be sure of the validity of the analysis for these fixed points in the perturbation theory of $h$ and $u$. The next critical value of $\epsilon$ for which the character of a fixed point changes is $\epsilon_{c'} \sim 0.44$, but this is likely too high to trust within our perturbative expansion in $\epsilon$. We investigate the robustness of our results in this section to changing the loop orders of the computation, as well as the effect of increasing $\epsilon$, in Appendix \ref{Robustsubsection}.

We note that the character of the non-Hermitian saddle fixed point $F_2$ seems to be preserved as we extend our analysis to $D = 3$ from above (and so $\epsilon \rightarrow 1$) with Pad\'e approximants. 

\section{Pad\'e approximants and the $D=3$ fixed point} \label{PadeSec}

The $\epsilon$ expansion is used in the study of critical phenomena~\cite{R11, R13a}, but its convergence is not understood in any systematic way. Although series using the $\epsilon$ expansion are readily generated, the series are generally divergent. Hence there is no radius of convergence $\epsilon_{R} $ such that the series is convergent for $\left| \epsilon \right|  <\epsilon_{R} $. If the perturbation series is singular, it diverges for all non-zero $\epsilon$. Pad\'e approximants can sometimes offer a way of summing such a series. The partial sums of the $\epsilon$ series cannot be summed directly, since for fixed $\epsilon$  the sequence of partial sums diverge.

If we have a formal power series $P(\epsilon)=\sum a_{n}\epsilon^{n} $  in $\epsilon$ then the Pad\'e approximant $P^{N}_{M}\left( \epsilon \right) $ is defined by
\begin{equation}
\label{Pade }
P^{N}_{M}\left( \epsilon \right)  =\frac{\sum^{N}_{n=0} A_{n}\epsilon^{n} }{\sum^{M}_{n=0} B_{n}\epsilon^{n} }. 
\end{equation}
Without loss of generality we take $B_{0}=1$ and the first $M+N+1$ coefficients of $\sum a_{n}\epsilon^{n} $ are used to determine the coefficients $A_{0},A_{1},\ldots ,A_{N},B_{1},B_{2},\ldots ,B_{M}$. $P^{N}_{N}\left( \epsilon \right)  $ is a diagonal Pad\'e sequence. All Pad\'e approximants have pole singularities from the denominator and zeros from the numerator. If there are poles in the neighbourhood of $\epsilon=1$ then an extrapolation to $\epsilon=1$ using  Pad\'e sequences is not viable. By checking for the consistent predictions of fixed points and their stability as  $N$ and $M$ are varied, we decide on the validity of our extrapolation~\cite{R15} to  $\epsilon=1$. This is a necessary (but not sufficient) criterion for a valid extrapolation to $D=3$.

We consider the cases where $P(\epsilon)$ is truncated to $\epsilon^{2n}$, for $n=4,\  5,\  6,\  7$; then we examine the corresponding diagonal Pad\'e approximants $P^{N}_{N}\left( \epsilon \right)$ for $N=4,\  5,\  6,\  7$, as well as off-diagonal Pad\'e sequences $P^{N+1}_{N-1}\left( \epsilon \right)$ and $P^{N-1}_{N+1}\left( \epsilon \right)$. The convergence of the various Pad\'e approximants for the fixed points $F_{\alpha }$ is only consistent for $F_{2 }$, a non-Hermitian fixed point. The resultant fixed point at $D=3$ is 
\begin{equation}
\label{D3 }
\left( h^{\ast },u^{\ast }\right) = \left( 17.6,-32.3\right)  
\end{equation}
whose linearised stability is characterised by eigenvalues $\Lambda_{1} = -1.16$ and $\Lambda_{2} = 1.08$. Hence the fixed point has saddle-like stability. The eigenvectors $\vec E_{j}$ associated with $\Lambda_{j}$, for $j=1,2$ are

\begin{equation}
\label{Evec1}
\vec E_{1} = (-0.0121, 1)
\end{equation}
and 
\begin{equation}
\label{Evec2}
\vec E_{2}=(-4.21, 1).
\end{equation}
As $\epsilon$ has increased from small values this fixed point has retained its non-Hermitian character and its Pad\'e approximants have been stable for diagonal and off-diagonal sequences. Hence these computations provide some confidence that this is a genuine non-perturbative fixed point for $D=3$. The putative fixed point may be relevant to studies of UV completions of the Nambu-Jona-Lasinio and Gross-Neveu models between $2$ and $4$ dimensions~\cite{Fei:2016sgs} and quantum phase transitions in electronic systems~\cite{Herbut:2023xgz,PhysRevB.84.205128}, which is beyond the scope of this paper. We examine the robustness of our conclusions in this section as we change the loop orders for the computations in Appendix \ref{Robustsubsection}. 

\section{Perspective on the perturbative calculations}
The methods we apply are used in the study of critical phenomena~\cite{amit84a, zinn2021quantum}.~It is widely recognised that they are applicable in the context of relativistic field theories in particle physics~\cite{Weinberg:1976xy}.~Although in this work we have focused on the emergence of a \cPT-symmetric field-theory description emerging from a Hermitian theory, this Hermitian theory is a prototype theory for axion physics. The role of relativistic fermions in such models certainly distinguishes them from the scalar field theories belonging to the Ising universality class, which are influential in critical phenomena. 

The presence of fermions necessitates revisiting discussions on the nature of perturbation series~\cite{Dyson:1952tj, LeGuillou:1990nq} and dimensional regularisation~\cite{RevModPhys.47.849, Jegerlehner:2000dz}. Our calculations raise some technical issues that appear in the presence of fermions, which we will discuss below. 

\subsection{The behaviour of higher orders of perturbation theory for our Yukawa model}

In examining our results from \ref{YukawaSec}, we ignore the high-$u$ fixed points (for the scalar self interaction), as we expect them to be untrustworthy in perturbation theory. Here we clarify our intuition on this point.

A naive expectation of perturbation theory in a coupling $u$, is that for a quantity $f(u)$ (such as a beta function or partition function), there exists a sequence
\be
f_{N}\left( u\right)  =\sum^{N}_{n=0} f_{n} u^{n}
\ee
which converges to $f(u)$ as $N \to \infty$. In a field theory where the perturbation is generated by Feynman diagrams, the number of diagrams increases with $n$. This increases the number of terms that contribute to $f_n$ and consequently $f_n$ is expected to increase with higher $n$ \cite{LeGuillou:1990nq}; however in order to understand the convergence it will be insufficient to just have bounds on $f_n$.

Major progress on estimating $f_n$ was made by Bender and Wu \cite{Bender:1969si} for the ground state energy of the anharmonic oscillator in $D=1$ dimensions (the $\phi^4$ field theory for quantum mechanics). The wavefunction for the energy level with energy $E$ satisfies the Schr{\"o}dinger equation
\be
-\frac{d^{2}}{dx^{2}} \psi \left( x\right)  +\left( \frac{x^{2}}{4} +u\frac{x^{4}}{4} -E\right)  \psi \left( x\right)  =0,\  \  \  \  \  \  \  \psi \left( \pm \infty \right)  =0.
\ee
For $E=E_0$ the ground state energy has $f_{n}\sim -\left( \frac{6}{\pi^{3} } \right)^{1/2}  \left( -3\right)^{n}  \Gamma \left( n+\frac{1}{2} \right).$ The resulting series is divergent and is an example of an asymptotic series, where \cite{R15} 
\be
f\left( u\right)  -f_{N}\left( u\right)  =O\left( u^{N+1}\right)  \    {\rm as}\; u\rightarrow 0 .
\ee
If $u$ is $\epsilon$ dependent, then $\epsilon$ is another control parameter that one can use to make $u$ small. This gives additional confidence in the resulting fixed points.

The extension of Bender and Wu's work to higher order terms in field theory is intimately related to the contributions of instantons in false vacuum decay in a semi-classical analysis of path integrals \cite{Coleman:1985rnk, Lipatov:1976ny, LeGuillou:1990nq}. The resulting estimates for the higher order terms are qualitatively similar to that of Bender and Wu.

This analysis has been extended to $D \geq 3$ for Yukawa field theories involving a single fermion and scalar in \cite{zinn2021quantum}. Qualitatively similar results were found as for the $\phi^4$ theory.
 
Hence any finite number of higher order terms in perturbation theory would not allow us to investigate putative high-$u$ fixed points for $D$ near 4. 

\subsection{Comparison with a standard-model inspired Yukawa theory}
There is some similarity of our work with another non-gauge Yukawa model (which we denote by M2) that is obtained from a simplification of the Standard Model in the leptonic sector~\cite{R9}. The fields in M2 are a left-handed fermion doublet (under $SU(2) $), a right-handed fermion $ SU(2) $ singlet  and a $SU(2$) scalar doublet. There is a Yukawa coupling of the fermions and scalars consistent with the $SU(2)$ structure. The fact that there are multi-component (flavour) fields in M2 contrasts with the single Dirac fermion and  pseudoscalar field in the axion model that we consider \cite{ deCesare:2014dga, R10Boss, R11Boss, Sarkar:2022odh, Mavromatos:2023bdx}. For two component pseudoscalar fields, for example, it is not possible to distinguish a parity transformation from a rotation. Therefore in the presence of multi-component fields it is not always possible to make a $\mathcal{PT}$ transformation. Our axion model is manifestly \cPT-symmetric when the couplings flow away from Hermitician values.

We have two types of \cPT-symmetric extensions of Hermitian theories in the axion model. One is in terms of a negative self-coupling and the other is in terms of an imaginary $g$ (or negative $h$)~\cite{R3.14}. Starting from a Hermitian value of $u$ the renormalisation group flow to negative $u$ is possible. Such a feature was noted in the model of M2 as a possibility but issues of $\mathcal{PT}$ symmetry were not discussed there~\cite{R9}. We have noted that renormalisation group flows do \emph{not} connect positive $h$ to negative $h$. However, the renormalisation group flows are symmetric about the axis $h=0$ in the $h-u$ plane. See Appendix $\ref{Negative_h_appendix}$ for more discussion.

\section{Conclusions}
\label{sec:RG consequences}

In terms of a simple renormalisable field theory relevant for axion physics involving a pseudoscalar field and a Dirac fermion,  the role of renormalisation in linking Hermitian and \cPT-symmetric Hamiltonians in $D=4-\epsilon$ has been explored in depth.
In order to carry out this investigation, it has been necessary to use path integrals, which in turn has depended on the complex deformations of path integrals within the context of steepest descent paths \cite{R3b}. This deformation can be regarded as a non-trivial change in the measure employed in the definition of the path integral.  It has been argued that on complexifying the bosonic path in the path integral and invoking $\mathcal{PT}$ symmetry, that it is possible to have a theory where Green's functions can be calculated in a weak coupling expansion \cite{R3a}. In this limit, the path integral is defined on a steepest descent contour (or its higher dimensional generalisation the Lefschetz thimble). Expansions around individual stationary points on  the contour give rise to asymptotic series, of which the trivial saddle point gives the dominant contribution.

The key to our analysis is the flow pattern between $\epsilon$-dependent fixed points which provides a degree of control over the perturbation series~\cite{R11} in terms of the renormalised coupling, together with calculations of the renormalisation group performed at higher loop. More recently, the possible emergence of unstable \cPT-symmetric potentials in the Standard Model due to renormalisation has been considered within the framework of $\mathcal{PT}$ symmetry \cite{R8, R3a} (but restricted to $D=1$). This treatment can be enhanced to address the issues for $D=4$ since we have clarified
\begin{itemize}
    \item the steepest descent-like paths in the path integral, and the role of the trivial saddle points in function space within the steepest descent path, together with the sub-dominant contributions from the non-trivial fixed points.
    \item renormalisation around the trivial fixed point and introduction of Wilson-Fisher $\epsilon$-dependent fixed points.
    \item the significance of beta functions from Feynman perturbation theory and the renormalisation group flows of couplings.
    \item the usefulness of RGBeta, a program in the symbolic language program  Mathematica, which can handle complex values of couplings.

    \end{itemize}
    
 Our analysis has found that Hermitian to non-Hermitian flows occur only in terms of the quartic self-couplings. These flows have been observed previously in the context of Hermitian theories, but can now be \emph{reinterpreted in the context of $\mathcal{PT}$-symmetric theory} with full justification. We conjecture that renormalisation and the emergence of \cPT-symmetric theory starting with a Hermitian theory may well occur  in other field theories. This conjecture is related to the possibility of square-root type singularities in the coupling appearing generically in other field theories (just as in the Lee model). The robustness of these findings in other renormalisable field theories is worthy of further study.

\section*{Acknowledgements}

L.C. is supported by King’s College London through an NMES funded studentship. The work of S.S. is supported in part by the UK Science and Technology Facilities Research Council (STFC) under the research grant ST/T000759/1 and EPSRC grant EP/V002821/1. We would like to thank Wen-Yuan Ai, Carl Bender, Nick Mavromatos, Alex Soto and Andy Stergiou for discussions.

\appendix
\section{Data for fixed points and their stability for $\epsilon \neq 0$}
\label{nonzeroeps}
In this appendix, we give the series results in $\epsilon$ for the fixed points and their linear stability eigenvalues and eigenvectors. Here, we provide these results to three decimal places (unless exact, or where this would give no significant figures). 
\begin{itemize}
    \item $F_{0h} = 0$, $F_{0u} = 0$. This is the trivial Hermitian fixed point. The stability matrix has degenerate eigenvalues: $\Lambda_{0 ,1}=\Lambda_{0 ,2}=-\epsilon $.  For $\epsilon \ne 0$ (and sufficiently small), this is a UV-stable stellar node (so that trajectories which begin near $F_{0}$ approach $F_{0}$ on straight lines). 

    \item $F_{1h} = 0$, \\ $F_{1u} = 52.638 \epsilon+33.142 \epsilon^2+41.735 \epsilon^3+65.694 \epsilon^4+115.816 \epsilon^5+218.763 \epsilon^6+432.896 \epsilon^7+885.833 \epsilon^8+1859.156 \epsilon^9+3979.970 \epsilon^{10}+8656.771 \epsilon^{11}+19076.958 \epsilon^{12}$. \\ The stability matrix has eigenvalues $\Lambda_{1 ,1} = \epsilon-0.630 \epsilon^2-0.793 \epsilon^3-1.248 \epsilon^4-2.200 \epsilon^5-4.156 \epsilon^6-8.224 \epsilon^7-16.829 \epsilon^8-35.320 \epsilon^9-75.610 \epsilon^{10}-164.459 \epsilon^{11}-362.419 \epsilon^{12}$ \\ and $\Lambda_{1, 2} = - \epsilon+0.019 \epsilon^2+0.019 \epsilon^3+0.028 \epsilon^4+0.048 \epsilon^5+0.090 \epsilon^6+0.176 \epsilon^7+0.359 \epsilon^8+0.750 \epsilon^9+1.601 \epsilon^{10}+3.472 \epsilon^{11}+7.636 \epsilon^{12}$, \\ with corresponding eigenvectors $\vec E_{1,1} = \left( \begin{matrix}0\\ 1\end{matrix} \right)$ and $\vec E_{1,2} = \left( \begin{matrix}A_{1,2}\\ 1\end{matrix} \right)$, \\ with $A_{1,2} = -0.750+0.340 \epsilon+0.383 \epsilon^2+0.566 \epsilon^3+0.960 \epsilon^4+1.765 \epsilon^5+3.425 \epsilon^6+6.905 \epsilon^7+14.326 \epsilon^8+30.386 \epsilon^9+65.590 \epsilon^{10}+143.623 \epsilon^{11}+318.258 \epsilon^{12}$. \\ For $\epsilon \ne 0$ (and sufficiently small), this is a Hermitian saddle fixed point.

    \item $F_{2h} = 15.791 \epsilon+1.819 \epsilon^2+1.646 \epsilon^3-0.757 \epsilon^4+0.405 \epsilon^5-1.241 \epsilon^6+0.643 \epsilon^7-1.430 \epsilon^8+1.411 \epsilon^9-1.983 \epsilon^{10}+2.625 \epsilon^{11}-3.393 \epsilon^{12}$, \\ $F_{2u} = -58.121 \epsilon+16.812 \epsilon^2-8.154 \epsilon^3+16.338 \epsilon^4-9.360 \epsilon^5+17.343 \epsilon^6-16.587 \epsilon^7+23.178 \epsilon^8-28.866 \epsilon^9+37.721 \epsilon^{10}-50.784 \epsilon^{11}+67.832 \epsilon^{12}$. \\ The stability matrix has eigenvalues $\Lambda_{2 ,1} = -2.408 \epsilon-0.601 \epsilon^2+1.301 \epsilon^3-0.089 \epsilon^4+1.006 \epsilon^5-0.593 \epsilon^6+0.986 \epsilon^7-1.204 \epsilon^8+1.462 \epsilon^9-2.076 \epsilon^{10}+2.641 \epsilon^{11}-3.691 \epsilon^{12}$ \\ and $\Lambda_{2, 2} = \epsilon-0.115 \epsilon^2-0.159 \epsilon^3+0.186 \epsilon^4-0.072 \epsilon^5+0.255 \epsilon^6-0.173 \epsilon^7+0.336 \epsilon^8-0.383 \epsilon^9+0.546 \epsilon^{10}-0.752 \epsilon^{11}+1.030 \epsilon^{12}$, \\ with corresponding eigenvectors $\vec E_{2,1} = \left( \begin{matrix}A_{2,1}\\ 1\end{matrix} \right)$ and $\vec E_{2,2} = \left( \begin{matrix}A_{2,2}\\ 1\end{matrix} \right)$, \\ with $A_{2,1} = 0.015 \epsilon-0.013 \epsilon^2+0.001 \epsilon^3-0.013 \epsilon^4+0.006 \epsilon^5-0.015 \epsilon^6+0.014 \epsilon^7-0.022 \epsilon^8+0.029 \epsilon^9-0.039 \epsilon^{10}+0.055 \epsilon^{11}-0.077 \epsilon^{12}$ \\ and $A_{2,2} = -0.272-0.188 \epsilon-0.075 \epsilon^2-0.159 \epsilon^3-0.087 \epsilon^4-0.178 \epsilon^5-0.080 \epsilon^6-0.209 \epsilon^7-0.051 \epsilon^8-0.262 \epsilon^9+0.014 \epsilon^{10}-0.362 \epsilon^{11}+0.293 \epsilon^{12}$. \\ For $\epsilon \ne 0$ (and sufficiently small), this is a non-Hermitian saddle fixed point.

    \item $F_{3h} = 15.791 \epsilon+6.749 \epsilon^2-3.314 \epsilon^3-12.829 \epsilon^4-11.559 \epsilon^5+9.263 \epsilon^6+37.770 \epsilon^7+28.770 \epsilon^8-64.624 \epsilon^9-196.697 \epsilon^{10}-156.077 \epsilon^{11}+274.654 \epsilon^{12}$, \\ $F_{3u} = 68.648 \epsilon+29.392 \epsilon^2+2.112 \epsilon^3-11.144 \epsilon^4+26.493 \epsilon^5+143.046 \epsilon^6+300.979 \epsilon^7+383.667 \epsilon^8+347.310 \epsilon^9+566.087 \epsilon^{10}+2056.631 \epsilon^{11}+5955.454 \epsilon^{12}$. \\ The stability matrix has eigenvalues $\Lambda_{3 ,1} = \epsilon-0.427 \epsilon^2+0.785 \epsilon^3+1.460 \epsilon^4+0.700 \epsilon^5-1.668 \epsilon^6-2.969 \epsilon^7+2.758 \epsilon^8+20.656 \epsilon^9+48.759 \epsilon^{10}+86.232 \epsilon^{11}+188.086 \epsilon^{12}$ \\ and $\Lambda_{3, 2} = 2.408 \epsilon-2.406 \epsilon^2-3.775 \epsilon^3-2.340 \epsilon^4+1.815 \epsilon^5+4.386 \epsilon^6-3.621 \epsilon^7-28.393 \epsilon^8-59.880 \epsilon^9-72.951 \epsilon^{10}-78.896 \epsilon^{11}-238.428 \epsilon^{12}$, \\ with corresponding eigenvectors $\vec E_{3,1} = \left( \begin{matrix}A_{3,1}\\ 1\end{matrix} \right)$ and $\vec E_{3,2} = \left( \begin{matrix}A_{3,2}\\ 1\end{matrix} \right)$, \\ with $A_{3,1} = 0.230-0.000 \epsilon-0.244 \epsilon^2-0.768 \epsilon^3-1.951 \epsilon^4-4.607 \epsilon^5-10.748 \epsilon^6-25.330 \epsilon^7-60.213 \epsilon^8-143.193 \epsilon^9-339.680 \epsilon^{10}-806.636 \epsilon^{11}-1998.394 \epsilon^{12}$ \\ and $A_{3,2} = -0.018 \epsilon+0.019 \epsilon^2+0.060 \epsilon^3+0.141 \epsilon^4+0.332 \epsilon^5+0.867 \epsilon^6+2.439 \epsilon^7+6.966 \epsilon^8+19.714 \epsilon^9+55.425 \epsilon^{10}+156.092 \epsilon^{11}+441.899 \epsilon^{12}$. \\ For $\epsilon \ne 0$ (and sufficiently small), this is a Hermitian IR-stable fixed point.

    \item $F_{4h} = 0$, \\ $F_{4u} = 83.601-52.638 \epsilon-33.142 \epsilon^2-41.735 \epsilon^3-65.694 \epsilon^4-115.816 \epsilon^5-218.763 \epsilon^6-432.896 \epsilon^7-885.833 \epsilon^8-1859.156 \epsilon^9-3979.970 \epsilon^{10}-8656.771 \epsilon^{11}-19076.958 \epsilon^{12}$. \\ The stability matrix has eigenvalues $\Lambda_{4 ,1} = -1.588+3.000 \epsilon+0.630 \epsilon^2+0.793 \epsilon^3+1.248 \epsilon^4+2.200 \epsilon^5+4.156 \epsilon^6+8.224 \epsilon^7+16.829 \epsilon^8+35.320 \epsilon^9+75.610 \epsilon^{10}+164.459 \epsilon^{11}+362.419 \epsilon^{12}$ \\ and $\Lambda_{4, 2} = 0.028-1.024 \epsilon-0.019 \epsilon^2-0.019 \epsilon^3-0.028 \epsilon^4-0.048 \epsilon^5-0.090 \epsilon^6-0.176 \epsilon^7-0.359 \epsilon^8-0.750 \epsilon^9-1.601 \epsilon^{10}-3.472 \epsilon^{11}-7.636 \epsilon^{12}$, \\ with corresponding eigenvectors $\vec E_{4,1} = \left( \begin{matrix}0\\ 1\end{matrix} \right)$ and $\vec E_{4,2} = \left( \begin{matrix}A_{4,2}\\ 1\end{matrix} \right)$, \\ with $A_{4,2} = 1.854-7.949 \epsilon+14.292 \epsilon^2-28.867 \epsilon^3+53.621 \epsilon^4-107.027 \epsilon^5+199.582 \epsilon^6-398.417 \epsilon^7+740.733 \epsilon^8-1486.571 \epsilon^9+2742.767 \epsilon^{10}-5559.741 \epsilon^{11}+10127.112 \epsilon^{12}$. \\ For $\epsilon \ne 0$ (and sufficiently small), this is a Hermitian saddle fixed point.

\end{itemize}

\section{Robustness of the loop analysis}
\label{Robustsubsection}
In this appendix, we examine the consistency of our results for the fixed points found in \ref{stabnonzeroeps}, by varying the orders of loops \footnote{This procedure has also been advocated in \cite{R9}.}. In \ref{stabnonzeroeps}, we gave, for example, the renormalisation group flows for $\epsilon = 0.01$ as a representative flow for the case $3+2$ where the $3$ refers to calculation of beta functions to $3$ loops in the Yukawa coupling and $2$ refers to $2$ loops in the scalar self-coupling.  

We report on the sensitivity of our results to loop order.
%in terms of the finite orders we apply in the perturbation theory, both in the coupling constants $g$ and $u$, and also the engineering dimension $\epsilon$. k, for $\epsilon = 0.01$, how robust are the results to changing the loop order of the couplings? 
The package RGBeta allows changes to the order of the loops.
 We  compare the results for different loop orders: $1+1$, $2+1$, $2+2$ and $3+2$ in the Figure \ref{0.01Loops} and focus on the fixed points that spawn from  the origin in coupling constant space as $\epsilon$ is turned on~\footnote{The other fixed points are too large for perturbation theory to be reliable.}. 

\begin{figure*}[ht]
        \centering
        \begin{subfigure}[t]{0.475\textwidth}
            \centering
            \includegraphics[scale=0.6]{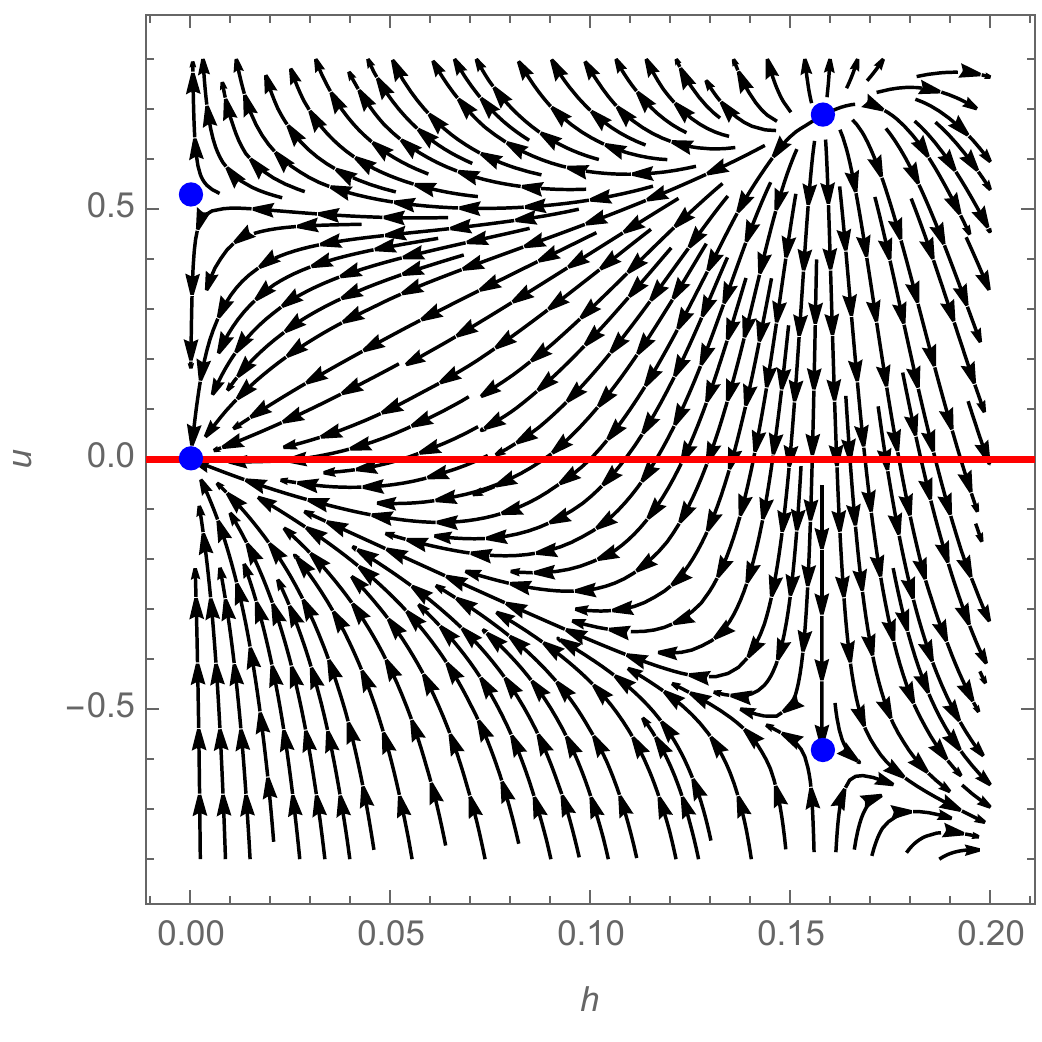}
            \caption[]%
            {{\small $1+1$ loop.}}    
            \label{0.01_1+1}
        \end{subfigure}
        \hfill
        \begin{subfigure}[t]{0.475\textwidth}  
            \centering 
            \includegraphics[scale=0.585]{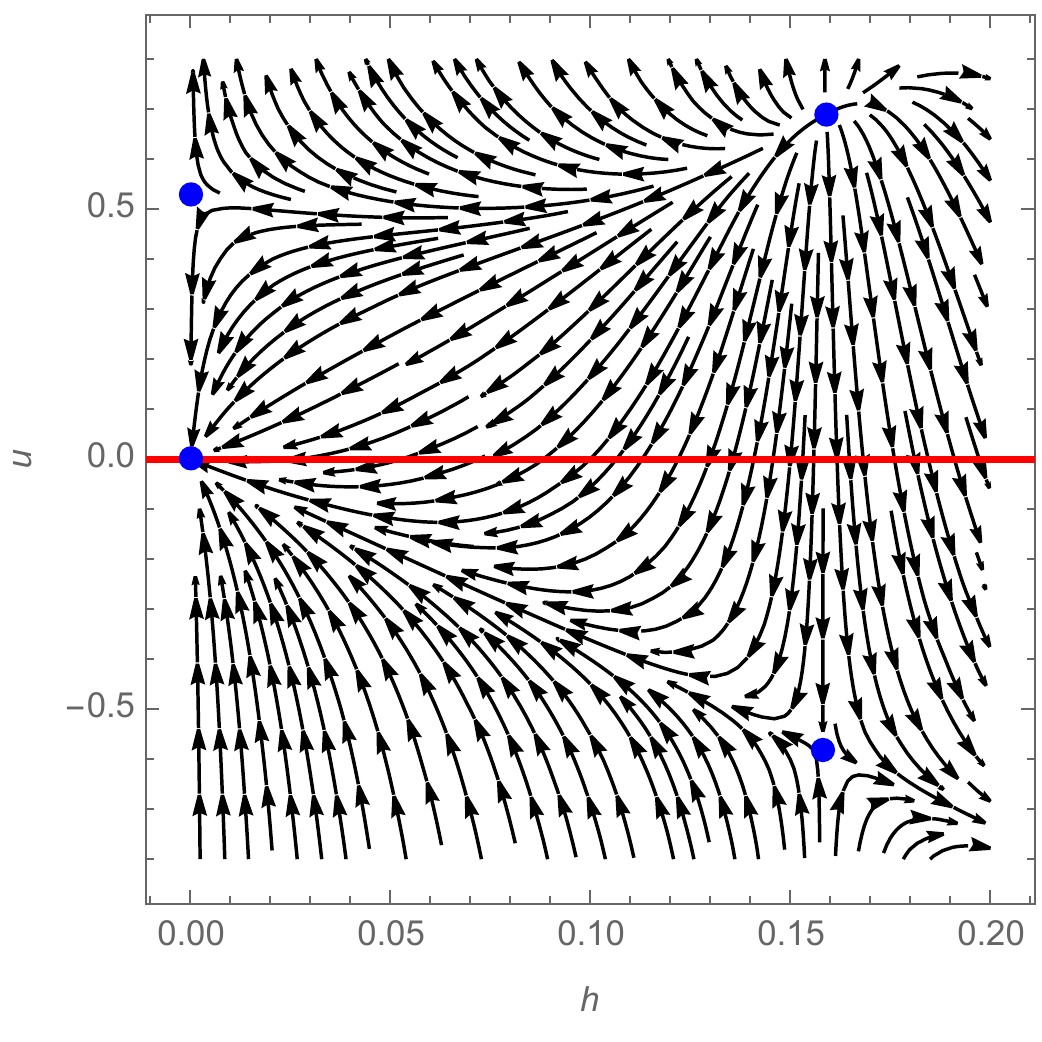}
            \caption[]%
            {{\small $2+1$ loops.}}    
            \label{0.01_2+1}
        \end{subfigure}
        \vskip\baselineskip
        \begin{subfigure}[t]{0.475\textwidth}   
            \centering 
            \includegraphics[scale=0.6]{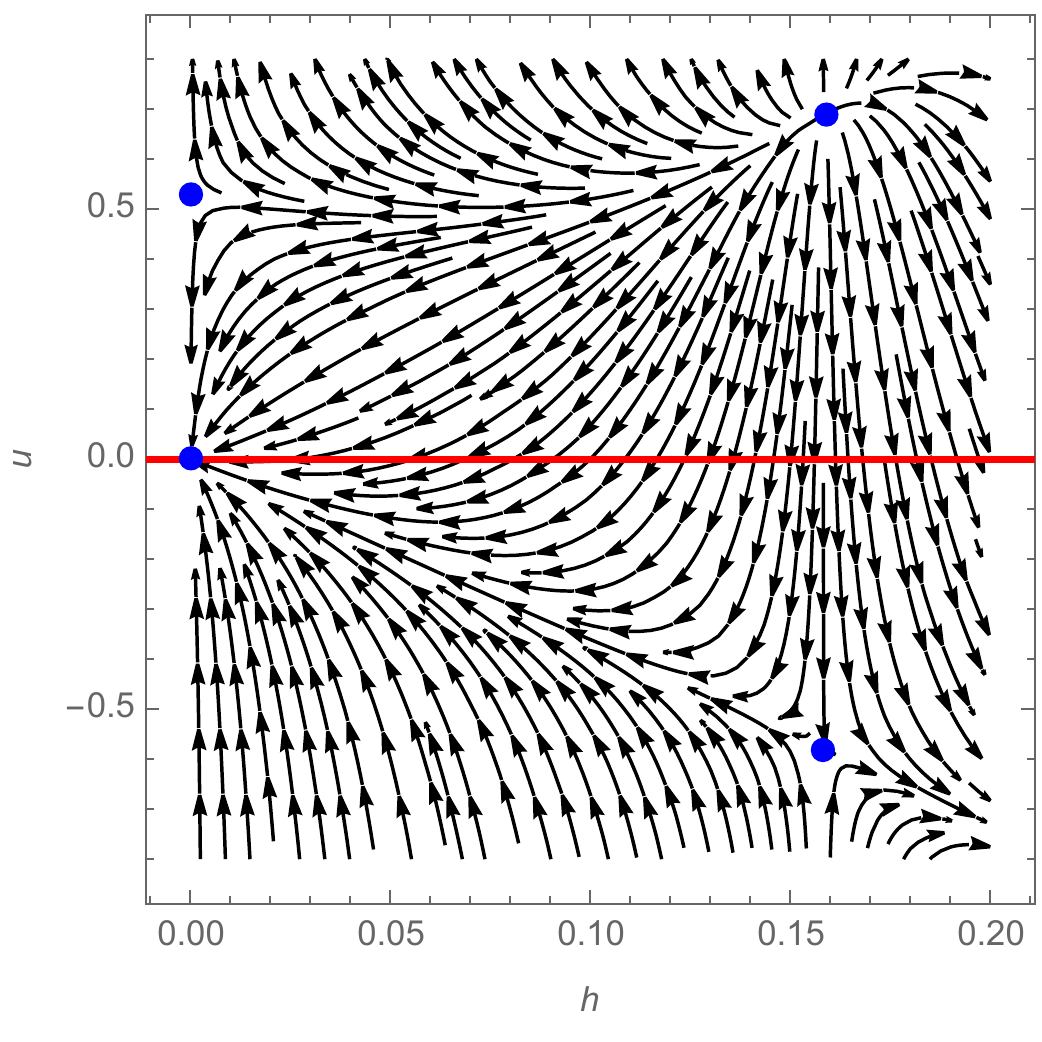}
            \caption[]%
            {{\small $2+2$ loops.}}    
            \label{0.01_2+2}
        \end{subfigure}
        \hfill
        \begin{subfigure}[t]{0.475\textwidth}   
            \centering 
            \includegraphics[scale=0.585]{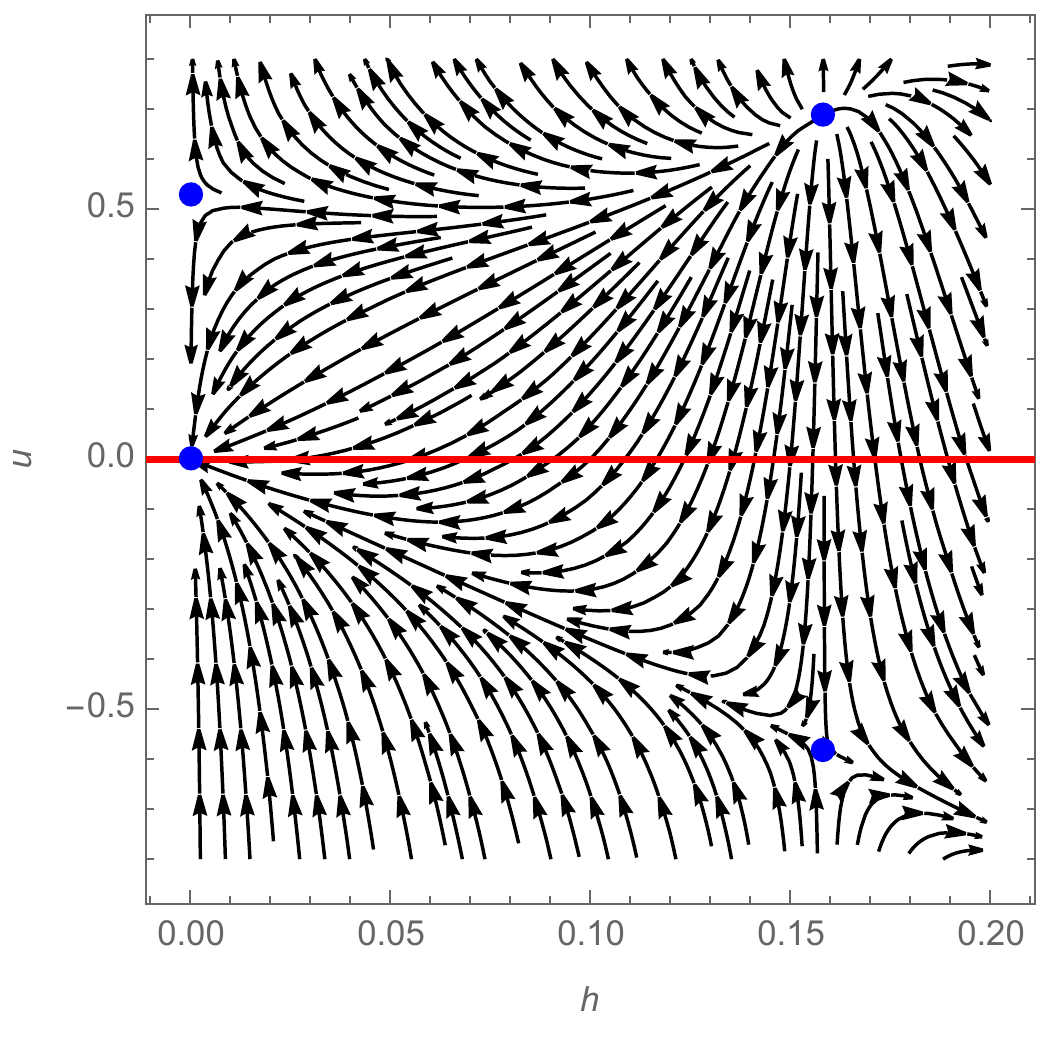}
            \caption[]%
            {{\small $3+2$ loops.}}    
            \label{0.01_3+2}
        \end{subfigure}
        \caption[]
        {\small Flows (at the labelled loop orders) near the fixed points that spawn from the origin as $\epsilon$ is introduced, for $\epsilon = 0.01$.} 
        \label{0.01Loops}
    \end{figure*}

\eject

The resulting flows for $\epsilon = 0.01$ for the aforementioned loop orders are plotted in Figure \ref{0.01Loops}. Qualitatively, we observe that the flow diagrams in Figure \ref{0.01Loops} appear very similar on changing the loop order. Quantitatively, in terms of $h$ and $u$, the fixed points only vary at most with 1\% relative difference, as we change the loop orders in the manner prescribed above. Since the magnitudes of the coupling constants at the fixed points are  small, it is consistent that an increase of loop order only leads to small changes, i.e. the additional terms that enter into the beta functions are subdominant at this level. The changes of the fixed point couplings are more significant at the  lower end of the loop orders (or equivalently the coupling constant values at the fixed points are more stable at the higher end of the loop orders). 

 Furthermore we can check whether this feature continues to hold as we begin to increase $\epsilon$. To probe this, we consider $\epsilon = 0.1$ and perform the same analysis (through changing the loop orders) as above. The resulting flows are shown in Figure \ref{0.1Loops}.
\begin{figure*}[ht]
        \centering
        \begin{subfigure}[t]{0.475\textwidth}
            \centering
            \includegraphics[scale=0.6]{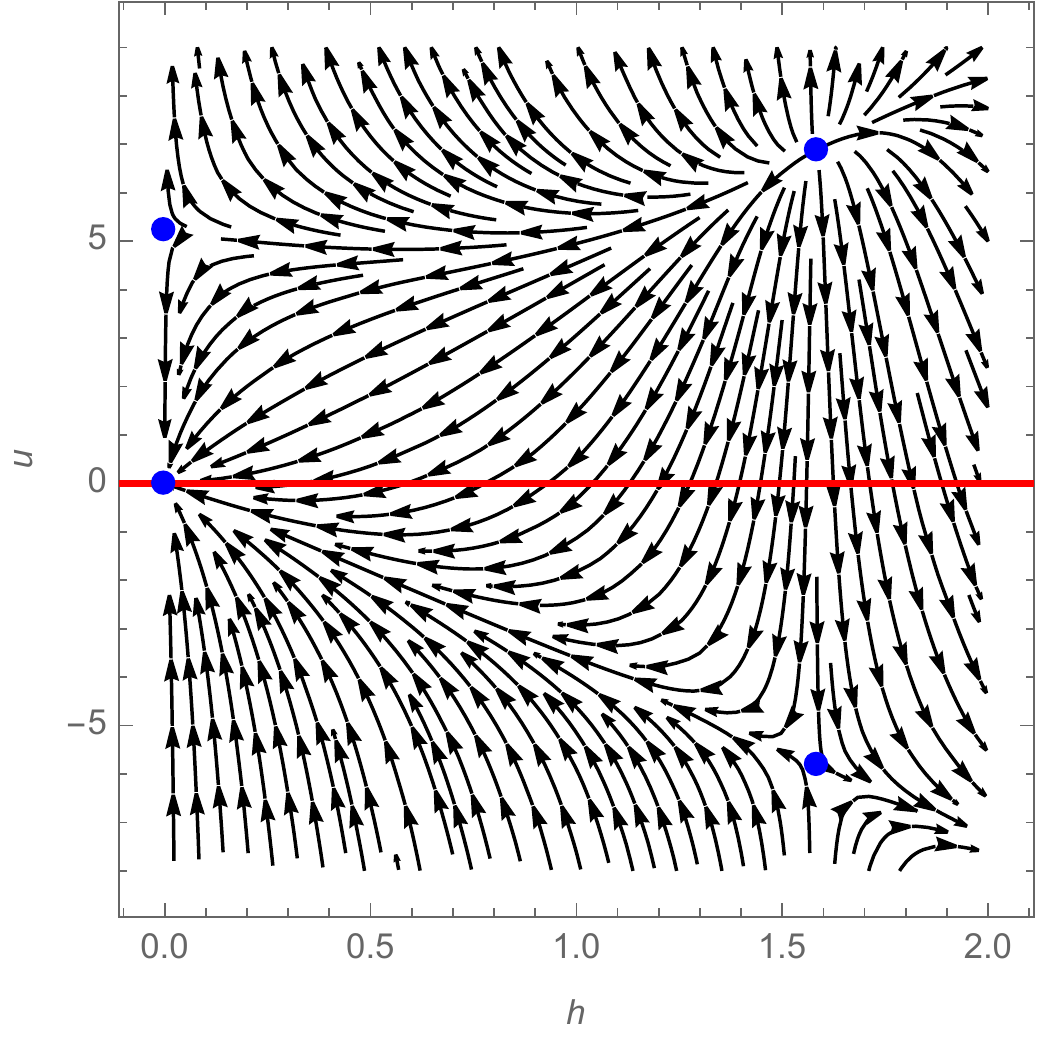}
            \caption[]%
            {{\small $1+1$ loop.}}    
            \label{0.1_1+1}
        \end{subfigure}
        \hfill
        \begin{subfigure}[t]{0.475\textwidth}  
            \centering 
            \includegraphics[scale=0.585]{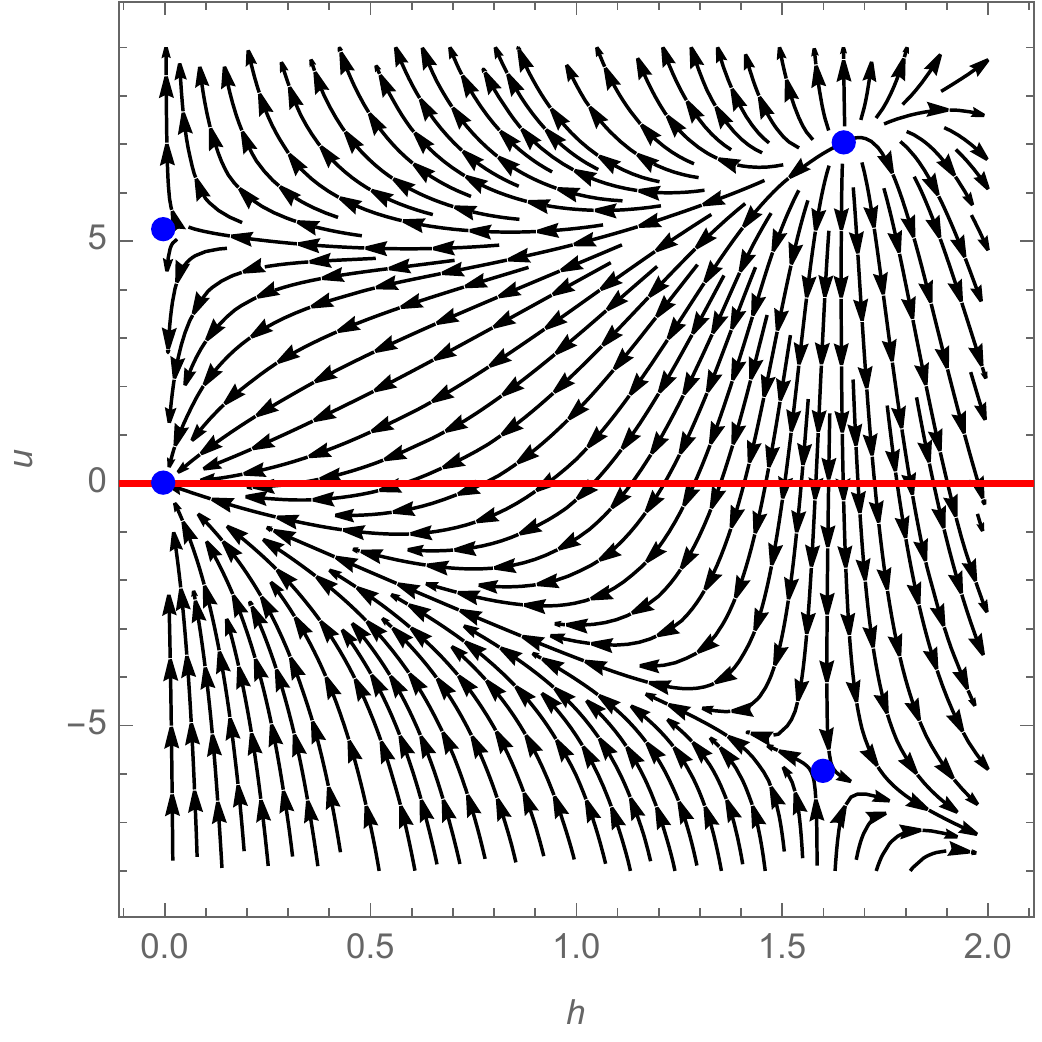}
            \caption[]%
            {{\small $2+1$ loops.}}    
            \label{0.1_2+1}
        \end{subfigure}
        \vskip\baselineskip
        \begin{subfigure}[t]{0.475\textwidth}   
            \centering 
            \includegraphics[scale=0.6]{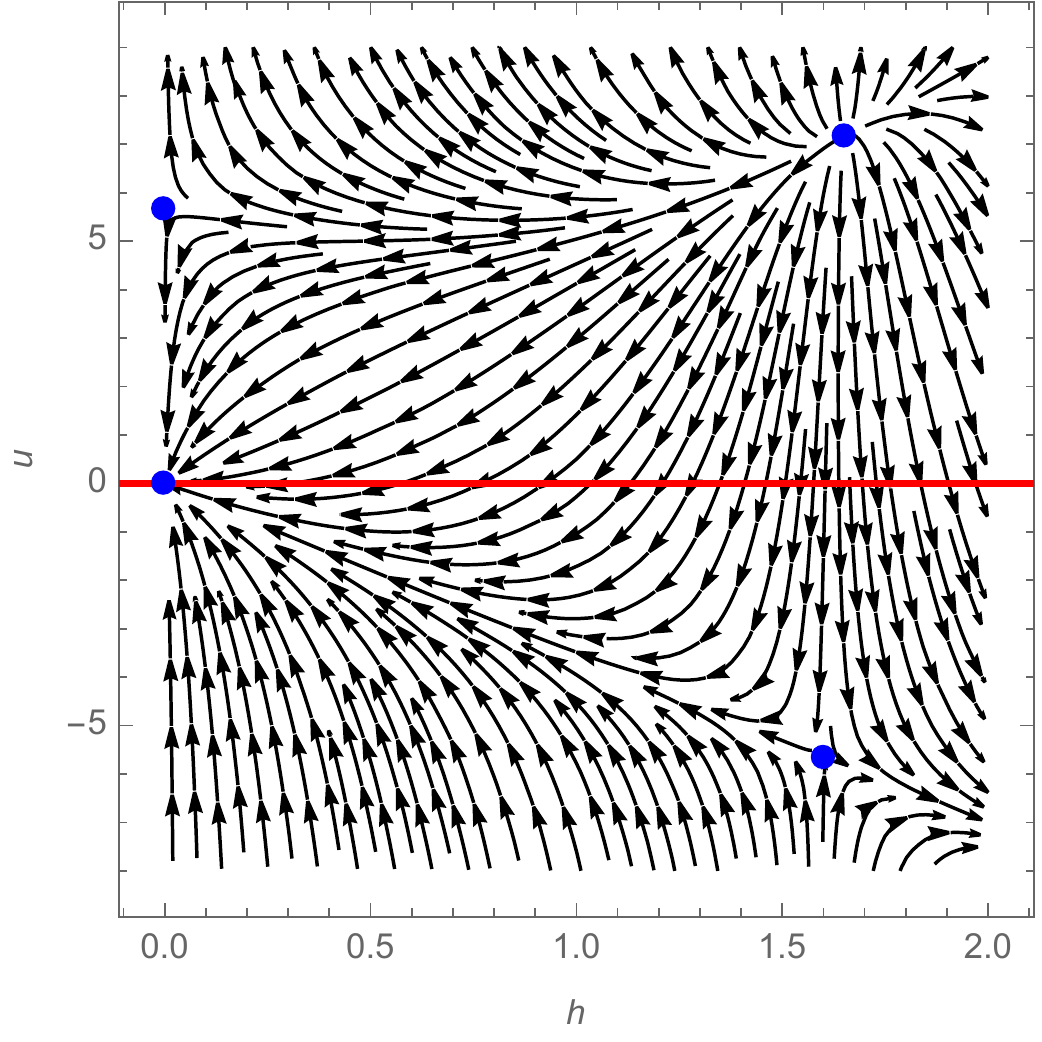}
            \caption[]%
            {{\small $2+2$ loops.}}    
            \label{0.1_2+2}
        \end{subfigure}
        \hfill
        \begin{subfigure}[t]{0.475\textwidth}   
            \centering 
            \includegraphics[scale=0.585]{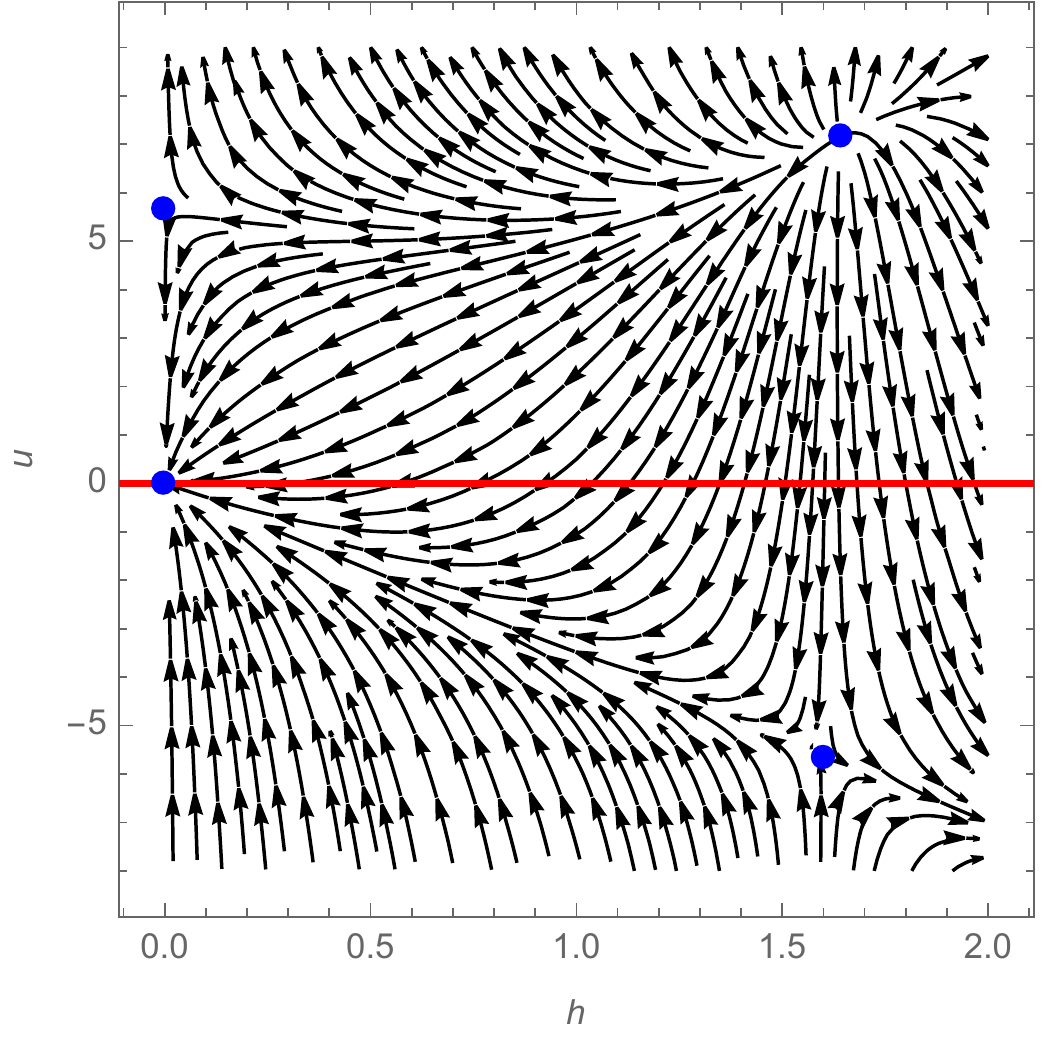}
            \caption[]%
            {{\small $3+2$ loops.}}    
            \label{0.1_3+2}
        \end{subfigure}
        \caption[]
        {\small Flows (at the labelled loop orders) near the fixed points that spawn from the origin as $\epsilon$ is introduced, for $\epsilon = 0.1$.} 
        \label{0.1Loops}
    \end{figure*} The  flow diagrams in Figure \ref{0.1Loops} remain similar as we change the loop order. The relative difference of the fixed point values vary at most by 5\%, as we change the loop orders. This maximum relative difference is moderately strong for $\epsilon = 0.1$ compared to the corresponding result for $\epsilon = 0.01$, and so indeed the higher loop corrections to the beta function become more significant at larger $\epsilon$ (as we would expect). As before, the changes are more significant at the lower end of the loop orders.

We could continue increasing $\epsilon$, but, as noted in Section \ref{stabnonzeroeps}, there is a critical value $\epsilon_{c'} \sim 0.44$ beyond which the character of one of the $\epsilon$-dependent fixed points change. By this point, the value of $\epsilon$ is likely too large to trust in the perturbative expansion in $\epsilon$; and simultaneously the resulting $\epsilon$-dependent fixed points spawning from the origin also become too large in magnitude to trust the perturbation theory.  

Therefore,~from these tests, we conclude that  within the region of parameter space for which perturbation theory is likely to be valid, the results from Section \ref{stabnonzeroeps} are robust.  

We can also consider the robustness of the non-perturbative results in Section \ref{PadeSec}. There, we find a putative non-perturbative $D=3$ fixed point, which is non-Hermitian with saddle stability. Since we perform a Pad\'e analysis and set $\epsilon = 1$, we can only consider the robustness of these results for different loop orders, taking $1+1$, $2+1$, $2+2$ and $3+2$ loops as above. We give the values of the coupling constants $\left( h^{\ast },u^{\ast }\right)$, the eigenvalues $\Lambda_{1, 2}$ and the eigenvectors $E_{1, 2}$, in each case as
\begin{itemize}
    \item 1+1 loops: $\left( h^{\ast },u^{\ast }\right) = \left( 15.8,-58.1\right)$, $\Lambda_{1} = -2.41$, $\Lambda_{2} = 1.00$, $E_{1} = (0, 1)$, $E_{2}=(-0.272, 1)$.
    \item 2+1 loops: $\left( h^{\ast },u^{\ast }\right) = \left( 17.9,-69.4\right)$, $\Lambda_{1} = -3.00$, $\Lambda_{2} = 0.873$, $E_{1} = (0.0190, 1)$, $E_{2}=(-0.250, 1)$.
    \item 2+2 loops: $\left( h^{\ast },u^{\ast }\right) = \left( 25.9,-76.7\right)$, $\Lambda_{1} = -0.953$, $\Lambda_{2} = 1.82$, $E_{1} = (0.0858, 1)$, $E_{2}=(-0.0826, 1)$.
    \item 3+2 loops: $\left( h^{\ast },u^{\ast }\right) = \left( 17.6,-32.3\right)$, $\Lambda_{1} = -1.16$, $\Lambda_{2} = 1.08$, $E_{1} = (-0.0121, 1)$, $E_{2}=(-4.21, 1)$.
\end{itemize}
The values of the coupling constants, and the eigenvalues and eigenvectors do not appear to be converging as the loop orders increase.~We should bear in mind that the application to epsilon expansions of Pad\'e approximants is long known not to be rigorous~\cite{R11}. However, in the results the fixed points do remain of the same character in each case: Hermitian in $h$, but non-Hermitian in $u$. In all of the loop cases that we consider here, the relevant $D=3$ fixed point is a non-Hermitian saddle. Hence this analysis is suggestive that there is a non-Hermitian fixed point which is of saddle type at $D=3$. Only a non-perturbative analysis, perhaps using the functional renormalisation group, can  prove the existence of such a fixed point rigorously.

\section{Discussion of $h<0$}
\label{Negative_h_appendix}
In the main text, we only consider $h \geq 0$ in our renormalisation group analysis, where $h = g^2$, and $g$ is the quartic self-coupling in the Lagrangian \eqref{e1}. In particular, the beta functions given in \eqref{betah} and \eqref{betau} are only valid for $h \geq 0$. When $h < 0$, $g$ is imaginary, which causes alterations of the beta functions through their dependence generally on $g$ and its complex conjugate $\bar{g}$. Instead, the beta functions in the $h < 0$ case are
\be
\begin{split}
\beta_{h} \left( h,u\right) = & - \epsilon h - \frac{1}{(4\pi)^2} 10h^2 + \frac{1}{(4\pi)^4} \left(-\frac{57}{2} h^3 + 4 h^2 u + \frac{1}{6} h u^2 \right) \\ & + \frac{1}{(4\pi)^6} \left(\left[\frac{339}{8} - 222 \; \zeta(3)\right] h^4 + 72 h^3 u - \frac{61}{24} h^2 u^2 - \frac{1}{8} h u^3 \right) 
\end{split}
\ee
and
\be
\beta_{u} \left( h,u\right) = - \epsilon u + \frac{1}{(4\pi)^{2} } \left( -48h^{2} - 8 h u + 3 u^{2}\right)  +\frac{1}{(4\pi)^{4} } \left( -384 h^{3} + 28 h^{2}u + 12 hu^{2} - \frac{17}{3} u^{3} \right).
\ee
These $h < 0$ beta functions are identical to those given in \eqref{betah} and \eqref{betau} for $h \geq 0$, except for the relative signs between terms. However, the signs work out such that the differential equations \eqref{basicbeta} governing the renormalisation group flows for $h<0$ are the same as those for $h>0$, but with $h \rightarrow -h$. This ultimately causes a $h \rightarrow -h$ reflection symmetry in the results.

\begin{figure*}[ht]
        \centering
        \begin{subfigure}[t]{0.3\textwidth}
            \centering
            \includegraphics[scale=0.5]{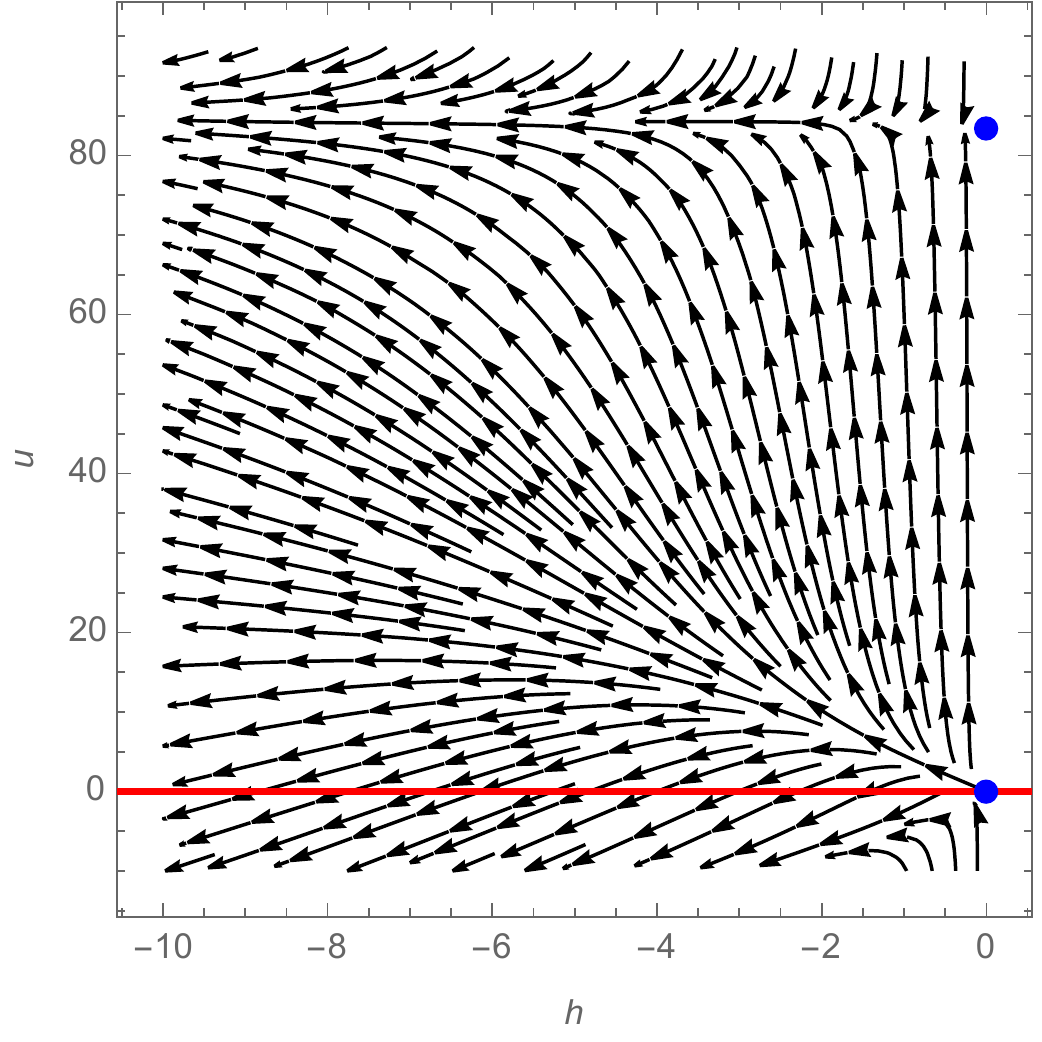}
            \caption[]%
            {{\small The global flow for $\epsilon = 0$.}}  
            \label{flow13}
        \end{subfigure}
        \hfill
        \begin{subfigure}[t]{0.3\textwidth}
            \centering 
            \includegraphics[scale=0.5]{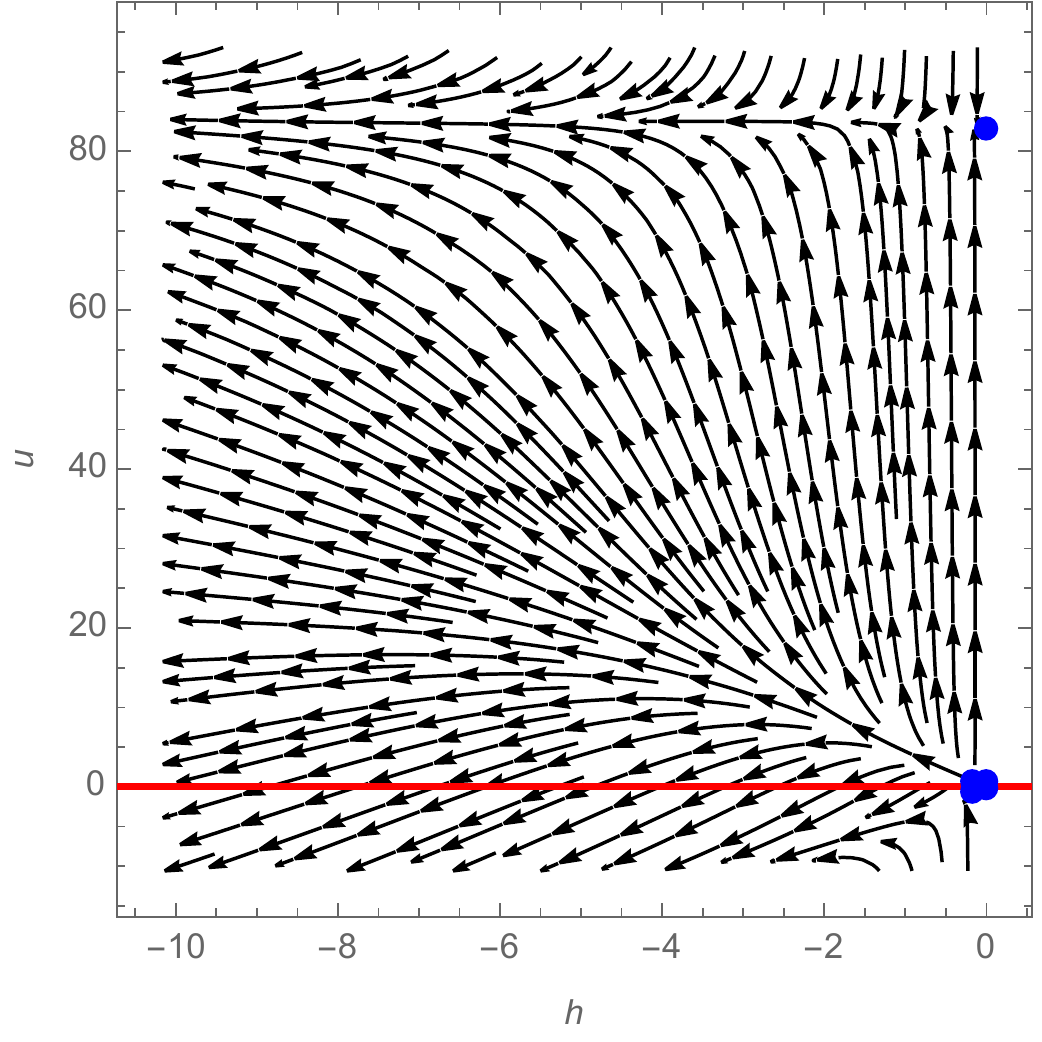}
            \caption[]%
            {{\small The global flow for $\epsilon = 0.01$.}}
            \label{flow14}
        \end{subfigure}
        \hfill
        \begin{subfigure}[t]{0.3\textwidth}
            \centering
            \includegraphics[scale=0.508]{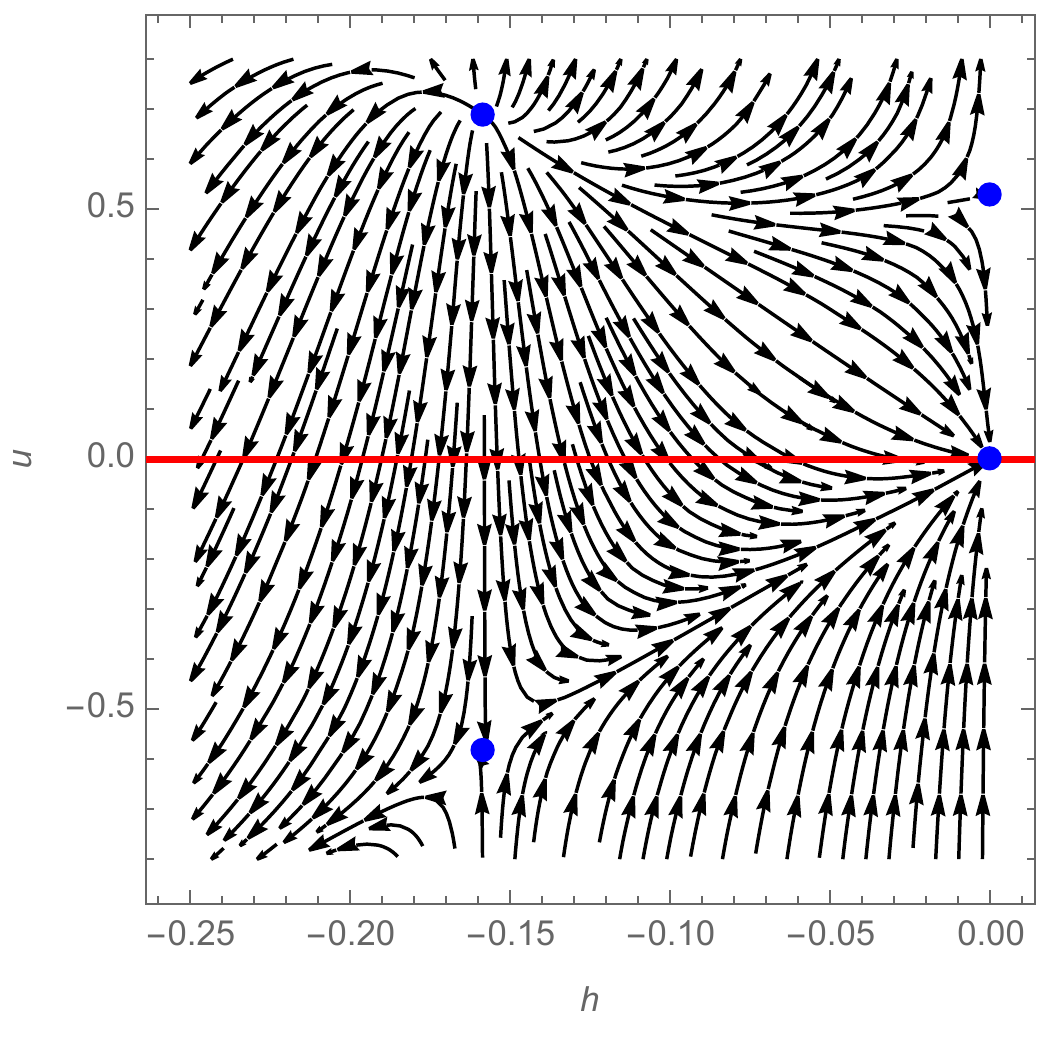}
            \caption[]%
            {{\small The flows around the group of fixed points near the origin for $\epsilon = 0.01$.}}  
            \label{flow15}
        \end{subfigure}
        \vskip\baselineskip
        \caption[]
        {\small Some flow diagrams for $h < 0$.}
        \label{h_negative_figures}
    \end{figure*}

We illustrate this in Figure \ref{h_negative_figures}, showing the $h<0$ results for
\begin{itemize}
    \item The global flow for $\epsilon = 0$.
    \item The global flow for $\epsilon = 0.01$.
    \item The flows around the group of fixed points near the origin for $\epsilon = 0.01$.
\end{itemize}
which are the analogues of the $h \geq 0$ results presented in Figures \ref{flow1}, \ref{flow4} and \ref{flow5}, respectively. Indeed, the figures show identical results to the aforementioned $h \geq 0$ counterparts, except reflected in the vertical $u$-axis. Furthermore, no flows cross the vertical $u$-axis, so the $h \geq 0$ sector can essentially be considered independently of the $h < 0$ sector (and there is no flow from the Hermitian to non-Hermitian values of $h$, or vice-versa).
For each fixed point with $h \geq 0$, there is an identical one with $h < 0$, with the same $|h|$, but opposite sign. The nature and stability of these fixed points are also preserved. For brevity, we therefore restrict to $h \geq 0$ in the main text. 

However, some non-trivial comments should be made:
\begin{itemize}
    \item In the case of $\epsilon = 0.01$, there are two fixed points with $h \neq 0$, shown in the last plot in Figure \ref{h_negative_figures}. These are therefore both non-Hermitian fixed points in $h$. Of particular interest is the point with $h<0$ and $u>0$, which is non-Hermitian but also IR stable, which may be significant for dynamical mass generation~\cite{R3a}.
    \item The symmetry gives rise to another non-Hermitian (now both in $g$ and $u$) saddle in the $D=3$ Pad\'e analysis with $\left( h^{\ast },u^{\ast }\right) = \left( -17.6,-32.3\right)$.
\end{itemize}

We further note that the possibility of negative $h$ in effective theories has been motivated previously ~\cite{R3.14} in terms of a microscopic picture. The picture is string inspired and is motivated by a mathematical ambiguity in continuing from an Euclidean to a Minkowski formulation. ~After compactification to four dimensions, the closed string sector of heterotic superstring theory~\cite{Green:2012oqa, Green:2012pqa} consists of spin 0 dilaton field $\Phi$, spin 2 graviton field $g_{\mu\nu}$ and spin 1 anti-symmetric gauge field tensor $B_{\mu \nu}$, the Kalb-Ramond field.~To lowest order in the string Regge slope $\alpha'$, the Euclidean effective action of the closed bosonic string is
\begin{equation}
\label{Eqn3}
S_{B}=-\int d^{4}x\  \sqrt{-g} \left( \frac{1}{2\kappa^{2} } R+\frac{1}{6} \mathcal{H}_{\lambda \mu \nu }\mathcal{H}^{\lambda \mu \nu }+\ldots \right)  
\end{equation}
where 
\begin{equation}
\label{Eqn2}
\mathcal{H}_{\mu \nu \rho }\left( x\right)  =\partial_{[\mu } B_{\nu \rho ]},
\end{equation}
$R$ is the Ricci scalar, $\kappa =\frac{\sqrt{8\pi } }{M_{P}} $, $M_{P}$ is the Planck mass, and $g$ is the determinant of $g_{\mu\nu}$. To this order in the expansion in $\alpha'$, $S_B$ can be interpreted as a modified gravity theory with torsion~\cite{Gross:1986mw, Metsaev:1987zx} where the usual metric based connection $\Gamma^{\rho }_{\  \mu \nu }$ is replaced by
\begin{equation}
\label{Eqn4}
\bar{\Gamma }^{\rho }_{\  \mu \nu } =\Gamma^{\rho }_{\  \mu \nu } +\frac{\kappa }{\sqrt{3} } \mathcal{H}^{\rho }_{\  \mu \nu }\neq \bar{\Gamma }^{\rho }_{\  \nu \mu } . 
\end{equation}
For the heterotic string the Bianchi identity is 
\begin{equation}
\label{Eqn6}
\epsilon^{\mu \nu \rho \sigma } \mathcal{H}_{[\nu \rho \sigma ;\mu ]}=\frac{\alpha^{\prime } }{32\kappa } \sqrt{-g} \left( R_{\mu \nu \rho \sigma }\tilde{R}^{\mu \nu \rho \sigma } -F^{a}_{\mu \nu }\tilde{F}^{a\mu \nu } \right)  \equiv \sqrt{-g} \, \mathcal{G}\left( \omega ,{\bf{A}}\right) 
\end{equation}
where ${\bf{A}}^a$ is a Yang-Mills gauge field with a Latin group index $a$ and
\begin{equation}
\label{Eqn8}
\tilde{R}_{\mu \nu \rho \sigma } =\frac{1}{2} \epsilon_{\mu \nu \lambda \pi } R^{\lambda \pi }_{\  \  \  \rho \sigma },\  \  \tilde{F^{a}}_{\mu \nu } =\frac{1}{2} \epsilon_{\mu \nu \lambda \pi } F^{a\lambda \pi }.\  \ \end{equation} 
with 
\begin{equation}
\label{Eqn7}
\epsilon^{\mu \nu \rho \sigma } =\frac{\text{sgn} \left( g\right)  }{\sqrt{-g} } \eta^{\mu \nu \rho \sigma } 
\end{equation}
and $\eta^{\mu \nu \rho \sigma } $ is the \emph{flat space} Levi-Civita symbol with $\eta^{0 1 2 3 } =1$. The Bianchi identity is implemented in the path integral $Z_B$ through a delta function:
\begin{equation}
\label{Eqn9}
Z_{B}=\int D\mathcal{H}\  \exp \left( -S_{B}\right) \prod_{x} \delta \left(\eta^{\mu \nu \rho \sigma } \mathcal{H}_{[\nu \rho \sigma ;\mu ]}\left(x \right) -\mathcal{G}\left( \omega ,{\bf{A}}\right)\right). 
\end{equation}
The axion field $b(x)$ appears as a Lagrange multiplier field implementing the delta function
\begin{equation}
\label{Eqn10B}
\int Db\exp \left[ -i\int d^{4}x\sqrt{-g(x)} \left( \frac{1}{\sqrt{3} } \partial^{\mu } b(x)\eta_{\mu \nu \rho \sigma } H^{\nu \rho \sigma }(x)+\frac{b}{\sqrt{3} }\mathcal{G}\left( \omega ,{\bf{A}}\right) \right)  \right]  
\end{equation}
On integrating over $\mathcal{H}$, $Z_B$ becomes
\begin{equation}
\label{Eqn11}
Z_{B}=\int db\exp \left( -\int d^{4}x\sqrt{g^{\left( E\right)  }} \left\{ \frac{1}{2\kappa^{2} } R+\frac{1}{12} \eta^{\left( E\right)  }_{\mu \nu \rho \lambda } \eta^{\mu \nu \rho \sigma \left( E\right)  } \partial^{\lambda } b\partial_{\sigma } b + \, \frac{b}{\sqrt{3} }\mathcal{G}\left( \omega ,{\bf{A}}\right)\right\} \right).  
\end{equation}
The Euclidean formulation is emphasised by using the superscript $(E)$. There is an ambiguity (or ordering issue)~\cite{Giddings:1987cg} on continuing back from Euclidean to Minkowski space. In~\cite{R3.14} it was stressed that one has two choices: 
\begin{enumerate}
  \item Before continuing back to Minkowski space  we can replace $\eta^{\left( E\right)  }_{\mu \nu \rho \lambda }\eta^{\mu \nu \rho \sigma \left( E\right)  } $  with $6\delta^{\sigma }_{\lambda } $.
  \item After continuing back to Minkowski space we can replace $\eta^{\left( E\right)  }_{\mu \nu \rho \lambda }\eta^{\mu \nu \rho \sigma \left( E\right)  } $  with $-6\delta^{\sigma }_{\lambda }(=\eta_{\mu \nu \rho \lambda } \eta^{\mu \nu \rho \sigma   }) $ and also redefine the phase of $b$ by $\pi/2$ in order to get the canonical sign for the kinetic term. This leads to the redefinition $b\rightarrow ib$.  A Hermitian $b$  transforms as $\mathcal{T}: b \to -b$ \cite{Bjorken:100769}; hence with the field redefinition we get the transformation in \eqref{par}.

On introducing fermions the above ambiguity leads to a Yukawa term
\begin{align}\label{bF}
\mathcal S_{\rm b-F} = {\rm const}\, \times \, \int d^4 x\, \sqrt{-g}\, i^\xi  \, b(x) \, \nabla_\mu \Big( \overline \psi \, \gamma^5 \, \gamma^\mu \, \psi \Big)~\,, 
\end{align}
with $\xi=0 ~\rm or~ 1$, depending on the way we analytically continue. 
Consequently it is not surprising that we did not find any renormalisation group flow between the Hermitian and non-Hermitian sectors of the Yukawa coupling constant $g$.
\end{enumerate}





\bibliography{bibli.bib}

\begin{thebibliography}{96}
\expandafter\ifx\csname natexlab\endcsname\relax\def\natexlab#1{#1}\fi
\expandafter\ifx\csname bibnamefont\endcsname\relax
  \def\bibnamefont#1{#1}\fi
\expandafter\ifx\csname bibfnamefont\endcsname\relax
  \def\bibfnamefont#1{#1}\fi
\expandafter\ifx\csname citenamefont\endcsname\relax
  \def\citenamefont#1{#1}\fi
\expandafter\ifx\csname url\endcsname\relax
  \def\url#1{\texttt{#1}}\fi
\expandafter\ifx\csname urlprefix\endcsname\relax\def\urlprefix{URL }\fi
\providecommand{\bibinfo}[2]{#2}
\providecommand{\eprint}[2][]{\url{#2}}

\bibitem[{\citenamefont{Alexandre et~al.}(2020{\natexlab{a}})\citenamefont{Alexandre, Ellis, and Millington}}]{R3.1}
\bibinfo{author}{\bibfnamefont{J.}~\bibnamefont{Alexandre}}, \bibinfo{author}{\bibfnamefont{J.}~\bibnamefont{Ellis}}, \bibnamefont{and} \bibinfo{author}{\bibfnamefont{P.}~\bibnamefont{Millington}}, \bibinfo{journal}{Phys. Rev. D} \textbf{\bibinfo{volume}{102}}, \bibinfo{pages}{125030} (\bibinfo{year}{2020}{\natexlab{a}}), \eprint{2006.06656}.

\bibitem[{\citenamefont{Alexandre et~al.}(2020{\natexlab{b}})\citenamefont{Alexandre, Ellis, Millington, and Seynaeve}}]{R3.2}
\bibinfo{author}{\bibfnamefont{J.}~\bibnamefont{Alexandre}}, \bibinfo{author}{\bibfnamefont{J.}~\bibnamefont{Ellis}}, \bibinfo{author}{\bibfnamefont{P.}~\bibnamefont{Millington}}, \bibnamefont{and} \bibinfo{author}{\bibfnamefont{D.}~\bibnamefont{Seynaeve}}, \bibinfo{journal}{Phys. Rev. D} \textbf{\bibinfo{volume}{101}}, \bibinfo{pages}{035008} (\bibinfo{year}{2020}{\natexlab{b}}), \eprint{1910.03985}.

\bibitem[{\citenamefont{Alexandre et~al.}(2019)\citenamefont{Alexandre, Ellis, Millington, and Seynaeve}}]{R3.3}
\bibinfo{author}{\bibfnamefont{J.}~\bibnamefont{Alexandre}}, \bibinfo{author}{\bibfnamefont{J.}~\bibnamefont{Ellis}}, \bibinfo{author}{\bibfnamefont{P.}~\bibnamefont{Millington}}, \bibnamefont{and} \bibinfo{author}{\bibfnamefont{D.}~\bibnamefont{Seynaeve}}, \bibinfo{journal}{Phys. Rev. D} \textbf{\bibinfo{volume}{99}}, \bibinfo{pages}{075024} (\bibinfo{year}{2019}), \eprint{1808.00944}.

\bibitem[{\citenamefont{Alexandre et~al.}(2018)\citenamefont{Alexandre, Ellis, Millington, and Seynaeve}}]{R3.4}
\bibinfo{author}{\bibfnamefont{J.}~\bibnamefont{Alexandre}}, \bibinfo{author}{\bibfnamefont{J.}~\bibnamefont{Ellis}}, \bibinfo{author}{\bibfnamefont{P.}~\bibnamefont{Millington}}, \bibnamefont{and} \bibinfo{author}{\bibfnamefont{D.}~\bibnamefont{Seynaeve}}, \bibinfo{journal}{Phys. Rev. D} \textbf{\bibinfo{volume}{98}}, \bibinfo{pages}{045001} (\bibinfo{year}{2018}), \eprint{1805.06380}.

\bibitem[{\citenamefont{Mannheim}(2021)}]{R3.5}
\bibinfo{author}{\bibfnamefont{P.~D.} \bibnamefont{Mannheim}} (\bibinfo{year}{2021}), \eprint{2109.08714}.

\bibitem[{\citenamefont{Mannheim}(2019)}]{R3.6}
\bibinfo{author}{\bibfnamefont{P.~D.} \bibnamefont{Mannheim}}, \bibinfo{journal}{Phys. Rev. D} \textbf{\bibinfo{volume}{99}}, \bibinfo{pages}{045006} (\bibinfo{year}{2019}), \eprint{1808.00437}.

\bibitem[{\citenamefont{Fring and Taira}(2021)}]{R3.7}
\bibinfo{author}{\bibfnamefont{A.}~\bibnamefont{Fring}} \bibnamefont{and} \bibinfo{author}{\bibfnamefont{T.}~\bibnamefont{Taira}}, \bibinfo{journal}{J. Phys. Conf. Ser.} \textbf{\bibinfo{volume}{2038}}, \bibinfo{pages}{012010} (\bibinfo{year}{2021}), \eprint{2103.13519}.

\bibitem[{\citenamefont{Fring and Taira}(2020{\natexlab{a}})}]{R3.8}
\bibinfo{author}{\bibfnamefont{A.}~\bibnamefont{Fring}} \bibnamefont{and} \bibinfo{author}{\bibfnamefont{T.}~\bibnamefont{Taira}}, \bibinfo{journal}{Phys. Lett. B} \textbf{\bibinfo{volume}{807}}, \bibinfo{pages}{135583} (\bibinfo{year}{2020}{\natexlab{a}}), \eprint{2006.02718}.

\bibitem[{\citenamefont{Fring and Taira}(2022)}]{R3.9}
\bibinfo{author}{\bibfnamefont{A.}~\bibnamefont{Fring}} \bibnamefont{and} \bibinfo{author}{\bibfnamefont{T.}~\bibnamefont{Taira}}, \bibinfo{journal}{Eur. Phys. J. Plus} \textbf{\bibinfo{volume}{137}}, \bibinfo{pages}{716} (\bibinfo{year}{2022}), \eprint{2004.00723}.

\bibitem[{\citenamefont{Fring and Taira}(2020{\natexlab{b}})}]{R3.10}
\bibinfo{author}{\bibfnamefont{A.}~\bibnamefont{Fring}} \bibnamefont{and} \bibinfo{author}{\bibfnamefont{T.}~\bibnamefont{Taira}}, \bibinfo{journal}{Phys. Rev. D} \textbf{\bibinfo{volume}{101}}, \bibinfo{pages}{045014} (\bibinfo{year}{2020}{\natexlab{b}}), \eprint{1911.01405}.

\bibitem[{\citenamefont{Fring and Taira}(2020{\natexlab{c}})}]{R3.11}
\bibinfo{author}{\bibfnamefont{A.}~\bibnamefont{Fring}} \bibnamefont{and} \bibinfo{author}{\bibfnamefont{T.}~\bibnamefont{Taira}}, \bibinfo{journal}{Nucl. Phys. B} \textbf{\bibinfo{volume}{950}}, \bibinfo{pages}{114834} (\bibinfo{year}{2020}{\natexlab{c}}), \eprint{1906.05738}.

\bibitem[{\citenamefont{Alexandre et~al.}(2017)\citenamefont{Alexandre, Millington, and Seynaeve}}]{R3.12}
\bibinfo{author}{\bibfnamefont{J.}~\bibnamefont{Alexandre}}, \bibinfo{author}{\bibfnamefont{P.}~\bibnamefont{Millington}}, \bibnamefont{and} \bibinfo{author}{\bibfnamefont{D.}~\bibnamefont{Seynaeve}}, \bibinfo{journal}{Phys. Rev. D} \textbf{\bibinfo{volume}{96}}, \bibinfo{pages}{065027} (\bibinfo{year}{2017}), \eprint{1707.01057}.

\bibitem[{\citenamefont{Mavromatos and Soto}(2021)}]{R3.13}
\bibinfo{author}{\bibfnamefont{N.~E.} \bibnamefont{Mavromatos}} \bibnamefont{and} \bibinfo{author}{\bibfnamefont{A.}~\bibnamefont{Soto}}, \bibinfo{journal}{Nucl. Phys. B} \textbf{\bibinfo{volume}{962}}, \bibinfo{pages}{115275} (\bibinfo{year}{2021}), \eprint{2006.13616}.

\bibitem[{\citenamefont{Mavromatos}(2020)}]{R3.14}
\bibinfo{author}{\bibfnamefont{N.~E.} \bibnamefont{Mavromatos}}, \bibinfo{journal}{J. Phys. Conf. Ser.} \textbf{\bibinfo{volume}{2038}}, \bibinfo{pages}{012019} (\bibinfo{year}{2020}), \eprint{2010.15790}.

\bibitem[{\citenamefont{Grinstein et~al.}(2008)\citenamefont{Grinstein, O'Connell, and Wise}}]{R3.15}
\bibinfo{author}{\bibfnamefont{B.}~\bibnamefont{Grinstein}}, \bibinfo{author}{\bibfnamefont{D.}~\bibnamefont{O'Connell}}, \bibnamefont{and} \bibinfo{author}{\bibfnamefont{M.~B.} \bibnamefont{Wise}}, \bibinfo{journal}{Phys. Rev. D} \textbf{\bibinfo{volume}{77}}, \bibinfo{pages}{025012} (\bibinfo{year}{2008}), \eprint{0704.1845}.

\bibitem[{\citenamefont{Bender and Boettcher}(1998)}]{R1}
\bibinfo{author}{\bibfnamefont{C.~M.} \bibnamefont{Bender}} \bibnamefont{and} \bibinfo{author}{\bibfnamefont{S.}~\bibnamefont{Boettcher}}, \bibinfo{journal}{Phys. Rev. Lett.} \textbf{\bibinfo{volume}{80}}, \bibinfo{pages}{5243} (\bibinfo{year}{1998}), \eprint{physics/9712001}.

\bibitem[{\citenamefont{Bender}(2019)}]{R2}
\bibinfo{author}{\bibfnamefont{C.}~\bibnamefont{Bender}}, \emph{\bibinfo{title}{PT Symmetry}} (\bibinfo{publisher}{WORLD SCIENTIFIC (EUROPE)}, \bibinfo{year}{2019}), \urlprefix\url{https://www.worldscientific.com/doi/abs/10.1142/q0178}.

\bibitem[{\citenamefont{Bender et~al.}(2002)\citenamefont{Bender, Brody, and Jones}}]{R17}
\bibinfo{author}{\bibfnamefont{C.~M.} \bibnamefont{Bender}}, \bibinfo{author}{\bibfnamefont{D.~C.} \bibnamefont{Brody}}, \bibnamefont{and} \bibinfo{author}{\bibfnamefont{H.~F.} \bibnamefont{Jones}}, \bibinfo{journal}{Phys. Rev. Lett.} \textbf{\bibinfo{volume}{89}}, \bibinfo{pages}{270401} (\bibinfo{year}{2002}), \bibinfo{note}{[Erratum: Phys.Rev.Lett. 92, 119902 (2004)]}, \eprint{quant-ph/0208076}.

\bibitem[{\citenamefont{Mavromatos et~al.}(2022)\citenamefont{Mavromatos, Sarkar, and Soto}}]{R3a}
\bibinfo{author}{\bibfnamefont{N.~E.} \bibnamefont{Mavromatos}}, \bibinfo{author}{\bibfnamefont{S.}~\bibnamefont{Sarkar}}, \bibnamefont{and} \bibinfo{author}{\bibfnamefont{A.}~\bibnamefont{Soto}}, \bibinfo{journal}{Phys. Rev. D} \textbf{\bibinfo{volume}{106}}, \bibinfo{pages}{015009} (\bibinfo{year}{2022}), \eprint{2111.05131}.

\bibitem[{\citenamefont{Mavromatos et~al.}(2023)\citenamefont{Mavromatos, Sarkar, and Soto}}]{R3c}
\bibinfo{author}{\bibfnamefont{N.~E.} \bibnamefont{Mavromatos}}, \bibinfo{author}{\bibfnamefont{S.}~\bibnamefont{Sarkar}}, \bibnamefont{and} \bibinfo{author}{\bibfnamefont{A.}~\bibnamefont{Soto}}, \bibinfo{journal}{Nucl. Phys. B} \textbf{\bibinfo{volume}{986}}, \bibinfo{pages}{116048} (\bibinfo{year}{2023}), \eprint{2208.12436}.

\bibitem[{\citenamefont{Thomsen}(2021)}]{R10}
\bibinfo{author}{\bibfnamefont{A.~E.} \bibnamefont{Thomsen}}, \bibinfo{journal}{Eur. Phys. J. C} \textbf{\bibinfo{volume}{81}}, \bibinfo{pages}{408} (\bibinfo{year}{2021}), \eprint{2101.08265}.

\bibitem[{\citenamefont{Pickering et~al.}(2001)\citenamefont{Pickering, Gracey, and Jones}}]{Pickering:2001aq}
\bibinfo{author}{\bibfnamefont{A.~G.~M.} \bibnamefont{Pickering}}, \bibinfo{author}{\bibfnamefont{J.~A.} \bibnamefont{Gracey}}, \bibnamefont{and} \bibinfo{author}{\bibfnamefont{D.~R.~T.} \bibnamefont{Jones}}, \bibinfo{journal}{Phys. Lett. B} \textbf{\bibinfo{volume}{510}}, \bibinfo{pages}{347} (\bibinfo{year}{2001}), \bibinfo{note}{[Erratum: Phys.Lett.B 535, 377 (2002)]}, \eprint{hep-ph/0104247}.

\bibitem[{\citenamefont{Poole and Thomsen}(2019)}]{Poole:2019kcm}
\bibinfo{author}{\bibfnamefont{C.}~\bibnamefont{Poole}} \bibnamefont{and} \bibinfo{author}{\bibfnamefont{A.~E.} \bibnamefont{Thomsen}}, \bibinfo{journal}{JHEP} \textbf{\bibinfo{volume}{09}}, \bibinfo{pages}{055} (\bibinfo{year}{2019}), \eprint{1906.04625}.

\bibitem[{\citenamefont{Bednyakov and Pikelner}(2021)}]{Bednyakov:2021qxa}
\bibinfo{author}{\bibfnamefont{A.}~\bibnamefont{Bednyakov}} \bibnamefont{and} \bibinfo{author}{\bibfnamefont{A.}~\bibnamefont{Pikelner}}, \bibinfo{journal}{Phys. Rev. Lett.} \textbf{\bibinfo{volume}{127}}, \bibinfo{pages}{041801} (\bibinfo{year}{2021}), \eprint{2105.09918}.

\bibitem[{\citenamefont{Davies et~al.}(2022)\citenamefont{Davies, Herren, and Thomsen}}]{Davies:2021mnc}
\bibinfo{author}{\bibfnamefont{J.}~\bibnamefont{Davies}}, \bibinfo{author}{\bibfnamefont{F.}~\bibnamefont{Herren}}, \bibnamefont{and} \bibinfo{author}{\bibfnamefont{A.~E.} \bibnamefont{Thomsen}}, \bibinfo{journal}{JHEP} \textbf{\bibinfo{volume}{01}}, \bibinfo{pages}{051} (\bibinfo{year}{2022}), \eprint{2110.05496}.

\bibitem[{\citenamefont{Rivers}(2011)}]{Rivers:2011zz}
\bibinfo{author}{\bibfnamefont{R.~J.} \bibnamefont{Rivers}}, \bibinfo{journal}{Int. J. Mod. Phys. D} \textbf{\bibinfo{volume}{20}}, \bibinfo{pages}{919} (\bibinfo{year}{2011}).

\bibitem[{\citenamefont{Ai et~al.}(2022)\citenamefont{Ai, Bender, and Sarkar}}]{R3b}
\bibinfo{author}{\bibfnamefont{W.-Y.} \bibnamefont{Ai}}, \bibinfo{author}{\bibfnamefont{C.~M.} \bibnamefont{Bender}}, \bibnamefont{and} \bibinfo{author}{\bibfnamefont{S.}~\bibnamefont{Sarkar}}, \bibinfo{journal}{Phys. Rev. D} \textbf{\bibinfo{volume}{106}}, \bibinfo{pages}{125016} (\bibinfo{year}{2022}), \eprint{2209.07897}.

\bibitem[{\citenamefont{Rivers}(1988)}]{Rivers:1987hi}
\bibinfo{author}{\bibfnamefont{R.~J.} \bibnamefont{Rivers}}, \emph{\bibinfo{title}{{PATH INTEGRAL METHODS IN QUANTUM FIELD THEORY}}}, Cambridge Monographs on Mathematical Physics (\bibinfo{publisher}{Cambridge University Press}, \bibinfo{year}{1988}), ISBN \bibinfo{isbn}{978-0-521-36870-4, 978-1-139-24186-1}.

\bibitem[{\citenamefont{Bjorken and Drell}(1965)}]{Bjorken:100769}
\bibinfo{author}{\bibfnamefont{J.~D.} \bibnamefont{Bjorken}} \bibnamefont{and} \bibinfo{author}{\bibfnamefont{S.~D.} \bibnamefont{Drell}}, \emph{\bibinfo{title}{{Relativistic quantum fields}}}, International series in pure and applied physics (\bibinfo{publisher}{McGraw-Hill}, \bibinfo{address}{New York, NY}, \bibinfo{year}{1965}).

\bibitem[{\citenamefont{Swanson}(1992)}]{Swanson:1992cz}
\bibinfo{author}{\bibfnamefont{M.~S.} \bibnamefont{Swanson}}, \emph{\bibinfo{title}{{Path integrals and quantum processes}}} (\bibinfo{year}{1992}).

\bibitem[{\citenamefont{Dowker et~al.}(2010)\citenamefont{Dowker, Johnston, and Sorkin}}]{Dowker:2010ng}
\bibinfo{author}{\bibfnamefont{F.}~\bibnamefont{Dowker}}, \bibinfo{author}{\bibfnamefont{S.}~\bibnamefont{Johnston}}, \bibnamefont{and} \bibinfo{author}{\bibfnamefont{R.~D.} \bibnamefont{Sorkin}}, \bibinfo{journal}{J. Phys. A} \textbf{\bibinfo{volume}{43}}, \bibinfo{pages}{275302} (\bibinfo{year}{2010}), \eprint{1002.0589}.

\bibitem[{\citenamefont{Bender et~al.}(2005{\natexlab{a}})\citenamefont{Bender, Brandt, Chen, and Wang}}]{R2a}
\bibinfo{author}{\bibfnamefont{C.~M.} \bibnamefont{Bender}}, \bibinfo{author}{\bibfnamefont{S.~F.} \bibnamefont{Brandt}}, \bibinfo{author}{\bibfnamefont{J.-H.} \bibnamefont{Chen}}, \bibnamefont{and} \bibinfo{author}{\bibfnamefont{Q.-h.} \bibnamefont{Wang}}, \bibinfo{journal}{Phys. Rev. D} \textbf{\bibinfo{volume}{71}}, \bibinfo{pages}{065010} (\bibinfo{year}{2005}{\natexlab{a}}), \eprint{hep-th/0412316}.

\bibitem[{\citenamefont{Jones and Rivers}(2009)}]{Jones:2009br}
\bibinfo{author}{\bibfnamefont{H.~F.} \bibnamefont{Jones}} \bibnamefont{and} \bibinfo{author}{\bibfnamefont{R.~J.} \bibnamefont{Rivers}}, \bibinfo{journal}{Phys. Lett. A} \textbf{\bibinfo{volume}{373}}, \bibinfo{pages}{3304} (\bibinfo{year}{2009}), \eprint{0905.3522}.

\bibitem[{\citenamefont{Bender et~al.}(2006)\citenamefont{Bender, Chen, and Milton}}]{Bender_2006}
\bibinfo{author}{\bibfnamefont{C.~M.} \bibnamefont{Bender}}, \bibinfo{author}{\bibfnamefont{J.-H.} \bibnamefont{Chen}}, \bibnamefont{and} \bibinfo{author}{\bibfnamefont{K.~A.} \bibnamefont{Milton}}, \bibinfo{journal}{J. Phys. A} \textbf{\bibinfo{volume}{39}}, \bibinfo{pages}{1657} (\bibinfo{year}{2006}), \eprint{hep-th/0511229}.

\bibitem[{\citenamefont{Bender et~al.}(2005{\natexlab{b}})\citenamefont{Bender, Brandt, Chen, and Wang}}]{R5}
\bibinfo{author}{\bibfnamefont{C.~M.} \bibnamefont{Bender}}, \bibinfo{author}{\bibfnamefont{S.~F.} \bibnamefont{Brandt}}, \bibinfo{author}{\bibfnamefont{J.-H.} \bibnamefont{Chen}}, \bibnamefont{and} \bibinfo{author}{\bibfnamefont{Q.-h.} \bibnamefont{Wang}}, \bibinfo{journal}{Phys. Rev. D} \textbf{\bibinfo{volume}{71}}, \bibinfo{pages}{025014} (\bibinfo{year}{2005}{\natexlab{b}}), \eprint{hep-th/0411064}.

\bibitem[{\citenamefont{Bender et~al.}(2021)\citenamefont{Bender, Felski, Klevansky, and Sarkar}}]{R4}
\bibinfo{author}{\bibfnamefont{C.~M.} \bibnamefont{Bender}}, \bibinfo{author}{\bibfnamefont{A.}~\bibnamefont{Felski}}, \bibinfo{author}{\bibfnamefont{S.~P.} \bibnamefont{Klevansky}}, \bibnamefont{and} \bibinfo{author}{\bibfnamefont{S.}~\bibnamefont{Sarkar}}, \bibinfo{journal}{J. Phys. Conf. Ser.} \textbf{\bibinfo{volume}{2038}}, \bibinfo{pages}{012004} (\bibinfo{year}{2021}), \eprint{2103.14864}.

\bibitem[{\citenamefont{Lee}(1954)}]{R6.1}
\bibinfo{author}{\bibfnamefont{T.~D.} \bibnamefont{Lee}}, \bibinfo{journal}{Phys. Rev.} \textbf{\bibinfo{volume}{95}}, \bibinfo{pages}{1329} (\bibinfo{year}{1954}).

\bibitem[{\citenamefont{Streater and Wightman}(1989)}]{Streater:1989vi}
\bibinfo{author}{\bibfnamefont{R.~F.} \bibnamefont{Streater}} \bibnamefont{and} \bibinfo{author}{\bibfnamefont{A.~S.} \bibnamefont{Wightman}}, \emph{\bibinfo{title}{{PCT, spin and statistics, and all that}}} (\bibinfo{year}{1989}), ISBN \bibinfo{isbn}{978-0-691-07062-9}.

\bibitem[{\citenamefont{Sher}(1989)}]{R7.4}
\bibinfo{author}{\bibfnamefont{M.}~\bibnamefont{Sher}}, \bibinfo{journal}{Phys. Rept.} \textbf{\bibinfo{volume}{179}}, \bibinfo{pages}{273} (\bibinfo{year}{1989}).

\bibitem[{\citenamefont{Isidori et~al.}(2001)\citenamefont{Isidori, Ridolfi, and Strumia}}]{Isidori:2001bm}
\bibinfo{author}{\bibfnamefont{G.}~\bibnamefont{Isidori}}, \bibinfo{author}{\bibfnamefont{G.}~\bibnamefont{Ridolfi}}, \bibnamefont{and} \bibinfo{author}{\bibfnamefont{A.}~\bibnamefont{Strumia}}, \bibinfo{journal}{Nucl. Phys. B} \textbf{\bibinfo{volume}{609}}, \bibinfo{pages}{387} (\bibinfo{year}{2001}), \eprint{hep-ph/0104016}.

\bibitem[{\citenamefont{Behtash et~al.}(2017)\citenamefont{Behtash, Dunne, Sch\"afer, Sulejmanpasic, and \"Unsal}}]{Behtash:2015loa}
\bibinfo{author}{\bibfnamefont{A.}~\bibnamefont{Behtash}}, \bibinfo{author}{\bibfnamefont{G.~V.} \bibnamefont{Dunne}}, \bibinfo{author}{\bibfnamefont{T.}~\bibnamefont{Sch\"afer}}, \bibinfo{author}{\bibfnamefont{T.}~\bibnamefont{Sulejmanpasic}}, \bibnamefont{and} \bibinfo{author}{\bibfnamefont{M.}~\bibnamefont{\"Unsal}}, \bibinfo{journal}{Ann. Math. Sci. Appl.} \textbf{\bibinfo{volume}{02}}, \bibinfo{pages}{95} (\bibinfo{year}{2017}), \eprint{1510.03435}.

\bibitem[{\citenamefont{Witten}(2010)}]{R32w}
\bibinfo{author}{\bibfnamefont{E.}~\bibnamefont{Witten}} (\bibinfo{year}{2010}), \eprint{1009.6032}.

\bibitem[{\citenamefont{Wilson and Kogut}(1974)}]{R11}
\bibinfo{author}{\bibfnamefont{K.~G.} \bibnamefont{Wilson}} \bibnamefont{and} \bibinfo{author}{\bibfnamefont{J.~B.} \bibnamefont{Kogut}}, \bibinfo{journal}{Phys. Rept.} \textbf{\bibinfo{volume}{12}}, \bibinfo{pages}{75} (\bibinfo{year}{1974}).

\bibitem[{\citenamefont{Peskin and Schroeder}(1995)}]{R14a}
\bibinfo{author}{\bibfnamefont{M.~E.} \bibnamefont{Peskin}} \bibnamefont{and} \bibinfo{author}{\bibfnamefont{D.~V.} \bibnamefont{Schroeder}}, \emph{\bibinfo{title}{{An Introduction to Quantum Field Theory}}} (\bibinfo{publisher}{Westview Press}, \bibinfo{year}{1995}), \bibinfo{note}{reading, USA: Addison-Wesley (1995) 842 p}.

\bibitem[{\citenamefont{Kallen and Pauli}(1955)}]{R6.2}
\bibinfo{author}{\bibfnamefont{A.~O.~G.} \bibnamefont{Kallen}} \bibnamefont{and} \bibinfo{author}{\bibfnamefont{W.}~\bibnamefont{Pauli}}, \bibinfo{journal}{Kong. Dan. Vid. Sel. Mat. Fys. Med.} \textbf{\bibinfo{volume}{30}}, \bibinfo{pages}{1} (\bibinfo{year}{1955}).

\bibitem[{\citenamefont{Bjorken and Drell}(1964)}]{R20}
\bibinfo{author}{\bibfnamefont{J.~D.} \bibnamefont{Bjorken}} \bibnamefont{and} \bibinfo{author}{\bibfnamefont{S.~D.} \bibnamefont{Drell}}, \emph{\bibinfo{title}{{Relativistic quantum mechanics}}}, International series in pure and applied physics (\bibinfo{publisher}{McGraw-Hill}, \bibinfo{address}{New York, NY}, \bibinfo{year}{1964}), \urlprefix\url{https://cds.cern.ch/record/100769}.

\bibitem[{\citenamefont{Leibbrandt}(1975)}]{RevModPhys.47.849}
\bibinfo{author}{\bibfnamefont{G.}~\bibnamefont{Leibbrandt}}, \bibinfo{journal}{Rev. Mod. Phys.} \textbf{\bibinfo{volume}{47}}, \bibinfo{pages}{849} (\bibinfo{year}{1975}), \urlprefix\url{https://link.aps.org/doi/10.1103/RevModPhys.47.849}.

\bibitem[{\citenamefont{Breitenlohner and Maison}(1977{\natexlab{a}})}]{Breitenlohner:1977hr}
\bibinfo{author}{\bibfnamefont{P.}~\bibnamefont{Breitenlohner}} \bibnamefont{and} \bibinfo{author}{\bibfnamefont{D.}~\bibnamefont{Maison}}, \bibinfo{journal}{Commun. Math. Phys.} \textbf{\bibinfo{volume}{52}}, \bibinfo{pages}{11} (\bibinfo{year}{1977}{\natexlab{a}}).

\bibitem[{\citenamefont{Bender et~al.}(2004)\citenamefont{Bender, Brod, Refig, and Reuter}}]{Bender:2004zz}
\bibinfo{author}{\bibfnamefont{C.~M.} \bibnamefont{Bender}}, \bibinfo{author}{\bibfnamefont{J.}~\bibnamefont{Brod}}, \bibinfo{author}{\bibfnamefont{A.}~\bibnamefont{Refig}}, \bibnamefont{and} \bibinfo{author}{\bibfnamefont{M.}~\bibnamefont{Reuter}}, \bibinfo{journal}{J. Phys. A} \textbf{\bibinfo{volume}{37}}, \bibinfo{pages}{10139} (\bibinfo{year}{2004}), \eprint{quant-ph/0402026}.

\bibitem[{\citenamefont{Coleman and Weinberg}(1973)}]{Coleman:1973jx}
\bibinfo{author}{\bibfnamefont{S.~R.} \bibnamefont{Coleman}} \bibnamefont{and} \bibinfo{author}{\bibfnamefont{E.~J.} \bibnamefont{Weinberg}}, \bibinfo{journal}{Phys. Rev. D} \textbf{\bibinfo{volume}{7}}, \bibinfo{pages}{1888} (\bibinfo{year}{1973}).

\bibitem[{\citenamefont{Ellis et~al.}(2020)\citenamefont{Ellis, Quevillon, Vuong, You, and Zhang}}]{Ellis:2020ivx}
\bibinfo{author}{\bibfnamefont{S.~A.~R.} \bibnamefont{Ellis}}, \bibinfo{author}{\bibfnamefont{J.}~\bibnamefont{Quevillon}}, \bibinfo{author}{\bibfnamefont{P.~N.~H.} \bibnamefont{Vuong}}, \bibinfo{author}{\bibfnamefont{T.}~\bibnamefont{You}}, \bibnamefont{and} \bibinfo{author}{\bibfnamefont{Z.}~\bibnamefont{Zhang}}, \bibinfo{journal}{JHEP} \textbf{\bibinfo{volume}{11}}, \bibinfo{pages}{078} (\bibinfo{year}{2020}), \eprint{2006.16260}.

\bibitem[{\citenamefont{Manohar and Nardoni}(2021)}]{Manohar:2020nzp}
\bibinfo{author}{\bibfnamefont{A.~V.} \bibnamefont{Manohar}} \bibnamefont{and} \bibinfo{author}{\bibfnamefont{E.}~\bibnamefont{Nardoni}}, \bibinfo{journal}{JHEP} \textbf{\bibinfo{volume}{04}}, \bibinfo{pages}{093} (\bibinfo{year}{2021}), \eprint{2010.15806}.

\bibitem[{\citenamefont{Fainberg and Iofa}(1980)}]{R21}
\bibinfo{author}{\bibfnamefont{V.~Y.} \bibnamefont{Fainberg}} \bibnamefont{and} \bibinfo{author}{\bibfnamefont{M.~Z.} \bibnamefont{Iofa}}, \bibinfo{journal}{Nucl. Phys. B} \textbf{\bibinfo{volume}{168}}, \bibinfo{pages}{495} (\bibinfo{year}{1980}).

\bibitem[{\citenamefont{Hollowood}(2009)}]{Hollowood:2009eh}
\bibinfo{author}{\bibfnamefont{T.~J.} \bibnamefont{Hollowood}}, in \emph{\bibinfo{booktitle}{{38th British Universities Summer School in Theoretical Elementary Particle Physics}}} (\bibinfo{year}{2009}), \eprint{0909.0859}.

\bibitem[{\citenamefont{M\o{}lgaard and Shrock}(2014)}]{R9}
\bibinfo{author}{\bibfnamefont{E.}~\bibnamefont{M\o{}lgaard}} \bibnamefont{and} \bibinfo{author}{\bibfnamefont{R.}~\bibnamefont{Shrock}}, \bibinfo{journal}{Phys. Rev. D} \textbf{\bibinfo{volume}{89}}, \bibinfo{pages}{105007} (\bibinfo{year}{2014}), \eprint{1403.3058}.

\bibitem[{\citenamefont{Longhi and Della~Valle}(2012)}]{PhysRevA.85.012112}
\bibinfo{author}{\bibfnamefont{S.}~\bibnamefont{Longhi}} \bibnamefont{and} \bibinfo{author}{\bibfnamefont{G.}~\bibnamefont{Della~Valle}}, \bibinfo{journal}{Phys. Rev. A} \textbf{\bibinfo{volume}{85}}, \bibinfo{pages}{012112} (\bibinfo{year}{2012}), \urlprefix\url{https://link.aps.org/doi/10.1103/PhysRevA.85.012112}.

\bibitem[{\citenamefont{Witten}(2011)}]{Witten:2010cx}
\bibinfo{author}{\bibfnamefont{E.}~\bibnamefont{Witten}}, \bibinfo{journal}{AMS/IP Stud. Adv. Math.} \textbf{\bibinfo{volume}{50}}, \bibinfo{pages}{347} (\bibinfo{year}{2011}), \eprint{1001.2933}.

\bibitem[{\citenamefont{Callan and Coleman}(1977)}]{R34cc}
\bibinfo{author}{\bibfnamefont{C.~G.} \bibnamefont{Callan}, \bibfnamefont{Jr.}} \bibnamefont{and} \bibinfo{author}{\bibfnamefont{S.~R.} \bibnamefont{Coleman}}, \bibinfo{journal}{Phys. Rev. D} \textbf{\bibinfo{volume}{16}}, \bibinfo{pages}{1762} (\bibinfo{year}{1977}).

\bibitem[{\citenamefont{Coleman}(1977)}]{PhysRevD.15.2929}
\bibinfo{author}{\bibfnamefont{S.}~\bibnamefont{Coleman}}, \bibinfo{journal}{Phys. Rev. D} \textbf{\bibinfo{volume}{15}}, \bibinfo{pages}{2929} (\bibinfo{year}{1977}), \urlprefix\url{https://link.aps.org/doi/10.1103/PhysRevD.15.2929}.

\bibitem[{\citenamefont{Bender and Orszag}(1978)}]{R15}
\bibinfo{author}{\bibfnamefont{C.~M.} \bibnamefont{Bender}} \bibnamefont{and} \bibinfo{author}{\bibfnamefont{S.~A.} \bibnamefont{Orszag}}, \emph{\bibinfo{title}{{Advanced Mathematical Methods for Scientists and Engineers}}} (\bibinfo{publisher}{McGraw-Hill}, \bibinfo{year}{1978}).

\bibitem[{\citenamefont{Di~Pietro and Stamou}(2018)}]{DiPietro:2017vsp}
\bibinfo{author}{\bibfnamefont{L.}~\bibnamefont{Di~Pietro}} \bibnamefont{and} \bibinfo{author}{\bibfnamefont{E.}~\bibnamefont{Stamou}}, \bibinfo{journal}{Phys. Rev. D} \textbf{\bibinfo{volume}{97}}, \bibinfo{pages}{065007} (\bibinfo{year}{2018}), \eprint{1708.03739}.

\bibitem[{\citenamefont{Jegerlehner}(2001)}]{Jegerlehner:2000dz}
\bibinfo{author}{\bibfnamefont{F.}~\bibnamefont{Jegerlehner}}, \bibinfo{journal}{Eur. Phys. J. C} \textbf{\bibinfo{volume}{18}}, \bibinfo{pages}{673} (\bibinfo{year}{2001}), \eprint{hep-th/0005255}.

\bibitem[{\citenamefont{Breitenlohner and Maison}(1977{\natexlab{b}})}]{cmp/1103900439}
\bibinfo{author}{\bibfnamefont{P.}~\bibnamefont{Breitenlohner}} \bibnamefont{and} \bibinfo{author}{\bibfnamefont{D.}~\bibnamefont{Maison}}, \bibinfo{journal}{Communications in Mathematical Physics} \textbf{\bibinfo{volume}{52}}, \bibinfo{pages}{11 } (\bibinfo{year}{1977}{\natexlab{b}}).

\bibitem[{\citenamefont{'t~Hooft and Veltman}(1972)}]{THOOFT1972189}
\bibinfo{author}{\bibfnamefont{G.}~\bibnamefont{'t~Hooft}} \bibnamefont{and} \bibinfo{author}{\bibfnamefont{M.~J.~G.} \bibnamefont{Veltman}}, \bibinfo{journal}{Nucl. Phys. B} \textbf{\bibinfo{volume}{44}}, \bibinfo{pages}{189} (\bibinfo{year}{1972}).

\bibitem[{\citenamefont{Bonneau}(1990)}]{Bonneau:1990xu}
\bibinfo{author}{\bibfnamefont{G.}~\bibnamefont{Bonneau}}, \bibinfo{journal}{Int. J. Mod. Phys. A} \textbf{\bibinfo{volume}{5}}, \bibinfo{pages}{3831} (\bibinfo{year}{1990}).

\bibitem[{\citenamefont{Schubert}(1989)}]{Schubert:1989xu}
\bibinfo{author}{\bibfnamefont{C.}~\bibnamefont{Schubert}}, \bibinfo{journal}{Nucl. Phys. B} \textbf{\bibinfo{volume}{323}}, \bibinfo{pages}{478} (\bibinfo{year}{1989}).

\bibitem[{\citenamefont{Le~Guillou and Zinn-Justin}(1990)}]{LeGuillou:1990nq}
\bibinfo{editor}{\bibfnamefont{J.~C.} \bibnamefont{Le~Guillou}} \bibnamefont{and} \bibinfo{editor}{\bibfnamefont{J.}~\bibnamefont{Zinn-Justin}}, eds., \emph{\bibinfo{title}{{Large order behavior of perturbation theory}}} (\bibinfo{year}{1990}).

\bibitem[{\citenamefont{Dyson}(1952{\natexlab{a}})}]{PhysRev.85.631}
\bibinfo{author}{\bibfnamefont{F.~J.} \bibnamefont{Dyson}}, \bibinfo{journal}{Phys. Rev.} \textbf{\bibinfo{volume}{85}}, \bibinfo{pages}{631} (\bibinfo{year}{1952}{\natexlab{a}}), \urlprefix\url{https://link.aps.org/doi/10.1103/PhysRev.85.631}.

\bibitem[{\citenamefont{Dunne}(2002)}]{Dunne:2002rq}
\bibinfo{author}{\bibfnamefont{G.~V.} \bibnamefont{Dunne}}, in \emph{\bibinfo{booktitle}{{Continuous Advances in QCD 2002 / ARKADYFEST (honoring the 60th birthday of Prof. Arkady Vainshtein)}}} (\bibinfo{year}{2002}), pp. \bibinfo{pages}{478--505}, \eprint{hep-th/0207046}.

\bibitem[{\citenamefont{Bender and Wu}(1969)}]{Bender:1969si}
\bibinfo{author}{\bibfnamefont{C.~M.} \bibnamefont{Bender}} \bibnamefont{and} \bibinfo{author}{\bibfnamefont{T.~T.} \bibnamefont{Wu}}, \bibinfo{journal}{Phys. Rev.} \textbf{\bibinfo{volume}{184}}, \bibinfo{pages}{1231} (\bibinfo{year}{1969}).

\bibitem[{\citenamefont{Simon and Dicke}(1970)}]{Simon:1970mc}
\bibinfo{author}{\bibfnamefont{B.}~\bibnamefont{Simon}} \bibnamefont{and} \bibinfo{author}{\bibfnamefont{A.}~\bibnamefont{Dicke}}, \bibinfo{journal}{Annals Phys.} \textbf{\bibinfo{volume}{58}}, \bibinfo{pages}{76} (\bibinfo{year}{1970}).

\bibitem[{\citenamefont{Simon}(1991)}]{Simon1991FiftyYO}
\bibinfo{author}{\bibfnamefont{B.}~\bibnamefont{Simon}}, \bibinfo{journal}{Bulletin of the American Mathematical Society} \textbf{\bibinfo{volume}{24}}, \bibinfo{pages}{303} (\bibinfo{year}{1991}).

\bibitem[{\citenamefont{Glendinning}(1994)}]{R12a}
\bibinfo{author}{\bibfnamefont{P.}~\bibnamefont{Glendinning}}, \emph{\bibinfo{title}{Stability, Instability and Chaos: An Introduction to the Theory of Nonlinear Differential Equations}}, Cambridge Texts in Applied Mathematics (\bibinfo{publisher}{Cambridge University Press}, \bibinfo{year}{1994}).

\bibitem[{\citenamefont{Hubbard and West}(2013)}]{R12b}
\bibinfo{author}{\bibfnamefont{J.~H.} \bibnamefont{Hubbard}} \bibnamefont{and} \bibinfo{author}{\bibfnamefont{B.~H.} \bibnamefont{West}}, \emph{\bibinfo{title}{Differential Equations: A Dynamical Systems Approach}}, Texts in Applied Mathematics (\bibinfo{publisher}{Springer New York, NY}, \bibinfo{year}{2013}).

\bibitem[{\citenamefont{Degrassi et~al.}(2012)\citenamefont{Degrassi, Di~Vita, Elias-Miro, Espinosa, Giudice, Isidori, and Strumia}}]{Degrassi:2012ry}
\bibinfo{author}{\bibfnamefont{G.}~\bibnamefont{Degrassi}}, \bibinfo{author}{\bibfnamefont{S.}~\bibnamefont{Di~Vita}}, \bibinfo{author}{\bibfnamefont{J.}~\bibnamefont{Elias-Miro}}, \bibinfo{author}{\bibfnamefont{J.~R.} \bibnamefont{Espinosa}}, \bibinfo{author}{\bibfnamefont{G.~F.} \bibnamefont{Giudice}}, \bibinfo{author}{\bibfnamefont{G.}~\bibnamefont{Isidori}}, \bibnamefont{and} \bibinfo{author}{\bibfnamefont{A.}~\bibnamefont{Strumia}}, \bibinfo{journal}{JHEP} \textbf{\bibinfo{volume}{08}}, \bibinfo{pages}{098} (\bibinfo{year}{2012}), \eprint{1205.6497}.

\bibitem[{\citenamefont{Bender et~al.}(2016)\citenamefont{Bender, Hook, Mavromatos, and Sarkar}}]{R8}
\bibinfo{author}{\bibfnamefont{C.~M.} \bibnamefont{Bender}}, \bibinfo{author}{\bibfnamefont{D.~W.} \bibnamefont{Hook}}, \bibinfo{author}{\bibfnamefont{N.~E.} \bibnamefont{Mavromatos}}, \bibnamefont{and} \bibinfo{author}{\bibfnamefont{S.}~\bibnamefont{Sarkar}}, \bibinfo{journal}{J. Phys. A} \textbf{\bibinfo{volume}{49}}, \bibinfo{pages}{45LT01} (\bibinfo{year}{2016}), \eprint{1506.01970}.

\bibitem[{\citenamefont{De~Cesare et~al.}(2021)\citenamefont{De~Cesare, Di~Pietro, and Serone}}]{R13a}
\bibinfo{author}{\bibfnamefont{F.}~\bibnamefont{De~Cesare}}, \bibinfo{author}{\bibfnamefont{L.}~\bibnamefont{Di~Pietro}}, \bibnamefont{and} \bibinfo{author}{\bibfnamefont{M.}~\bibnamefont{Serone}}, \bibinfo{journal}{Phys. Rev. D} \textbf{\bibinfo{volume}{104}}, \bibinfo{pages}{105015} (\bibinfo{year}{2021}), \eprint{2107.00342}.

\bibitem[{\citenamefont{Fei et~al.}(2016)\citenamefont{Fei, Giombi, Klebanov, and Tarnopolsky}}]{Fei:2016sgs}
\bibinfo{author}{\bibfnamefont{L.}~\bibnamefont{Fei}}, \bibinfo{author}{\bibfnamefont{S.}~\bibnamefont{Giombi}}, \bibinfo{author}{\bibfnamefont{I.~R.} \bibnamefont{Klebanov}}, \bibnamefont{and} \bibinfo{author}{\bibfnamefont{G.}~\bibnamefont{Tarnopolsky}}, \bibinfo{journal}{PTEP} \textbf{\bibinfo{volume}{2016}}, \bibinfo{pages}{12C105} (\bibinfo{year}{2016}), \eprint{1607.05316}.

\bibitem[{\citenamefont{Herbut}(2023)}]{Herbut:2023xgz}
\bibinfo{author}{\bibfnamefont{I.~F.} \bibnamefont{Herbut}} (\bibinfo{year}{2023}), \eprint{2304.07654}.

\bibitem[{\citenamefont{Esaki et~al.}(2011)\citenamefont{Esaki, Sato, Hasebe, and Kohmoto}}]{PhysRevB.84.205128}
\bibinfo{author}{\bibfnamefont{K.}~\bibnamefont{Esaki}}, \bibinfo{author}{\bibfnamefont{M.}~\bibnamefont{Sato}}, \bibinfo{author}{\bibfnamefont{K.}~\bibnamefont{Hasebe}}, \bibnamefont{and} \bibinfo{author}{\bibfnamefont{M.}~\bibnamefont{Kohmoto}}, \bibinfo{journal}{Phys. Rev. B} \textbf{\bibinfo{volume}{84}}, \bibinfo{pages}{205128} (\bibinfo{year}{2011}), \urlprefix\url{https://link.aps.org/doi/10.1103/PhysRevB.84.205128}.

\bibitem[{\citenamefont{Amit}(1984)}]{amit84a}
\bibinfo{author}{\bibfnamefont{D.}~\bibnamefont{Amit}}, \emph{\bibinfo{title}{Field Theory, the renormalization Group, and Critical Phenomena}} (\bibinfo{publisher}{World Scientific}, \bibinfo{year}{1984}), \bibinfo{note}{iSBN: 9971-966-10-7;9971-966-11-5}.

\bibitem[{\citenamefont{Zinn-Justin}(2021)}]{zinn2021quantum}
\bibinfo{author}{\bibfnamefont{J.}~\bibnamefont{Zinn-Justin}}, \emph{\bibinfo{title}{Quantum field theory and critical phenomena}}, vol. \bibinfo{volume}{171} (\bibinfo{publisher}{Oxford university press}, \bibinfo{year}{2021}).

\bibitem[{\citenamefont{Weinberg}(1976)}]{Weinberg:1976xy}
\bibinfo{author}{\bibfnamefont{S.}~\bibnamefont{Weinberg}}, in \emph{\bibinfo{booktitle}{{14th International School of Subnuclear Physics: Understanding the Fundamental Constitutents of Matter}}} (\bibinfo{year}{1976}).

\bibitem[{\citenamefont{Dyson}(1952{\natexlab{b}})}]{Dyson:1952tj}
\bibinfo{author}{\bibfnamefont{F.~J.} \bibnamefont{Dyson}}, \bibinfo{journal}{Phys. Rev.} \textbf{\bibinfo{volume}{85}}, \bibinfo{pages}{631} (\bibinfo{year}{1952}{\natexlab{b}}).

\bibitem[{\citenamefont{Coleman}(1985)}]{Coleman:1985rnk}
\bibinfo{author}{\bibfnamefont{S.}~\bibnamefont{Coleman}}, \emph{\bibinfo{title}{{Aspects of Symmetry}: {Selected Erice Lectures}}} (\bibinfo{publisher}{Cambridge University Press}, \bibinfo{address}{Cambridge, U.K.}, \bibinfo{year}{1985}), ISBN \bibinfo{isbn}{978-0-521-31827-3}.

\bibitem[{\citenamefont{Lipatov}(1977)}]{Lipatov:1976ny}
\bibinfo{author}{\bibfnamefont{L.~N.} \bibnamefont{Lipatov}}, \bibinfo{journal}{Sov. Phys. JETP} \textbf{\bibinfo{volume}{45}}, \bibinfo{pages}{216} (\bibinfo{year}{1977}).

\bibitem[{\citenamefont{de~Cesare et~al.}(2015)\citenamefont{de~Cesare, Mavromatos, and Sarkar}}]{deCesare:2014dga}
\bibinfo{author}{\bibfnamefont{M.}~\bibnamefont{de~Cesare}}, \bibinfo{author}{\bibfnamefont{N.~E.} \bibnamefont{Mavromatos}}, \bibnamefont{and} \bibinfo{author}{\bibfnamefont{S.}~\bibnamefont{Sarkar}}, \bibinfo{journal}{Eur. Phys. J. C} \textbf{\bibinfo{volume}{75}}, \bibinfo{pages}{514} (\bibinfo{year}{2015}), \eprint{1412.7077}.

\bibitem[{\citenamefont{Bossingham et~al.}(2018)\citenamefont{Bossingham, Mavromatos, and Sarkar}}]{R10Boss}
\bibinfo{author}{\bibfnamefont{T.}~\bibnamefont{Bossingham}}, \bibinfo{author}{\bibfnamefont{N.~E.} \bibnamefont{Mavromatos}}, \bibnamefont{and} \bibinfo{author}{\bibfnamefont{S.}~\bibnamefont{Sarkar}}, \bibinfo{journal}{Eur. Phys. J. C} \textbf{\bibinfo{volume}{78}}, \bibinfo{pages}{113} (\bibinfo{year}{2018}), \eprint{1712.03312}.

\bibitem[{\citenamefont{Bossingham et~al.}(2019)\citenamefont{Bossingham, Mavromatos, and Sarkar}}]{R11Boss}
\bibinfo{author}{\bibfnamefont{T.}~\bibnamefont{Bossingham}}, \bibinfo{author}{\bibfnamefont{N.~E.} \bibnamefont{Mavromatos}}, \bibnamefont{and} \bibinfo{author}{\bibfnamefont{S.}~\bibnamefont{Sarkar}}, \bibinfo{journal}{Eur. Phys. J. C} \textbf{\bibinfo{volume}{79}}, \bibinfo{pages}{50} (\bibinfo{year}{2019}), \eprint{1810.13384}.

\bibitem[{\citenamefont{Sarkar}(2022)}]{Sarkar:2022odh}
\bibinfo{author}{\bibfnamefont{S.}~\bibnamefont{Sarkar}}, \bibinfo{journal}{PoS} \textbf{\bibinfo{volume}{DISCRETE2020-2021}}, \bibinfo{pages}{039} (\bibinfo{year}{2022}), \eprint{2206.05203}.

\bibitem[{\citenamefont{Mavromatos and sarkar}(2023)}]{Mavromatos:2023bdx}
\bibinfo{author}{\bibfnamefont{N.~E.} \bibnamefont{Mavromatos}} \bibnamefont{and} \bibinfo{author}{\bibfnamefont{S.}~\bibnamefont{sarkar}}, \bibinfo{journal}{Eur. Phys. J. C} \textbf{\bibinfo{volume}{83}}, \bibinfo{pages}{866} (\bibinfo{year}{2023}), \eprint{2306.02122}.

\bibitem[{\citenamefont{Green et~al.}(2012{\natexlab{a}})\citenamefont{Green, Schwarz, and Witten}}]{Green:2012oqa}
\bibinfo{author}{\bibfnamefont{M.~B.} \bibnamefont{Green}}, \bibinfo{author}{\bibfnamefont{J.~H.} \bibnamefont{Schwarz}}, \bibnamefont{and} \bibinfo{author}{\bibfnamefont{E.}~\bibnamefont{Witten}}, \emph{\bibinfo{title}{{Superstring Theory Vol. 1}: {25th Anniversary Edition}}}, Cambridge Monographs on Mathematical Physics (\bibinfo{publisher}{Cambridge University Press}, \bibinfo{year}{2012}{\natexlab{a}}), ISBN \bibinfo{isbn}{978-1-139-53477-2, 978-1-107-02911-8}.

\bibitem[{\citenamefont{Green et~al.}(2012{\natexlab{b}})\citenamefont{Green, Schwarz, and Witten}}]{Green:2012pqa}
\bibinfo{author}{\bibfnamefont{M.~B.} \bibnamefont{Green}}, \bibinfo{author}{\bibfnamefont{J.~H.} \bibnamefont{Schwarz}}, \bibnamefont{and} \bibinfo{author}{\bibfnamefont{E.}~\bibnamefont{Witten}}, \emph{\bibinfo{title}{{Superstring Theory Vol. 2}: {25th Anniversary Edition}}}, Cambridge Monographs on Mathematical Physics (\bibinfo{publisher}{Cambridge University Press}, \bibinfo{year}{2012}{\natexlab{b}}), ISBN \bibinfo{isbn}{978-1-139-53478-9, 978-1-107-02913-2}.

\bibitem[{\citenamefont{Gross and Sloan}(1987)}]{Gross:1986mw}
\bibinfo{author}{\bibfnamefont{D.~J.} \bibnamefont{Gross}} \bibnamefont{and} \bibinfo{author}{\bibfnamefont{J.~H.} \bibnamefont{Sloan}}, \bibinfo{journal}{Nucl. Phys. B} \textbf{\bibinfo{volume}{291}}, \bibinfo{pages}{41} (\bibinfo{year}{1987}).

\bibitem[{\citenamefont{Metsaev and Tseytlin}(1987)}]{Metsaev:1987zx}
\bibinfo{author}{\bibfnamefont{R.~R.} \bibnamefont{Metsaev}} \bibnamefont{and} \bibinfo{author}{\bibfnamefont{A.~A.} \bibnamefont{Tseytlin}}, \bibinfo{journal}{Nucl. Phys. B} \textbf{\bibinfo{volume}{293}}, \bibinfo{pages}{385} (\bibinfo{year}{1987}).

\bibitem[{\citenamefont{Giddings and Strominger}(1988)}]{Giddings:1987cg}
\bibinfo{author}{\bibfnamefont{S.~B.} \bibnamefont{Giddings}} \bibnamefont{and} \bibinfo{author}{\bibfnamefont{A.}~\bibnamefont{Strominger}}, \bibinfo{journal}{Nucl. Phys. B} \textbf{\bibinfo{volume}{306}}, \bibinfo{pages}{890} (\bibinfo{year}{1988}).

\end{thebibliography}

\end{document}